\documentclass[10pt,oneside]{amsart}                
\usepackage{amsfonts,amssymb,latexsym,xspace,pifont,calc,amsmath,
            enumerate,epic,pstricks}
\numberwithin{equation}{subsection}

\makeatletter
\def\draft{\textheight=10.5truein \textwidth=7.5truein \parindent=8pt
           \voffset=-1truein \topmargin=0Truein
           \ifcase \@ptsize \hoffset=-1.5truein \or \hoffset=-1.35truein
                        \or \hoffset=-1.15truein \fi}
\def\quality{\textheight=240mm \textwidth=150mm \topmargin=0Truein
             \ifcase \@ptsize \hoffset=-23mm
                     \or \hoffset=-20mm \or \hoffset=-15mm \fi}
\makeatother \quality
\newcommand\myparwidth{\textwidth-2cm}

\newtheorem{Thm}{Theorem}
\newtheorem{lem}{Lemma}[subsection]

\newtheorem{sublem}[lem]{Sub-lemma}
\newtheorem{prop}[lem]{Proposition}

\newtheorem{defin}[lem]{Definition}

\newtheorem{exmp}[lem]{Example}
\newtheorem{rem}[lem]{Remark}
\newtheorem{alem}{Lemma}[section]
\newtheorem{arem}[alem]{Remark}
\newtheorem{asublem}[alem]{Sub-Lemma}
\newcommand\Cal[1]{{\mathcal #1}}
\newcommand\B{{\Cal B}}
\newcommand\Co{{\Cal C}}
\newcommand\D{{\Cal D}}
\newcommand\F{{\Cal F}}
\newcommand\Lp{{\Cal L}}
\newcommand\M{{\Cal M}}

\newcommand\T{{\Cal T}}

\newcommand\V{{\Cal V}}
\newcommand\W{{\Cal W}}
\newcommand\Ac{{\mathcal A}}

\def\BV{{\rm BV}}
\newcommand\A{{\mathbb A}}
\newcommand\E{{\mathbb E}}
\newcommand\N{{\mathbb N}}
\newcommand\R{{\mathbb R}}
\newcommand\Z{{\mathbb Z}}
\newcommand\C{{\mathbb C}}
\newcommand\To{{\mathbb T}}
\newcommand{\Amatrix}{\A}
\newcommand{\Vmat}{{\mathbb V}}
\newcommand\vf{\varphi}
\newcommand\nn{\nonumber}
\newcommand\ve{\varepsilon}

\newcommand{\spectr}{\ensuremath{\text{sp}}} 

\newcommand\CoT{{\Co^{3}}}
\newcommand\CoJ{{\Co^{1+\beta}}}
\newcommand{\cnorm}[1]{\|#1\|_{\Co^1\to(\Co^1)^*}}
\newcommand{\newnorm}[1]{\|#1\|^{\raisebox{0.4ex}[0cm][0cm]{-}}}
\newcommand{\tildenorm}[1]{\|#1\|^{{\sim}}}
\newcommand{\manif}{\M}
\newcommand{\measures}{M}
\newcommand{\loc}{{\mbox{{\scriptsize loc}}}}
\newcommand{\rank}{\operatorname{rank}}
\newcommand{\sign}{\operatorname{sign}}
\newcommand{\tr}{\operatorname{trace}}
\newcommand{\divg}{\operatorname{div}}
\newcommand{\diam}{\operatorname{diam}}
\newcommand{\supp}{\operatorname{supp}}
\newcommand{\dist}{\operatorname{dist}}
\newcommand{\const}{\operatorname{const}}
\newcommand{\vol}{\operatorname{vol}}
\newcommand{\Id}{\operatorname{Id}}
\newcommand{\eins}{\operatorname{{\bf 1}}}
\newcommand{\Eins}{\operatorname{{\bf 1}}}
\newcommand{\limn}{\lim_{n\to\infty}}
\newcommand{\eg}{\emph{e.g.}\xspace}
\newcommand{\ie}{\emph{i.e.}\xspace}
\newcommand{\cf}{\emph{cf.}\xspace}
\newcommand{\etc}{\emph{etc.}\xspace}

\def\mlbscale{1pt} 
\def\bline(#1,#2)(#3,#4)(#5){\put(#1,#2){\line(#3,#4){#5}}}

\def\Bfig(#1,#2)#3#4{\begin{figure} \begin{center}
    \setlength{\unitlength}{\mlbscale} \begin{picture}(#1,#2)#3
    \end{picture} \end{center} \caption{ }#4 \end{figure}}

\def\bpic(#1,#2)#3{\setlength{\unitlength}{\mlbscale}
    \begin{picture}(#1,#2) #3 \end{picture}}

\def\bline(#1,#2)(#3,#4)(#5){\put(#1,#2){\line(#3,#4){#5}}}

\begin{document}

\title{Ruelle-Perron-Frobenius spectrum for Anosov maps}
\author{Michael Blank}
\author{Gerhard Keller}
\author{Carlangelo Liverani}
\address{Michael Blank\\
Institute for Information Transmission Problems,
B.Karetnij Per. 19, Moscow, Russia}
\email{{\tt blank@obs-nice.fr}}
\address{Gerhard Keller\\
Mathematisches Institut\\
Universit\"at Erlangen\\
Bismarckstr. $1\frac 12$, 91054 Erlangen, Germany}
\email{{\tt keller@mi.uni-erlangen.de}}
\address{Carlangelo Liverani\\
Dipartimento di Matematica\\
Universit\`{a} di Roma ``Tor Vergata''\\
Via della Ricerca Scientifica, 00133 Roma, Italy}
\email{{\tt liverani@mat.uniroma2.it}}
\thanks{We acknowledge the partial support of the ESF Programme
PRODYN and the DFG priority research program DANSE. 
The work of M.B. was partially supported by DFG, CRDF and RFBR grants. 
We wish to thank V.Baladi, M.Benedicks, U.Bessi, B.Hasselblatt, A.Katok, 
F.Ledrappier and J.Schmeling for helpful discussions. 
G.K. and C.L. acknowledge the hospitality of the I.H.E.S. where this 
paper was finished and thank D.Ruelle for making this possible.}
\date{March 2001}
\begin{abstract}
We extend a number of results from one dimensional dynamics based
on spectral properties of the Ruelle-Perron-Frobenius transfer 
operator to Anosov diffeomorphisms on compact manifolds. This allows 
to develop a direct operator approach to study ergodic properties of 
these maps. In particular, we show that it is possible to define 
Banach spaces on which the transfer operator is quasicompact.
(Information on the existence of an SRB measure, its smoothness
properties and statistical properties readily follow from such a
result.) In dimension $d=2$ we show that the transfer operator
associated to smooth random perturbations of the map is close, in
a proper sense, to the unperturbed transfer operator. This allows
to obtain easily very strong spectral stability results, which in 
turn imply spectral stability results for smooth deterministic 
perturbations as well. Finally, we are able to implement an Ulam 
type finite rank approximation scheme thus reducing the study of 
the spectral properties of the transfer operator to a finite 
dimensional problem.
\end{abstract}
\maketitle

\section{Introduction}
The aim of the present paper is to extend a number of results
from one dimensional dynamics based on spectral properties of the
Ruelle-Perron-Frobenius operator (transfer operator, for short)
to $\CoT$ Anosov diffeomorphisms on compact manifolds, which allows 
to develop a direct, and very powerful, operator approach to study 
quantitative ergodic properties of these maps.

Since the spectrum of an operator depends heavily on the space on 
which it is defined, the first task is to identify a relevant function 
space. Note that on function spaces where the norm measures essentially 
the size of the functions (like $\Co^0$ or $L^1$), the spectrum of 
a transfer operator carries only little dynamically relevant 
information: The $L^1$-spectrum is always contained in $\{|z|=1\}$, 
and for other $L^p$-spaces as well as for $\Co^0$ the same is true if 
the map is volume preserving.\footnote{\label{footnote:koopman}
Note, however, that the Koopman operator (which in the volume preserving 
case is the adjoint of the transfer operator on $L^2$) was studied 
thoroughly, and some qualitative dynamic information like ergodicity, 
weak mixing and mixing can be read off its spectral measure, see 
\eg \cite{FKS}.}
Spaces of more regular functions (like $\Co^1$ or $W^{1,1}$), though they
are well adapted to the dynamics of Anosov diffeomorphisms in unstable
directions, are unsuitable to capture the dynamics in stable directions.
In fact they would lead to unreasonable spectral radii larger than one.
These observations suggest to use strongly anisotropic Banach spaces of 
generalised functions (where the anisotropy is spatially inhomogeneous 
as soon as the diffeomorphism is nonlinear). Accordingly, our basic result is:
\begin{enumerate}
\item[(1)] We define inhomogeniously anisotropic Banach spaces
  of generalised functions on which the transfer operator is quasicompact 
  and give an explicit bound for the essential spectral radius of the 
  transfer operator in terms of expansion and contraction rates of the 
  diffeomorphism. (Information on the existence of an SRB measure, its 
  smoothness properties and its statistical properties readily follow 
  from such a result.)
\end{enumerate}
In dimension $d=2$ we also prove the following spectral perturbation results:
\begin{enumerate}
\item[(2)] Transfer operators associated to certain random perturbations 
  of the map are close, in a proper sense, to the unperturbed transfer 
  operator. This allows to obtain easily very strong spectral stability 
  results which imply, in particular, stability of the invariant measure, 
  of the rate of mixing, of the variance in the central limit theorem, \etc
\item[(3)] Transfer operators associated to maps that are close to each 
  other in the $\Co^{2}$ topology have ``close'' spectrum, although they
  are defined on different Banach spaces.
\item[(4)] It is possible to use an Ulam-like
  discretisation scheme that reduces the computation of the spectral
  data of the transfer operator to the study of a finite dimensional
  matrix. This allows to obtain precise quantitative information on
  the ergodic properties of a given map (e.g. ergodicity, invariant
  measures, mixing, rate of mixing etc.) and it is a partial, but rather
  satisfactory, solution to the so called {\em Ulam Conjecture}.
\end{enumerate}
The randomly perturbed operators from (2) and also the finite rank 
approximations from (4) are compact operators that can be studied as well 
on well known function spaces like $L^p$, $\Co^0$, $\Co^1$. On all these 
spaces they have the same pure point spectrum, and a way to interpret (2) 
and (4) is to say that the \emph{actually observed} (either in real 
systems where noise is always present or in numerical approximation schemes)
dynamically relevant spectrum consists of eigenvalues of finite multiplicity 
and is rather independent of the employed function space.

The precise choice of the Banach spaces of generalised functions is 
suggested by two observations: Restricted to an unstable fibre the
diffeomorphism acts like a (piecewise) expanding map. Therefore the
generalised densities should behave like functions of bounded
variation along such fibres. On the other hand, the action of the map 
on stable fibres is similar to the action of a contractive map. Thus 
the corresponding Banach space should be equipped with a norm related 
to the weak convergence of measures. This is the starting point of our 
construction and according to this idea the norm (\ref{eq:norms}) 
consists of two parts -- one related to the properties `along' the 
unstable foliation and another one related to the stable foliation. 
As a result the corresponding Banach space heavily depends on the 
structure of these foliations. This is one of the weaknesses of the 
approach, because generically different maps lead to different Banach 
spaces. Nevertheless, it is still possible to ``compare'' the 
corresponding transfer operators by means of a random smoothing, as we 
will see in the discussion of deterministic perturbations. 


The Banach spaces introduced are in fact indirectly related to the 
{\em cone method} used to investigate rates of mixing in \cite{Li}.
It is worth mentioning that the idea to study directly spectral
properties of smooth hyperbolic systems was proposed earlier by
Bakhtin \cite{Bachtin} in the analysis of space averagings and
also by Rugh \cite{Rugh} and Kitaev \cite{Kitaev} in their
investigation of dynamical (Ruelle) zeta-functions. On the other
hand, the analysis of stochastic stability of spectral properties
was earlier done only in the case of one-dimensional piecewise
expanding maps with convolution-like random perturbations by
Baladi and Young \cite{BaYo} and by Blank, Keller and Liverani
for more general classes of perturbations and maps in
\cite{BlKe,KL,Liverani-paderborn}. 
A more specific problem about the stochastic stability of the leading 
eigenfunction -- the density of the SRB measure -- was firstly studied 
by Kifer \cite{Kifer1} in the case of Gaussian-like random 
perturbations of smooth hyperbolic maps.
\medskip\\
\textbf{Outline of the work.}\quad
We start by discussing $\CoT$ Anosov maps on the $d$-dimensional torus. 
In this context we \emph{define a proper Banach space} in 
section~\ref{subsec:Banach-spaces} and prove 
\emph{quasicompactness of the transfer operator},
see section~\ref{subsec:quasicompactness}. The restriction to $\To^d$ is 
motivated only by our choice to present the method in the clearest form 
and not obscured by technical problems. Along the discussion we point 
out the trivial (for people acquainted with the basic facts of
differential geometry) modifications needed to extend the setting
to arbitrary compact Riemannian manifolds. The restriction to $\CoT$ 
Anosov maps could be slightly relaxed by requiring that $T$ and also 
its Jacobian are of class $\Co^2$. At present we do not know how to 
avoid the additional assumption on the Jacobian. We illustrate the 
general quasicompactness result by a
\emph{simple example}: We give an independent elementary proof of the 
fact that transfer operators of hyperbolic linear automorphisms of 
$\To^2$ have no eigenvalue (other than $1$) of modulus bigger than 
the contracting eigenvalue of the defining matrix.

The quasicompactness of the transfer operator allows to derive rather
directly the existence and (well known) essential 
\emph{properties of the Sinai-Ruelle-Bowen measure} of $T$, see
section~\ref{sec:SRB}.

In section~\ref{subsec:stability-smooth} we consider
\emph{small random perturbations of Anosov maps} of the
two-dimensional torus $\To^2$. Here, as above, the choice of the torus
rather than any two dimensional compact Riemannian manifold is
due to our effort not to cloud the ideas presented by unnecessary
technical details (yet, again, we will comment on how to treat
the more general setting). On the contrary, the restriction to
two dimensions is of a substantial nature. The reason is that in
two dimensions the stable and unstable distributions are known to
be smooth (at least $\Co^{1+\alpha}$) while in higher dimensions
distributions can be only H\"older, depending on the extreme
expansion rates in the stable and unstable bundle of the tangent
space (see \cite{KH} or the appendix for precise references). It is 
not hard to see that, if the distributions are smooth, then our results 
can easily be extended. For non smooth distributions the situation is
more complex. It is possible that such strong stability results
are unreasonable in this case and that only weaker results can be
obtained. In particular, the spectrum could be stable only far
away from the essential spectrum.

In section~\ref{subsec:deterministic-perturbations} we investigate 
\emph{smooth deterministic perturbations of Anosov maps} on $\To^2$. 
Here the problem is that the transfer operators are defined on 
different Banach spaces, hence no direct comparison is possible. 
The key to the comparison is given by the fact that the random 
perturbations are compact, hence their spectrum is rather robust 
when changing the underlying space. Therefore the strategy is to 
compare the spectrum of random perturbations (that can be viewed as
operators on the same space) and then to use the previous results to
conclude the argument.

\relax Finally, we turn to \emph{Ulam finite rank approximation schemes} 
for the transfer operator $\Lp$ associated with an Anosov map on $\To^2$. 
Here the first problem is that the original Ulam scheme consists in 
taking a conditional expectation with respect to some partition of the 
manifold. So the natural space of functions to consider must contain 
functions with discontinuities along unstable fibres, unlike the 
functions belonging to our Banach space. It turns out, however, that 
this is not the main problem. Indeed, in section \ref{subsec:remarks} 
we discuss an extension of our space which also contains piecewise 
constant functions and on which the transfer operator still has 
basically the same spectral properties as on the original space.
Nevertheless it can happen that the discretised operator of the
original Ulam scheme has ``large'' eigenvalues quite far from the
spectrum of the given transfer operator. An example of this pathology
is presented in section~\ref{subsec:Ulam}. The problem can be
overcome, similarly as in the previous section, by introducing 
additional smooth random perturbations (compare to the regularisation 
by random perturbations in \cite{Blank,DJ1,Kifer2}).    

In order to facilitate ``navigation'' in this paper, we provide a
detailed list of contents:
\tableofcontents
\vspace*{-1cm}\quad\\
\textbf{Open questions.}\quad
This article does not attempt to exhaust the possibilities of this 
new method but it rather exposes the method in some simple but non 
trivial cases. It seems clear to us that a considerable amount of 
work is still necessary to understand all the consequences of this 
approach. To extend the spectral stability results to higher 
dimensional systems and hence to cases where the stable and unstable 
foliation are only H\"older continuous is a first task. Then 
it would be interesting to investigate systems with discontinuities 
and different types of perturbations (e.g., pure Ulam approximation 
schemes). In addition, our Banach spaces do not have the amount of 
symmetry between the stable and the unstable direction that would 
seem natural. Also, we cannot take advantage of the higher smoothness 
of the map (in the spirit of Kitaev's work \cite{Kitaev}) and this 
certainly requires a deeper understanding.
\relax Finally, we would like to mention that, having at disposal 
a well defined operator, the question of the relation between its
isolated eigenvalues and the poles of the dynamical (Ruelle) zeta
function becomes quite a natural one.

\section{Statements and results}

\subsection{Banach spaces of generalised functions}
\label{subsec:Banach-spaces}

We will consider Anosov diffeomorphisms
\begin{quote}
$T:\manif\to\manif$ where $T$ is of class $\CoT$ and $\manif$ is 
a smooth compact Riemannian manifold. 
\end{quote}
Let us remind that by Anosov we mean (as usual) that there exist 
a direct sum decomposition of the tangent bundle $\T\manif$ into 
continuous sub-bundles $E^s$ and $E^u$, that is 
$\T_x\manif=E_x^s\oplus E_x^u$,
and constants $A\geq1$ and $0<\lambda_s<1<\lambda_u$ such
that\footnote{By $dT$ we denote the differential of $T$, clearly
$d_xT:\T_x\manif\to \T_{Tx}\manif$. Similarly, if $f:\manif\to\R$ is
differentiable, then $d_xf: \T_x\manif\to \R$.}
\begin{equation}
  \begin{split}
  \label{eq:hyperbolicity}
    (d_xT)(E_x^s)
    &=E_{Tx}^s,\quad\|(d_xT^n)_{|E_x^s}\|\leq A\lambda_s^n,\\
    (d_xT)(E_x^u)
    &=E_{Tx}^u,\quad\|(d_xT^{-n})_{|E_x^u}\|\leq A\lambda_u^{-n},
  \end{split}
\end{equation}
for all $x\in\manif$ and $n\geq0$. Let $d_{u/s}=\text{dim}(E^{u/s})$.
It is well known that for such maps there exist stable and unstable 
foliations $(W^s(x))_{x\in\manif}$ and $(W^u(x))_{x\in\manif}$.
Each single $W^{s/u}(x)$ is an immersed $\CoT$ sub-manifold of
$\manif$, and $\T_yW^{s/u}(x)=E_x^{s/u}$ for any $y\in W^{s/u}(x)$.
The dependence of $E^{s/u}(x)$ and $W^{s/u}(x)$ on $x$, however,
is only H\"older in general--see the appendix for further precise
information on the relevant properties used throughout the paper. 
We will denote by $\tau$ the optimal common H\"older-exponent for
both distributions. This exponent depends in a well understood
way on various contraction and expansion coefficients of the
transformation, and there are a number of cases where the
foliations are indeed $\Co^{1+\alpha}$ for some $\alpha>0$.

Some of the results of this paper are proved only under this stronger
$\Co^{1+\alpha}$-assumption. However, in order to make transparent 
where in the proofs we really use the $\Co^{1+\alpha}$-property, we 
will argue with the $\Co^\tau$-property with $0<\tau\leq1$ whenever 
this is possible without any trouble, and use the 
$\Co^{1+\alpha}$-property with $0<\alpha\leq1$ only in more delicate 
estimates. (This includes to set $\tau=1$ if necessary.) We believe 
that, with considerably more effort, one could weaken the
$\Co^{1+\alpha}$-assumption.

The map $T$ induces naturally a map on the Borel measures on $\manif$ 
defined by
\[
T^*\mu(f):=\mu(f\circ T)\quad \forall \mu\in \measures(\manif,\,\R) 
\hbox{ and } f\in\Co^0(\manif,\R).
\]
It is immediate to verify that, if $\mu$ is absolutely continuous with
respect to the Riemannian volume $m$, then $T^*\mu$ is absolutely
continuous with respect to $m$. Given this fact, it is possible to
define the evolution of the corresponding densities:
\[
\Lp\frac{d\mu}{d m}:=\frac{ d(T^*\mu)}{dm}.
\]
The above defined operator $\Lp$ is usually called the Perron-Frobenius 
or the Ruelle-Perron-Frobenius or the transfer operator. There are 
numerous papers where such operators are used to investigate statistical
properties of expanding systems. The aim of this paper is to show that
such a strategy can be successfully extended to Anosov systems.

A direct computation shows that $\Lp$ has the following representation:
\begin{equation}
  \label{eq:jacobian}
  \Lp f
  =f\circ T^{-1}\cdot\frac{d(T^*m)}{dm}
  =:f\circ T^{-1}\cdot g
  \quad\forall f\in L^1(\manif,m)
\end{equation}
where $g\in\Co^2(\manif)$.\footnote{Note that, if $\manif=\To^d$, then in
(\ref{eq:jacobian}) $g(x)=|\det(D_xT^{-1})|$.}

To study the spectrum of the operator $\Lp$ we need to specify the space
on which such an operator acts. Indeed it is quite clear that
the spectrum of $\Lp$ on $L^1(\manif,m)$ carries only little 
interesting information, \cf the discussion in the introduction
and footnote ${}^{\ref{footnote:koopman}}$.
Instead we construct a space of generalised functions which behave more 
or less like functions of bounded variation along unstable manifolds 
whereas they look like signed measures (equipped with a weak topology) 
along stable manifolds.

We start by defining a suitable set of test functions to control the 
stable direction.
\relax For points $x,y\in\manif$ with $y\in W^s(x)$ we define $d^s(x,y)$ 
as the distance between $x$ and $y$ within the Riemannian manifold $W^s(x)$
(which inherits its Riemannian structure from $\manif$).
\relax Fix some $\delta>0$. For $0<\beta\leq1$ and bounded
measurable $\vf:\manif\to\R$, we define\footnote{The $\delta$ in the
definition is fixed once and for all, yet it must satisfy various smallness 
requirements that will be specified as we proceed in the discussion.}
\begin{equation}\label{eq:testnorm}
  H^s_\beta(\vf):=\sup\limits_{d^s(x,y)\leq\delta}
  \frac{|\vf(x)-\vf(y)|}{d^s(x,y)^\beta}\,.
\end{equation}
Here the supremum is taken over all pairs of points $x$ and $y$ such that
$y\in W^s(x)$. Clearly, $H^s_\beta$ is a seminorm, and we use it to define
\begin{equation}
\label{eq:testfunct}
  \D_\beta:=\{\vf:\manif\to\R\,:\;\vf\hbox{ measurable, }|\vf|_\infty\leq 1,
  H^s_\beta(\vf)\leq 1\}.
\end{equation}

In order to control the unstable direction we provide a set $\V$ of 
measurable test vector fields $v:\manif\to\T\manif$ adapted to the unstable 
foliation in the sense that $v(x)\in E^u_x$ for all $x\in\manif$.

Given the fact that $x\mapsto E^u_x$ is, in general, only $\tau$-H\"older
for some $\tau<1$ we cannot ask the vector fields to be globally
more regular than that. By a slight abuse of notation we
define\footnote{To compute the difference between two tangent
vectors at different (close) points we parallel transport one of
them to the tangent space of the other along the geodesic. We
will not mention this explicitly, since it is completely trivial
on $\manif=\To^d$ and it is a routine operation on general Riemannian
manifolds, see \eg \cite[Section 2.3]{DoCa}.}
\begin{equation}
\label{eq:vectornorm}
H_\beta^s(v):=\sup\limits_{d(x,y)^s\leq\delta}
                  \frac{\|v(x)-v(y)\|}{d^s(x,y)^\beta} \ .
\end{equation}
Then we will consider the vector fields
\begin{equation}\label{eq:vectors}
  \V_\beta:=\{v\in \V\,:\; |v|_\infty\leq 1;\;
  H_\beta^s(v)\leq 1\}\ .
\end{equation}
\relax From now on we will always assume that 
\begin{displaymath}
0\;<\;\beta\;<\;\gamma\;\leq 1\ .
\end{displaymath}
Using the above defined classes of test functions and test vector fields 
we now define the norms that will describe our Banach spaces of 
generalised functions. For $f\in\Co^1(\manif,\R)$ let 
\begin{equation}\label{eq:norms}
  \begin{array}l
    \|f\|_s:=\sup\limits_{\vf\in\D_\beta}\int_\manif f\vf\,dm\\[6pt]
    \|f\|_u:=\sup\limits_{v\in\V_\beta}\int_\manif df(v)dm\\[6pt]
    \|f\|:=\|f\|_u+b\|f\|_s\\[6pt]
    \|f\|_w:=\sup\limits_{\vf\in\D_\gamma}\int_\manif f\vf\,dm .\\[6pt]
  \end{array}
\end{equation}
where $\int_\manif df(v)dm$ is short hand for $\int_\manif d_xf(v(x))m(dx)$.
The constant $b\geq 1$ will be specified later. Except for $\|\cdot\|_u$, 
which is only a seminorm, all these expressions define norms on 
$\Co^1(\manif,\R)$ and $\|f\|_w\leq\|f\|_s\leq b^{-1}\|f\|$. Note that the 
above norms are inhomogeniously anisotropic because the stable and unstable 
directions are treated differently and may change from point to point.
\begin{defin}
  $\B(\manif)$ and $\B_w(\manif)$ denote the completions of 
  $\Co^1(\manif,\R)$ w.r.t. the norms $\|\cdot\|$ and $\|\cdot\|_w$,
  respectively.
\end{defin}

Each $f\in\Co^1(\manif,\R)$ naturally gives rise to a bounded linear
functional on $\Co^1(\manif,\R)$ by virtue of
\begin{displaymath}
  \langle f,\vf\rangle:=\int_\manif f\vf\,dm\ .
\end{displaymath}
Obviously, $\|f\|_{\Co^1}^*\leq\|f\|_w\leq\|f\|\leq\|f\|_{\Co^1}$. 
Therefore there exist canonical continuous embedding (not necessarily 
one-to-one)
\begin{displaymath}
  \Co^1(\manif,\R)\to\B(\manif)\to\B_w(\manif)\to\Co^1(\manif,\R)^*\ .
\end{displaymath}
In fact, each
$f\in\B_w$ defines a bounded linear functional on
$\Co^1(\manif,\R)$ by 
$\langle f,\vf\rangle:=\limn\langle f_n,\vf\rangle$ 
where $f_n\in\Co^1(\manif,\R)$ and
$\limn\|f-f_n\|_w=0$.
In the same way one can embed $\B(\manif)$ into $\B_w(\manif)$.

\begin{rem}\label{rem:norm-seminorm}
Note that although the embedding of $\Co^1(\manif,\R)$ into
each of the other three spaces is one-to-one this need not be
true for the other embedding. 
We have a closer look at the embedding of $\B(\manif)$ into $\B_w(\manif)$,
but the same remarks apply to the embedding into $\Co^1(\manif,\R)^*$.

As a functional on the dense subspace $\Co^1(\manif,\R)$ of $\B(\manif)$
the map $f\mapsto\|f\|_w$ is Lipschitz continuous, so it extends to
all of $\B(\manif)$. This extension may be only
a seminorm, however. Therefore, if we consider $\B(\manif)$ as a linear
subspace of $\B_w(\manif)$, we really mean 
$\B(\manif)/N_w\hookrightarrow\B_w(\manif)$ where $N_w$ is the closed 
linear subspace $\{f\in\B(\manif):\|f\|_w=0\}$.

Nevertheless, with an abuse of notation, we write 
$\B(\manif)\subset\B_w(\manif)$, 
where the inclusion must be interpreted in the above sense. In particular, 
for $f\in\B(\manif)$, $\|f\|_w$ can denote the seminorm $\|\cdot\|_w$ of 
$f$ (which generally will be the case), or it can be the norm of the 
element $f+N_w\in\B_w(\manif)$. Both interpretations are equivalent for 
our purposes.
\end{rem}

\begin{rem}\label{rem:seprem}
Since the space $\Co^1(\manif,\R)$ is separable, since the norms
$\|\cdot\|$ and $\|\cdot\|_w$ are weaker than the
$\Co^1(\manif,\R)$ norm, and since $\Co^1(\manif,\R)$ is dense both in
$\B(\manif)$  and $\B_w(\manif)$ by construction, it follows
that the Banach spaces $\B(\manif)$ and $\B_w(\manif)$ are
separable.
\end{rem}

Note that the unstable seminorm is given by a formula that does not 
extend to all of $\B(\manif)$, since for general elements of that space
there is no way to make sense of the differential.
The obvious way to compute $\|f\|_u$ for general $f\in\B(\manif)$
would be to approximate $f$ by functions from $\Co^1(\manif,\R)$. Since
this is not always desirable, it is sometimes convenient to write 
$\|\cdot\|_u$ in a way that applies directly to any element of 
$\B(\manif)$. The following Lemma is step in this direction.

\begin{lem}\label{lem:weak-unstable}
Let $U\subset \manif$ be a sufficiently small open set. There are
$\tau$-H\"older vector fields $\{V_i\}_{i=1}^{d_u}$ in $U$ which, 
at each point $x\in U$, form a basis of $E^u(x)$ and which, restricted 
to each unstable manifold, are $\Co^1$ vector fields. 

These vector fields can be constructed such that there exist constants 
$C>0$ and $\beta'\in(0,\tau)$ with the following property:
Given any $\vf\in\Co^{\beta}(\manif,\R^{d_u})$ with
$\supp\vf\subset U$ which is $\Co^1$ when restricted to any 
unstable manifold, let $V_\vf:=\sum_i\vf_iV_i$. Then 
$C^{-1}V_\vf\in\V_\beta$, and there exists
a measurable function $A:\manif\to\R^{d_u}$ such that
$C^{-1}A_i\in\D_{\beta'}(\manif)$ and such that, for each
$f\in\Co^1(\manif,\R)$,
\[
\int_\manif df(V_\vf)\,dm
=
-\int_\manif f\sum_i d\vf_i(V_i)\,dm+\int_\manif f\sum_iA_i\vf_i\,dm
.
\]
Observe that $d\vf_i(V_i)$ is well defined because $V_i(x)\in E^u(x)$ 
for all $x$, although the $\vf_i$ need only be H\"older continuous in 
stable direction. This point is further clarified in the proof.   
\end{lem}

\begin{rem}\label{rem:divergence-remark}
On the torus $\To^2$ there is the normalised (and oriented) unstable 
vector field $V_1=v^u$ which is even of class $\Co^{1+\alpha}$ for some 
$\alpha>0$ (see the appendix for details). Hence $A_1=-\divg(V_1)$ in 
this case.
\end{rem}

Since it is easy to check that the choice of a different basis changes
only the functions $A_i$ in the above formula one could
define the unstable norm - locally - by
\begin{displaymath}
  \sup_{\vf\in \Co^{1}(U,\R^{d_u})}
       \frac{\int_\manif f \sum_i d\vf_i(V_i)\,dm}
        {|\vf|_\infty+H^s_\beta(\vf)}
\end{displaymath}
provided one is willing to choose $\beta\leq\beta'$.
Yet, for this reason and
since it would be a bit cumbersome to extend globally the above
definition, we prefer to
work with the norm (\ref{eq:norms}). 
\relax For the special case $\manif=\To^2$ a related problem is encountered 
in the proof of Lemma~\ref{lem:crucial}.

The proof of Lemma
\ref{lem:weak-unstable} is best done in local coordinates introduced in
section \ref{subsec:compactness-proof} and so it is postponed to
section \ref{subsec:unorm-proof}.

\subsection{Quasi-compactness of the transfer operator $\Lp$ on $\B(\manif)$}
\label{subsec:quasicompactness}

To simplify the exposition we will give precise statements and
complete proofs only for the case $\manif=\To^d$. Nevertheless we will 
indicate for the results of this subsection and in the course of their 
proofs the modifications necessary in order to deal with general manifolds.

Our assumption that $T$ is of class $\CoT$ implies that the \emph{Jacobian}
$g$ of $T$ (see (\ref{eq:jacobian})) is of class $\Co^2$; in fact, 
we will use only that is of Class $\CoJ$.

Our first result is a Lasota-Yorke type inequality.

\begin{lem}\label{lem:LY}
Suppose $\beta<\min\{\tau, 1\}$ and $\gamma\in (\beta,1]$. Then $\Lp$
extends naturally to a bounded linear operator on both $\B_w(\manif)$
and $\B(\manif)$. In addition, for each 
$\sigma>\max\{\lambda_u^{-1},\lambda^{\beta}_s\}$, 
we can choose constants $b$ and $\delta$
in (\ref{eq:testnorm}) - (\ref{eq:norms}) for which there exists
$B>0$ such that, for each $f\in \B(\manif)$, we have
\begin{displaymath}
    \|\Lp^n f\|_w\leq A\,
      \|f\|_w\qquad\text{and}\qquad
    \|\Lp^n f\|\leq 3A^2\sigma^n\|f\|+B\|f\|_w
    \qquad\text{for $n=1,2,\dots$}
\end{displaymath}
where $A$ is the constant from (\ref{eq:hyperbolicity}).
\end{lem}

This clearly implies that the spectral radius of $\Lp$ is bounded
by one both on $\B_w(\manif)$ and $\B(\manif)$. Thus
$\spectr(\Lp)\subseteq\{z\in\C\,:\;|z|\leq 1\}$.
To exploit Lemma \ref{lem:LY} further, we need the following fact.

\begin{prop} \label{prop:compactnes}
If $\gamma\cdot\min\{\tau,1\}>\beta$, then the ball 
$\B_1:=\{f\in\B(\manif):\; \|f\|\leq 1\}$ is relatively compact in 
$\B_w(\manif)$.\footnote{This assertion can be interpreted in either of 
the two equivalent ways discussed in Remark~\ref{rem:norm-seminorm}.}
\end{prop}
The proof of Lemma \ref{lem:LY} is given in section \ref{subsec:LYproof},
that of Proposition \ref{prop:compactnes} in section
\ref{subsec:compactness-proof}.

\relax From these two results the quasicompactness of $\Lp$ follows 
along classical lines \cite{ITM}. A more recent argument, which is based 
on the Nussbaum formula, also provides an estimate on the essential 
spectral radius:

\begin{Thm}\label{thm:quasicompact}
  Suppose that $\gamma\cdot\min\{\tau,1\}>\beta$. Then, for each 
  $\sigma>\max\{\lambda_u^{-1},\lambda_s^\beta\}$, the operator 
  $\Lp: \B(\manif)\to \B(\manif)$ has essential spectral radius 
  bounded by $\sigma$ and is thus quasicompact.
\end{Thm}

An immediate consequence of Theorem~\ref{thm:quasicompact} is that for each
constant $r\in(\sigma,1)$ the portion\footnote{By $A\setminus B$ we mean the
symmetric difference of the sets $A$ and $B$.}
\begin{displaymath}
  \spectr_r(\Lp):= \spectr(\Lp)\setminus\{z\in\C:\,|z|\leq r\}
\end{displaymath}
of the spectrum
consists of
finitely many eigenvalues $\lambda_1,\dots,\lambda_p$ of finite
multiplicity. Denote by $\Pi_1,\dots,\Pi_p$ the corresponding
spectral projectors and let $\Pi:=\Id_\B-\sum_{j=1}^p\Pi_j$. Then
$\rank(\Pi_j)<\infty$
$(j=1,\dots,p)$, and the spectral radius $\rho_{\spectr}(\Lp \Pi)\leq r$. 
The next lemma shows that $\spectr_r(\Lp)$ and the spectral projectors 
$\Pi_j$ are the same on $\B(\manif)$ and on $\B(\manif)/N_w$, \cf 
Remark~\ref{rem:norm-seminorm}. 

\begin{lem}\label{lem:norm-domination}
Restricted to $(\Id_\B-\Pi )(\B(\manif))$, the seminorm $\|\cdot\|_w$
dominates the norm $\|\cdot\|$ (so that its restriction is actually a norm). 
\end{lem}
\begin{proof}
Since $\|\Lp^{-n}(\Id_\B-\Pi)\|\leq\const\cdot r^{-n}$, we have
for $f=(\Id_\B-\Pi)f$ and each $n\in\N$
\begin{displaymath}
  \|f\|
  =
  \|\Lp^{-n}\Lp^{n}f\|
  \leq
  \const\cdot r^{-n}\,\|\Lp^nf\|
  \leq
  \const\cdot A^2\left(\frac\sigma r\right)^n\|f\|+\const\cdot r^{-n}B\|f\|_w
\end{displaymath}
where we used Lemma~\ref{lem:LY} for the last inequality.
Choosing $n$ such that 
$\const\cdot A^2\left(\frac\sigma r\right)^n\leq\frac 12$
it follows that $\|f\|\leq2\const\cdot r^{-n}B\|f\|_w$.
\end{proof}
\begin{rem}
\label{rem:equivalent-norms}
Although the norms $\|\cdot\|$ and $\|\cdot\|_w$ depend on constants $b$ 
and $\delta$ which will be determined in the course of the proofs, they are
equivalent for any two choices of these constants. So the spectral 
assertions in Theorem \ref{thm:quasicompact} do not depend on them.
\end{rem}

\begin{proof}[Proof of Theorem~\ref{thm:quasicompact}] 
  For the reader's convenience we recall the brief argument from
  \cite{He} which deduces the estimate on the essential spectral radius from
  Nussbaum's formula \cite{Nussbaum}. Let $\epsilon>0$. Since $\B_1$
  is relatively compact in $\B_w$, there are $f_1,\dots,f_N\in\B_1$ such that
  $\B_1\subseteq\bigcup_{i=1}^NU_\epsilon(f_i)$, where 
  $U_\epsilon(f_i)=\{f\in\B:\,\|f-f_i\|_w<\epsilon\}$. 
  For $f\in \B_1\cap U_\epsilon(f_i)$, Lemma~\ref{lem:LY} implies that
  \begin{displaymath}
    \|\Lp^n(f-f_i)\|\leq3A^2\sigma^n\,\|f-f_i\|
                       +\frac{3A^2B}{1-\sigma}\|f-f_i\|_w
  \leq 6A^2\sigma^n+\frac{3A^2B\epsilon}{1-\sigma}\ .
  \end{displaymath}
  Choosing $\epsilon=\sigma^n$ we can conclude that for each $n\in\N$ the 
  set $\Lp^n(\B_1)$ can be covered by a finite number of 
  $\|\cdot\|$--balls of radius $C\sigma^n$, $C=6A^2+\frac {3A^2B}{1-\sigma}$. 
  This is precisely what is needed to apply Nussbaum's formula, which 
  states that the essential spectral radius of $\Lp$ is at most 
  $\liminf_{n\to\infty}\sqrt[n]{C\sigma^n}=\sigma$.
\end{proof}

The above result allows to obtain a quite precise picture of the
spectrum of $\Lp$, yet some more can be easily seen. First of all
notice that $\Lp$ is invertible and $\Lp^{-1}$ is nothing else than
the transfer operator associated to the map $T^{-1}$. Then the
following bound (the  proof can be found at the end of section
\ref{subsec:LYproof}) holds.

\begin{figure}[ht]
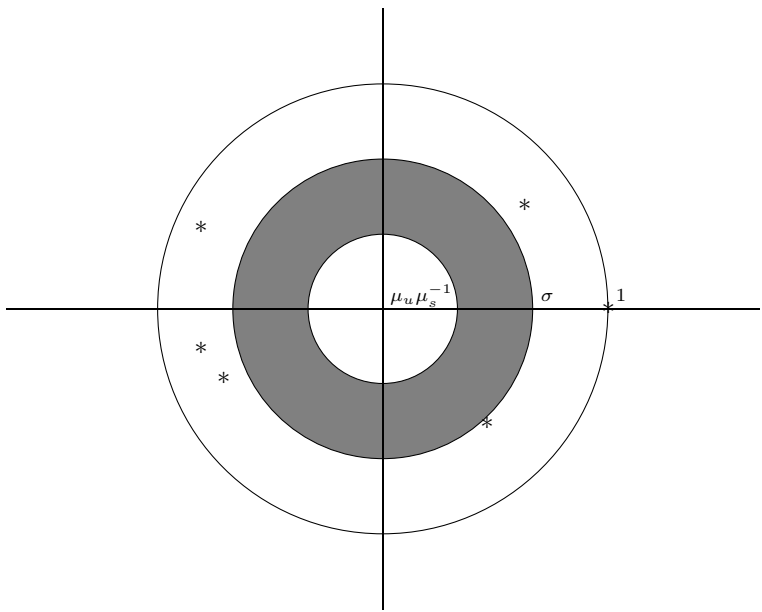
\ 
\psset{xunit=1mm, yunit=1mm}
\pscustom[linewidth=.4pt]{
\put(0,0){\pscircle{2}}
\fill[fillstyle=solid, fillcolor=gray]
}
\pscustom[linewidth=.4pt]{
\put(0,0){\pscircle{1}}
\fill[fillstyle=solid, fillcolor=white]
}
\pscustom[linewidth=.4pt]{
\put(0,0){\pscircle{3}}
}
\put(-50,0){\line(1,0){100}}
\put(0,-40){\line(0,1){80}}
\put(31,1){$\scriptstyle 1$}
\put(-25,-6){$*$}
\put(-22,-10){$*$}
\put(13,-16){$*$}
\put(18,13){$*$}
\put(-25,10){$*$}
\put(29.15,-.75){$*$}
\put(1,1){$\scriptstyle\mu_u\mu_s^{-1}$}
\put(21,1){$\scriptstyle\sigma$}
\caption{Region containing the spectrum of the transfer operator}
\label{fig:spectrum}
\end{figure}

\begin{lem}\label{lem:inverseLp}
Let $\mu_s,\,\mu_u$, $\mu_s\leq\lambda_s<1<\lambda_u\leq\mu_u$,
the maximal contraction and expansion rate of $T$, 
respectively.\footnote{By this we mean
$\|DT^{-n}\|\leq A\mu_s^n$, $\|DT^n\|\leq A\mu_u^n$.} Then
we have the following bound for the spectral radius of $\Lp^{-1}$:
\[
  \rho_{\spectr}(\Lp^{-1})\leq\mu_u\mu_s^{-1}\ .
\]
\end{lem}
\noindent Accordingly, the essential spectrum of $\Lp$ must be 
contained in the annulus
$\{z\in\C\,:\;\mu_u\mu_s^{-1}\leq|z|\leq \sigma\}$ as shown in figure
\ref{fig:spectrum}.  

\begin{exmp}[\em Eigenvalues of linear automorphisms of $\To^2$]
\label{exmp:linear-eigenvalues}
{\rm
Let $\Amatrix$ be a $2\times2$ integer matrix with $\det(\Amatrix)=1$ and trace
different from $\{-2,\dots,2\}$. 
Then $\Amatrix$ has eigenvalues $\lambda_\Amatrix$ and 
$\lambda_\Amatrix^{-1}$ with $|\lambda_\Amatrix|>1$, and 
$T(x):=\Amatrix x\mod\Z^2$ defines
a hyperbolic automorphism of $\To^2$. The stable and unstable fibres
are just the straight lines in direction of the eigenvectors for
$\lambda_\Amatrix^{-1}$ and $\lambda_\Amatrix$, respectively. 
Therefore $\tau=1+\alpha\geq 2$, and we can choose for the definition 
of our norms $\gamma=\alpha=1$ and $\beta\in(0,1)$ as close to one as 
we wish. Denote by $\Lp$ the transfer operator for $T$. We show here 
that for any choice of $\beta\in(0,1)$
\begin{equation}\label{eq:torus-ev}
  \text{eigenvalues}(\Lp)
  \subseteq
  \left\{z\in\C\,:\;|z|\leq \lambda^{-1}_\Amatrix\right\}\cup\{1\}
\end{equation}
and that $1$ is a simple eigenvalue. In view of the bound for the essential
spectral radius of $\Lp$ established in Theorem~\ref{thm:quasicompact}
it follows that
\begin{displaymath}
  \spectr(\Lp)
  \subseteq
  \left\{z\in\C\,:\;|z|\leq \lambda^{-\beta}_\Amatrix\right\}\cup\{1\}\ .
\end{displaymath}
This fact is not only a concrete illustration of our 
Theorem~\ref{thm:quasicompact}, but also its proof introduces in a highly
simplified setting some of the ideas that will be worked out later
in more generality. Therefore, in spite of the relative simplicity,
we provide many details.

So suppose that $\Lp f=\lambda f$ with $|\lambda|>\lambda_\Amatrix^{-1}$ and 
$f\in\B_{\beta_0}$ (where the index of $\B$ indicates the actual choice 
of $\beta$ we make). 
Since $\D_{\beta_0}\supseteq\D_\beta$ for each $\beta\in[\beta_0,1)$, 
we have $\B_{\beta_0}\subseteq\B_\beta$
and hence also $f\in\B_\beta$ for each such $\beta$. In particular we can
choose $\beta\in[\beta_0,1)$
such that $|\lambda|>\lambda_\Amatrix^{-\beta}$.

Since $\int \Lp h\,dm=\int h\,dm$ for all $h\in\Co^1(\To^2)$,
it follows that 
$\lambda\langle f,1\rangle=\langle\Lp f,1\rangle=\langle f,1\rangle$, so 
that either $\langle f,1\rangle=0$ or $\lambda=1$. In the latter case, 
if $1$ is not a simple eigenvalue, there nevertheless exists some 
$f\in\B(\To^2)$ with $\Lp f=f$ and $\langle f,1\rangle=0$. (Observe that 
because of $\|\Lp\|_w\leq1$ the eigenvalue $1$ is  semisimple.) So we 
will assume $\langle f,1\rangle=0$ from now on.

Our goal is to show that $\|f\|=0$. As a first step we consider $\|f\|_u$.
Let $v\in\V_\beta$ (see (\ref{eq:vectors})) and observe that 
$v$ is in direction of the unstable eigenvector of $\Amatrix$.
Observing the fact that $T$ leaves the Lebesgue measure on $\To^2$ 
invariant, we have for each $h\in\Co^1(\To^2)$
\begin{displaymath}
\begin{split}
  \int_{\To^2}d(\Lp h)(v)\,dm
  &=
  \int_{\To^2}d(h\circ T^{-1})(v)\,dm
  =
  \int_{\To^2}dh_{T^{-1}x}(\Amatrix^{-1}(v(x)))\,m(dx)\\
  &=
  \int_{\To^2}dh_{x}(\Amatrix^{-1}(v(Tx)))\,m(dx)
  =
  \lambda_\Amatrix^{-1}\int_{\To^2}dh(v\circ T)\,dm
\end{split}
\end{displaymath}
Since $|v\circ T|_\infty=|v|_\infty$ and 
since $H_\beta^s(v\circ T)\leq H_\beta^s(v)$ 
($T$ contracts in stable direction!), also $v\circ T\in\V_\beta$, and
it follows that 
$\|\Lp h\|_u\leq\lambda_\Amatrix^{-1}\|h\|_u$. Of course, this estimate extends
to all elements of $\B(\To^2)$ and in particular to $f$. So
$\|f\|_u=|\lambda^{-1}|\cdot\|\Lp f\|_u
\leq|\lambda\lambda_\Amatrix|^{-1}\cdot\|f\|_u$,
which implies $\|f\|_u=0$ because $|\lambda\lambda_\Amatrix|>1$ by assumption.

In order to show that also $\|f\|_s=0$, we use the flow $S_t$ which moves 
points at unit speed along unstable fibres. Then, for each $h\in\Co^1(\To^2)$ 
and $x\in\To^2$, integration by parts yields\footnote{Just integrate 
separately from $-r$ to $0$ and from $0$ to $+r$.}
\begin{displaymath}
  \frac 1{2r}\int_{-r}^{r}h(S_tx)\,dt
  =
  h(x)+\frac 1{2r}\int_{-r}^{r}h_1(S_tx)
    \,\theta(t,r)\,dt  
\end{displaymath}
where $h_1(x)=d_{x}h(v^u)$, $v^u$ denoting the normalised unstable 
eigenvector, and $\theta(t,r)=-t+r\cdot\sign(t)$. So, if $\vf\in\D_\beta$, 
then
\begin{equation}\label{eq:torus-example}
\begin{split}
  &\langle h,\vf\rangle\\
  &=
  \int_{\To^2} \left(\frac 1{2r}\int_{-r}^{r}h(S_tx)\,dt\right)
    \,\vf(x)\,m(dx)
  -\int_{\To^2} \left(\frac 1{2r}\int_{-r}^{r}h_1(S_tx)
    \,\theta(t,r)\,dt\right)\,\vf(x)\,m(dx)\\
  &=
  \frac 1{2r}\int_{-r}^{r}\left(\int_{\To^2}h(S_tx)\,\vf(x)\,m(dx)\right)dt
  -\frac 1{2r}\int_{-r}^{r}
      \left(\int_{\To^2}h_1(S_tx)\,\theta(t,r)\,\vf(x)\,m(dx)\right)dt\\
  &=
  \frac 1{2r}\int_{-r}^{r}\left(\int_{\To^2}h(x)\,\vf(S_{-t}x)\,m(dx)\right)dt
  -\frac 1{2r}\int_{-r}^{r}
      \left(\int_{\To^2}h_1(x)\,\theta(t,r)\,\vf(S_{-t}x)\,m(dx)\right)dt\\
  &=
  \int_{\To^2} \left(\frac 1{2r}\int_{-r}^{r}\vf(S_{-t}x)\,dt\right)
    \,h(x)\,m(dx)
  -2r\int_{\To^2} d_xh\left(\int_{-r}^{r}\vf(S_{-t}x)
    \,\frac 1{(2r)^2}\theta(t,r)\,dt\cdot v^u\right)\,m(dx)
\end{split}
\end{equation}
Suppose now that $w:\R\to\R$ is bounded with 
$\int_{-\infty}^{\infty}|w(t)|\,dt\leq 1$.
Define $\vf_w:\To^2\to\R$ by 
$\vf_w(x)=\int_{-\infty}^\infty\vf(S_{-t}x)w(t)\,dt$.
Then, if $y\in W^s(x)$ with $d(x,y)<\delta$,
\begin{displaymath}
  |\vf_w(y)-\vf_w(x)|
  \leq
  \int_{-\infty}^\infty|\vf(S_{-t}x)-\vf(S_{-t}y)|\,w(t)\,dt
  \leq  H_\beta^s(\vf)\,d(x,y)^\beta\ ,
\end{displaymath}
because $S_{-t}x\in W^s(S_{-t}y)$ and $d(S_{-t}x,S_{-t}y)=d(x,y)$.
Since also $|\vf_w|_\infty\leq |\vf|_\infty$, it follows that 
$\vf_w\in\D_\beta$.
Therefore, the integrals in brackets in the last line of
(\ref{eq:torus-example}) are, as functions of $x$, elements of $\D_\beta$.
So, the second one multiplied with $v^u$ is a vector field from $\V_\beta$.
Hence
\begin{displaymath}
  \left|\int_{\To^2} d_xh\left(\int_{-r}^{r}\vf(S_{-t}x)
    \,\frac 1{(2r)^2}\theta(t,r)\,dt\cdot v^u\right)\,m(dx)\right|
  \leq
  \|h\|_u .
\end{displaymath}
Accordingly, setting
\[
\vf_r:=\frac 1{2r}\int_{-r}^{r}\vf(S_{-t}x)\,dt,
\]
for each sequence $\{f_n\}\subset\Co^1$ such that $f_n\to f$ in $\B(\To^2)$,
holds
\begin{equation}
\label{eq:unif-est}
|\langle f,\vf\rangle|=|\lim_{n\to\infty}\langle f_n,\vf\rangle|
\leq \lim_{n\to\infty}|\langle
f_n,\vf_r\rangle|+2r\lim_{n\to\infty}\|f_n\|_u
=|\langle f,\vf_r\rangle|.
\end{equation}
On the other hand 
\[
\langle f,\vf_r\rangle=\lambda^{-n}\langle \Lp^n f,\vf_r\rangle=
\langle f,\lambda^{-n}\vf_r\circ T^n\rangle.
\]
To conclude just notice that
\begin{align*}
H^s_\beta(\lambda^{-n}\vf_r\circ T^n)\leq
|\lambda|^{-n}\lambda_\Amatrix^{-n\beta}\\
|D^u (\lambda^{-n}\vf_r\circ T^n)|
\leq \frac{|\lambda|^{-n}\lambda_\Amatrix^{n}}{r},
\end{align*}
where $D^u$ stands for the derivative along the unstable
direction. Since estimate (\ref{eq:unif-est}) is uniform in $r$ we can
choose $r=r_n:=\lambda_\Amatrix^{2n}$, which implies that the
$\beta$-H\"older constant (in all directions!) of 
$\hat\vf_{r_n}:=\lambda^{-n}\vf_{r_n}\circ T^n$ is at most
$2|\lambda|^{-n}\lambda_\Amatrix^{-n\beta}$. Accordingly
\[
\left|\hat\vf_{r_n}-\int_{\To^2}\hat\vf_{r_n}\,dm\right|_\infty\leq
2\delta^{-1+\beta}(|\lambda|\lambda_\Amatrix^\beta)^{-n}.
\]
The above estimate means that 
$\frac 12(\hat\vf_{r_n}-\int\hat\vf_{r_n}\,dm)
 \lambda^n\lambda_\Amatrix^{-n\beta}
 \delta^{1-\beta}\in\D_\beta$ for each $n\in\N$. 
Observing also that $\left|\int\hat\vf_{r_n}\,dm\right|<\infty$ for all $n$
and that $|\lambda|\lambda_\Amatrix^\beta>1$ we conclude
\[
|\langle f,\vf\rangle|
\leq 
|\langle f,\hat\vf_{r_n}\rangle|
\leq 
2\delta^{-1+\beta}(|\lambda|\lambda_\Amatrix^\beta)^{-n}\,\|f\|_s+
\underbrace{|\langle f,1\rangle|}_{=0} 
\left|\int_{\To^2}\hat\vf_{r_n}\,dm\right|
\to0
\]
as $n\to\infty$. Since this holds for any $\vf\in\D_\beta$,
this implies $\|f\|_s=0$ and finishes the proof.
}
\end{exmp}

\subsection{Peripheral spectrum and SRB measures}
\label{sec:SRB}

Recall that $\spectr_r(\Lp):= \spectr(\Lp)\setminus\{z\in\C:\,|z|\leq r\}$.
If $r$ is sufficiently close to $1$, then $\spectr_r(\Lp)$ contains only
eigenvalues of modulus $1$, and since 
$\|\Lp\|_w\leq 1$ (see Lemma~\ref{lem:LY}), these
eigenvalues are semisimple so that $\Lp \Pi_j=\lambda_j \Pi_j$ 
for $j=1,\dots,p$.
We
will call {\em peripheral spectrum} the set
$\{\lambda\in\spectr(\Lp):\;|\lambda|=1\}$.
This spectral decomposition allows to reproduce easily some well known facts 
from the ergodic theory of Anosov diffeomorphisms.

\relax First of all, if $|\lambda|=1$, then
  \begin{equation}
    \label{eq:P_lambda}
    \Pi_\lambda
    :=
    \limn\frac 1n\sum_{k=0}^{n-1}\lambda^{-k}\Lp^k
    =
    \sum_{j=1}^p\limn\frac 1n\sum_{k=0}^{n-1}
    \left(\frac{\lambda_j}\lambda\right)^k\Pi_j
    =
    \begin{cases}
      \Pi_j&\hbox{if $\lambda=\lambda_j$}\\
      0&\hbox{otherwise}
    \end{cases}
  \end{equation}
  exists in the uniform $\|\,\cdot \,\|$ - operator norm.
  In particular $\Pi_j=\Pi_{\lambda_j}$. Since, by Lemma~\ref{lem:LY},
  \begin{equation}\label{eq:PiBw-contraction}
    \|\Pi_{\lambda_j}f\|
    \leq
    \lim_{n\to\infty}\frac 1n\sum_{k=0}^{n-1}\|\Lp^kf\|
    \leq
    \lim_{n\to\infty}\frac 1n\sum_{k=0}^{n-1}
       \left(3A^2\sigma^k\|f\|+B\|f\|_w\right)
    =
    B\|f\|_w
  \end{equation}
  for all $f\in\Co^1(\manif,\R)$,  
  the finite rank operators 
  $\Pi_j$ extend continuously to an operator from $\B_w(\manif)$ to 
  $\B(\manif)$ and hence
  \emph{a fortiori}
  from $\Co^0(\manif,\R)$ to $\B(\manif)$ and also from 
  $L^1(\manif,\R)$ to $\B(\manif)$. 
  Since $\Co^1(\manif,\R)$ is a dense linear subspace of $\B_w(\manif)$   
  and since each $\Pi_j$
  has finite rank, we have 
  \begin{equation}\label{eq:same-ranks}
  \Pi_j(\B_w(\manif))=\Pi_j(\Co^1(\manif,\R))=\Pi_j(\B(\manif))\ .
  \end{equation}
  In the following proposition we characterise these projections more 
  precisely.

\begin{prop}\label{prop:SRB-existence}
\relax For each $h\in \Pi_j(\B(\manif))$ 
there is a finite signed Borel measure
$\mu_h$ on $\manif$ such that $\langle h,\vf\rangle=\int\vf\,d\mu_h$
for all $\vf\in\D_\gamma$. Moreover,
$\lambda_1:=1\in\spectr(\Lp)$, $\mu:=\mu_{\Pi_11}$ is a positive
measure, $\mu(\manif)=m(\manif)$, and all $\mu_h$ are absolutely
continuous with respect to $\mu$ and 
$|\frac{d\mu_h}{d\mu}|\leq\|f\|_\infty$ $\mu$-almost surely
for each $f\in\Co^1(\manif)$ such that $\Pi_1f=h$. Finally,   
\begin{enumerate}[a)]
\item $T^*\mu_h=\mu_h$ for all $h\in\Pi_1(\B(\manif))$ and
\item if $\lambda_1=1$ is a simple eigenvalue, then
  $\Pi_1f=\langle f,1\rangle\cdot \Pi_11$ for all $f\in\B_w(\manif)$.
\end{enumerate}
\end{prop}

In addition, it is possible to further characterise the invariant
measures. The  proofs of the above and the following proposition can be
found in section \ref{subsec:peripheral}.

\begin{prop}\label{prop:SRB-properties}
The measure $\mu$ is a \emph{Sinai-Ruelle-Bowen (SRB)} measure in the
following sense:
\begin{enumerate}[a)]
\item The measure $\mu$ is the weak Cesaro-limit of the Riemannian
  measure under the dynamics;
\item The ergodic decomposition of the measure $\mu$ is determined by
  the eigenspace of $\Lp$ associated to the eigenvalue 1. More precisely,
  there exists a
  finite basis $\{h_i\}$ of this eigenspace and measurable $T$-invariant sets
  $A_i\subseteq  \manif$ with $\mu(\bigcup_iA_i)=1$, such that the measures
  $\mu_i(A):=\mu(A\cap A_i)$ are ergodic and, for each 
  $\vf\in\D_\gamma$, 
  $\int\vf\,d\mu_i=\langle h_i,\vf\rangle$.
\item For all $\vf\in \Co^0(\manif,\R)$, the limit
  $\vf^+(x):=\lim\limits_{n\to\infty}\frac 1n\sum_{i=1}^{n-1}\vf\circ T^i(x)$ 
  exists $m$--a.e. and 
  takes only finitely many different values. If $\mu$ is ergodic, then
  $\vf^+(x)=\int\vf\,d\mu$ for $m$-almost every $x$.
\item \label{item:SRB-properties-d}
  For each $h\in\Pi_j(\B(\manif))$ 
  the measure $\mu_h$, locally conditioned to the
  unstable foliation, is absolutely continuous on $\mu$--almost every fibre
  with respect to the Riemannian measure on the
  fibre. 
  In addition, the density is a function of bounded variation.
\end{enumerate}
\end{prop}

Observe that although the spectral assertions on $\Lp$    
stated in Theorem~\ref{thm:quasicompact} do not depend
on the constants $\delta$ and $b$ entering the norm $\|\cdot\|$
(see Remark~\ref{rem:equivalent-norms}), quantities
like the norms of $\|\Lp\|$, $\|\Lp \Pi\|$ and
$\|\Pi_j\|$ are affected.

\subsection{Spectral stability -- Smooth random perturbations}
\label{subsec:stability-smooth}

It is well known that the Sinai-Ruelle-Bowen measure of a mixing
Anosov diffeomorphism is stable (in the sense of the weak
topology on measures) under sufficiently smooth small random
perturbations, see \eg \cite{Kifer1}, \cite{Babook} for the case
of Gaussian-like random perturbations. Further results and a list
of references on stochastic stability of SRB measures 
can be found in \cite{Blank}. In terms of our spectral
decomposition this means that the eigenvector $h=\Pi_11$ of $\Lp$
is stable under perturbations. In this section we assert much
more: The full spectral structure of $\Lp$ outside the disk with
radius $\sigma$ (see Lemma~\ref{lem:LY}) 
does not change much under small perturbations.
As we noted already in the introduction, the smoothness of the
stable and unstable foliations, which plays an essential r\^ole in the proofs, 
is automatically guaranteed only in the two-dimensional case where
the foliations are $\Co^{1+\alpha}$ for some $\alpha>0$.
Therefore, and in order to avoid some other technical problems,
we give the full proof only for the case $\manif=\To^2$. Nevertheless
we describe the setting and prove some partial results in greater
generality in order to point out more clearly where problems
due to higher dimensions do arise.

We start by introducing a class of small random perturbations.

\begin{defin}\label{def:kernel}    
\begin{enumerate}[a)]
\item
  A \emph{kernel family} is a family   
  of measurable functions
  $q_\ve:\To^2\times\To^2\to\R^+$, $\ve\in(0,1]$, such
  that
  \begin{displaymath}
      q_\ve(x,y)=0 \hbox{  if } |x-y|>\ve\ .
  \end{displaymath}
  For $(x,\xi)\in\To^2\times\R^2$ 
  we denote\footnote{This definition can easily be extended to general
    manifolds $\manif$. Just introduce, in a neighbourhood $U$ of $x$,
    normal coordinates via the exponential
    map $\exp_x$ and identify isometrically the tangent space $\T_x\manif$
    with $\R^d$. Then define:
    \[
    \tilde q_\ve(z,\xi):=q_\ve(z,\exp_x(\exp_x^{-1}(z)+\xi)).
    \]
    }
  \begin{displaymath}
    \tilde q_\ve(x,\xi):=q_\ve(x,x+\xi)\ .  
  \end{displaymath}
\item 
  A \emph{stochastic kernel family} is a kernel family satisfying additionally
  \begin{displaymath}
      \int_\manif q_\ve(x,y)\,m(dy)=1\quad\mbox{for all }x\in\To^2\ .
  \end{displaymath}
\item
  An \emph{admissible kernel family} is a kernel family
  of continuously differentiable functions $q_\ve$
  which satisfies some additional regularity properties.
  To state them we 
  denote
  $\partial_1\tilde q_\ve(x,\xi):=\frac{\partial}{\partial x}\tilde
q_\ve(x,\xi)$
  and
  $\partial_2\tilde q_\ve(x,\xi):=\frac{\partial}{\partial \xi}\tilde
q_\ve(x,\xi)$.
  We require that there is a stochastic kernel family 
  $(p_\ve)_{0<\ve\leq1}$ such that for each $\ve>0$
  the functions $\tilde q_\ve$, $\partial_1\tilde q_\ve$ and
  $\ve\cdot\partial_2\tilde q_\ve$
  are \emph{$\ve^{-1}$-dominated}
  by $\tilde p_{\ve}$ in the sense of the following
  definition.
\end{enumerate}
\end{defin}

\begin{defin}
  \label{def:Lipschitz-domination}
  Let $\tilde q_\ve:\To^2\times\To^2\to\R^k$, $k\geq 1$,
  and let $(p_\ve)_{0<\ve\leq 1}$ be a stochastic kernel family
  with the property that for each $0<\ve'\leq\ve$ there exists a constant
  $C>0$ such that $\tilde p_{\ve'}(x,\xi)\leq C\,\tilde p_\ve(x,\xi)$.
  We say that $\tilde q_\ve$ is \emph{$\ve^{-1}$-dominated} by 
  $\tilde p_{\ve}$ if
  there are some $a,M>0$ such that, for each $0<\ve\leq 1$,
  \begin{align}
    \label{eq:Lipschitz-domination-a}
      |\tilde q_\ve(x,\xi)|
      &\leq
      M\,\tilde p_{a\ve}(x,\xi)\\
    \label{eq:Lipschitz-domination-b}
      |\tilde q_\ve(x,\xi)-\tilde q_\ve(z,\xi)|
      &\leq
      M\,d(x,z)^\alpha\,\tilde p_{a\ve}(x,\xi)\\
    \label{eq:Lipschitz-domination-c}
      |\tilde q_\ve(x,\xi)-\tilde q_\ve(x,\zeta)|
      &\leq
      M\,\ve^{-1}\,d(x+\xi,x+\zeta)
      \cdot
      \big(\tilde p_{a\ve}(x,\xi)+\tilde p_{a\ve}(x,\zeta)\big)
  \end{align}
\end{defin}

\begin{rem}\label{rem:Lipschitz-domination}
Requirements (\ref{eq:Lipschitz-domination-a}) and 
(\ref{eq:Lipschitz-domination-b}) 
imply that $\tilde q_\ve(z,\xi)\leq 2M\tilde  p_{a\ve}(x,\xi)$ if
$d(x,z)\leq 1$.
\end{rem}
\begin{rem}\label{rem:perturbation}
  A typical example, of which the present setting constitutes an obvious
  generalisation, is the perturbation on $\To^2$ given by
  \begin{displaymath}
    q_\ve(x,y)=\ve^{-2}\bar q(\ve^{-1}(y-x))
  \end{displaymath}
  where $\bar q$ is nonnegative and continuously differentiable,
  $\supp(\bar q)\subset \{\xi\in\R^2\,:\;|\xi|\leq 1\}$,
  $\int\bar q(\xi)\,d\xi=1$ and
  $\inf\{\bar q(\xi):|\xi|\leq\frac 12\}=:\omega>0$. In this case
  $\tilde q_\ve(x,\xi)=\ve^{-2}\bar q(\ve^{-1}\xi)$,
  $\partial_1\tilde q_\ve=0$
  and $\partial_2\tilde q(x,\xi)=\ve^{-3}\bar q'(\ve^{-1}\xi)$ so that
  $\tilde q_\ve$, $\partial_1\tilde q_\ve$ and $\ve^{-1}\partial_2\tilde q_\ve$
  are $\ve^{-1}$-dominated by $\tilde q_{\ve}$ with $a=2$ and 
  $M=\omega^{-1}\|\bar q\|_{\Co^2}$.
\end{rem}

Combined with a deterministic transformation $T$, any stochastic
kernel 
$q_\ve$
gives rise to a Markov process $(X_n^\ve)_{n\geq0}$ on $\To^2$ where
$X_0$ can have any probability distribution and
\begin{displaymath}
  P(X_{n+1}^\ve\in A\ |\ X_n^\ve)=\int_A q_\ve(T(X_{n}^\ve),y)\,m(dy)\ .
\end{displaymath}

\begin{defin} A family 
$(X_n^\ve)_{n\geq 0}$ associated to an admissible stochastic 
kernel family is called 
a \emph{small random perturbation} of $T$.
\end{defin}    

If the distribution of $X_n^\ve$ has density $f$ with respect to $m$,
then the distribution of $X_{n+1}^\ve$ has density $Q_\ve\Lp f$ where
\begin{displaymath}
  Q_\ve f(y):=\int_{\To^2} f(x)q_\ve(x,y)\,m(dx)\ .
\end{displaymath}
Obviously $Q_\ve$ extends to $\B_w(\To^2)$ by continuity.
If $\vf\in \Co^0(\To^2,\R)$ is an ``observable'', then
the conditional expectation of $\vf(X_{n+1}^\ve)$ given $X_n^\ve$ is
$Q_\ve^*\vf(TX_n^\ve)$ where
\begin{displaymath}
  Q_\ve^*\vf(x)=\int_{\To^2}q_\ve(x,y)\vf(y)\,m(dy)\ .
\end{displaymath}
In particular, $Q_\ve^*1=1$.
Both operators are dual in the sense that
\begin{displaymath}
  \langle Q_\ve f,\vf\rangle=\langle f,Q_\ve^*\vf\rangle\ .
\end{displaymath}
We will call the operators $Q_\ve$ 
{\it smooth averaging operators}.\footnote{It is interesting to notice --
and it will be used in the following - that the product of smooth averaging 
operators is again a smooth averaging operator, \cf Lemma~\ref{lem:Qmult}.}

As already mentioned, the results of this section make much stronger
use of regularity properties of the foliations than did those of the 
previous section. Indeed, we assume that the unstable distribution is
$\Co^{\tau}$, for some
$\tau-1\ge\alpha\geq\gamma>\beta>0$.\footnote{This is always the
case if the unstable foliation has codimension one or if the
``gap'' between the weakest and the strongest contraction rates
in the stable sub-bundle is sufficiently small, see
the appendix for more details.}

The first basic estimate is stated in the following lemma.

\begin{lem}\label{lem:properties}
Suppose that the unstable foliation is $\Co^{\tau}$,
$\tau-1\ge\alpha\geq\gamma>\beta>0$. Then any smooth averaging operator 
can be extended in a unique way to a continuous operator on
$\B(\manif)$ and one can choose $\delta$
(small) and $b$ (large) in such a way that there exists $K>0$ with
\begin{displaymath}
  \|Q_\ve\|_w\leq K\qquad\text{and}\qquad
  \|Q_\ve f\|\leq 3\|f\| +K\|f\|_w
\end{displaymath}
for all sufficiently small $\ve>0$ and all $f\in\B(\manif)$.
\end{lem}

Remember from the previous discussion that the transfer operator
associated to the randomly perturbed system is defined by
\begin{displaymath}
  \Lp_\ve:=Q_\ve\Lp\ .
\end{displaymath}
The operators $\Lp_\ve$ satisfy a uniform Lasota-Yorke type inequality.

\begin{lem}\label{lem:propbound}
  Assume that $\manif=\To^2$. There is a constant $B'>0$ such that
  for sufficiently small $\ve>0$ and each $n\in\N$
  \begin{displaymath}
    \|\Lp^n_\ve\|_w\leq B'
    \quad\text{ and }\quad
    \|\Lp^n_\ve f\|\leq 9A^2\sigma^{n}\|f\|+B'\|f\|_w
  \end{displaymath}
for all $f\in\B(\manif)$.
\end{lem}

To state the next relevant fact it is helpful to introduce a special
norm for operators that involves the strong and the weak norm
and was used previously in \cite{Keller1,KL}.
\begin{displaymath}
  ||| Q_\ve |||:=\sup_{\{f\in \B(\manif)\,:\;\|f\|\leq 1\}}\|Q_\ve
  f\|_w .
\end{displaymath}

\begin{lem}\label{lem:three-norm}
Assume that $\manif=\To^2$.
There exists a constant $K>0$ such that
\begin{displaymath}
  |||Q_\ve -\Id|||\leq K\ve^{\gamma-\beta}.
\end{displaymath}
\end{lem}

Lemma~\ref{lem:three-norm} implies 
\begin{equation}\label{eq:tranpert}
  |||\Lp_\ve-\Lp|||\leq \|\Lp\|\,K\,\ve^{\gamma-\beta}\ .
\end{equation}

The proofs of the last three lemmas are furnished in section
\ref{subsec:smooth-random-proof}. Estimate (\ref{eq:tranpert})
together with Lemma~\ref{lem:propbound} allow to supplement Theorem
\ref{thm:quasicompact} with a rather strong result on spectral
stability under smooth random perturbations. Indeed, the
following is a direct consequence of \cite{KL}.

To formulate the result precisely we need some preparations.
\relax Fix any $r\in(\sigma,1)$
and denote by 
$\spectr(\Lp)$ the spectrum of $\Lp:\B(\manif)\to\B(\manif)$. 
Since the essential spectral radius of $\Lp$ does not exceed $\sigma$, 
the set $\spectr(\Lp)\cap\{z\in\C:|z|\geq r\}$ consists of a finite
number of eigenvalues $\lambda_1,\dots,\lambda_p$ of finite
multiplicity. Changing $r$ slightly we may assume that
$\spectr(\Lp)\cap\{z\in\C:\;|z|=r\}=\emptyset$. Hence there is
$\delta_*\in(0,r-\sigma)$ such that
\begin{gather*}
  |\lambda_i-\lambda_j|>\delta_*\quad(i\neq j)\\
  \dist(\spectr(\Lp),\{|z|=r\})>\delta_*
\end{gather*}
It is shown in \cite[Theorem 1]{KL} that, given $\delta\leq \delta_*$,
for sufficiently small $\ve$ the spectral projectors
\begin{equation}\label{eq:spectral-projectors}
  \begin{split}
    \Pi^{(j,\delta)}_\ve
    &:=
    \frac 1{2\pi\imath}\int_{\{|z-\lambda_j|
    =\delta\}}(z-\Lp_\ve)^{-1}\,dz\,\\
    \Pi^{(r)}_\ve
    &:=
    \frac 1{2\pi\imath}\int_{\{|z|=r\}}(z-\Lp_\ve)^{-1}\,dz\\
  \end{split}
\end{equation}
are well defined (and do not depend on $\delta$).
\relax Finally we set $\eta:=1-\frac{\log r}{\log\sigma}$ and observe that
$0<\eta<1$.

\relax Further results that follow directly from \cite{KL} 
can now be summarised as follows.

\begin{Thm}
  \label{thm:spectral-stability}
  Let $T:\To^2\to\To^2$ be a $\CoT$ Anosov diffeomorphism.
  For each $\delta\in (0,\delta_*]$ there exists
  $\ve_0=\ve_0(\delta,r)$ such that for $\ve\in[0,\ve_0]$ holds:
  \begin{enumerate}[a)]
  \item There is $K_1=K_1(\delta,r)>0$ such that
    $|||\Pi^{(j,\delta)}_\ve-\Pi^{(j,\delta)}_0|||\leq K_1\,
    \ve^{\eta(\gamma-\beta)}$.
  \item $\rank(\Pi^{(j,\delta)}_\ve)=\rank(\Pi^{(j,\delta)}_0)$
  \item $\lim_{\ve\to0}|||\Pi^{(r)}_\ve-\Pi^{(r)}_0|||=0$.
  \item There is $K_2=K_2(\delta,r)>0$ such that
    $\|\Lp_\ve^n\Pi^{(r)}_\ve\|\leq K_2\,r^n$ for all $n\in\N$.
  \end{enumerate}
\end{Thm}

\relax For later use we provide the key estimates from \cite{KL}
that are behind the previous theorem.

\begin{prop}\label{prop:resolvent-estimates}
  In the situation of Theorem~\ref{thm:spectral-stability}
  there are constants $\ve_0=\ve_0(\delta,r)$, $a=a(r)$,
  $b=b(\delta,r)$ and $c=c(\delta,r)$ such that for $\ve\in[0,\ve_0]$
  and $z\in\C$ with $r<|z|\leq 2$ and $\dist(z,\spectr(\Lp))>\delta$
  \begin{equation}\label{eq:resolvent-estimate}
    \|(z-\Lp_\ve)^{-1}f\|
    \leq
    a\,\|f\|+b\,\|f\|_w\quad\text{ for all }f\in\B(\manif)
  \end{equation}
  and
  \begin{equation}\label{eq:resolvent-difference}
    |||(z-\Lp_\ve)^{-1}-(z-\Lp)^{-1}|||
    \leq
    c\,\ve^{\eta(\gamma-\beta)}\,\|(z-\Lp)^{-1}\|\ .
  \end{equation}
\end{prop}

\begin{rem}\label{rem:resolvent-constants}
  The constants $\ve_0,a,b$ and $c$ are explicitly determined in the proofs
  of \cite{KL}.
  For the present setting one can deduce that
  $a$ and $c$ depend on the map $T$ only through
  the constants $\sigma$ and $B'$ from Lemma~\ref{lem:propbound}.
  (Working through the proof of that lemma one notices
  that $\sigma$ and $B'$ change continuously under small $\Co^2$-perturbations
  of $T$.) The constants $\ve_0$ and $b$ depend on $T$ through
  $\sigma$ and $B'$, and also through the quantity
  \begin{equation}\label{eq:H}
    H(\delta,r)
    :=
    \sup\left\{\|(z-\Lp)^{-1}\|:\;
      z\in\C, |z|>r,\dist(z,\spectr(\Lp))>\delta\right\}\ .
  \end{equation}
\end{rem}

\begin{rem}\label{rem:double-perturbation}
In the next section we will have to deal with perturbations of the
type $Q_\ve\Lp Q_\ve$. Proposition~\ref{prop:resolvent-estimates}
applies to these operators as well. To see this note that
\begin{displaymath}
  |||Q_\ve\Lp Q_\ve-\Lp|||
  \leq
  \|\Lp_\ve\|_w\cdot|||Q_\ve-\Id|||+|||Q_\ve-\Id|||\cdot\|\Lp\|
  \leq
  \const\cdot\ve^{\gamma-\beta}
\end{displaymath}
and that the estimates from Lemma~\ref{lem:propbound} carry over from 
$\Lp_\ve$ to $Q_\ve\Lp Q_\ve$ because
$(Q_\ve\Lp Q_\ve)^n=Q_\ve\Lp(Q_\ve^2\Lp)^{n-1}Q_\ve$, and
$Q_\ve^2$ is a again smooth averaging operator, \cf Lemma~\ref{lem:Qmult}.
\end{rem}

\medskip

\noindent \textbf{Quasi-compactness in dimension $d=2$. } 
We finish this subsection with a short proof of how to deduce the
quasicompactness and the estimate for the essential spectral radius
of $\Lp$ in the two-dimensional case without relying on 
Proposition~\ref{prop:compactnes} (compact embedding). 
Instead we use Lemmas
\ref{lem:properties} and \ref{lem:three-norm}
(hence assuming that $\tau-1\geq \gamma>\beta$).
Since, by Lemma~\ref{lem:LY},
$\|\Lp^n-\Lp^nQ_\ve\|\leq\sigma^n\|\Id-Q_\ve\|+\frac B{a-\sigma}
|||\Id-Q_\ve|||$,
the two above mentioned lemmas guarantee that one can choose $\ve$ so small
that $\|\Lp^n-\Lp^nQ_\ve\|\leq\const\,\sigma^n$.
But the $\Lp^nQ_\ve$ are compact:
\begin{equation}\label{eq:chain}
  \B(\manif)\overset{Q_\epsilon}\longrightarrow \Co^1(\manif,\R)
  \overset{\mbox{\scriptsize compact}}\hookrightarrow \Co^0(\manif,\R)
  \overset{\mbox{\scriptsize conts.}}\hookrightarrow \B_w(\manif)
  \overset{\Lp^n}\longrightarrow \B(\manif)\ ,
\end{equation}
so \cite[Lemma VIII.8.2]{DS} implies that $\Lp$ is quasicompact with 
essential spectral radius bounded by 
$\liminf_{n\to\infty}\sqrt[n]{\const\,\sigma^n}=\sigma$. 

\subsection{Spectral stability -- Deterministic perturbations.}
\label{subsec:deterministic-perturbations}

The next natural question concerns the relation between the
spectrum associated to two Anosov maps $T$ and $\tilde T$  when the two
maps are $\Co^1$ close. Clearly the associated operators $\Lp_T$ and 
$\Lp_{\tilde T}$
live on different Banach spaces $\B(\manif)$ and $\tilde\B(\manif)$,
respectively,
so their spectra cannot be
compared directly. 
However, denoting $R(z)=(z-\Lp_T)^{-1}:\B(\manif)\to\B(\manif)$ and 
$\tilde R(z)=(z-\Lp_{\tilde T})^{-1}:\tilde\B(\manif)\to\tilde\B(\manif)$
(at points $z$ where they are well defined), both these resolvents can 
be considered as linear operators from $\Co^1(\manif,\R)$ to
$\Co^1(\manif,\R)^*$, because\footnote{This statement must be interpreted 
in the sense of Remark~\ref{rem:norm-seminorm}.}
\begin{equation}\label{eq:four-norms} 
  \|f\|_{\Co^1}^*\leq\|f\|_w\leq\|f\|\leq\|f\|_{\Co^1}\quad\text{ and also }
  \quad\|f\|_{\Co^1}^*\leq\tildenorm{f}_w\leq\tildenorm{f}\leq\|f\|_{\Co^1}
\end{equation}
for each $f\in\Co^1$, 
where $\|\cdot\|_w$, $\|\cdot\|$ and $\tildenorm{\cdot}_w$, 
$\tildenorm{\cdot}$
denote the norms associated to $T$ and $\tilde T$, respectively.
By the same reason also the corresponding spectral projectors
(see (\ref{eq:spectral-projectors})) can be interpreted as operators from
$\Co^1(\manif,\R)$ to
$\Co^1(\manif,\R)^*$, and since their ranges are finite-dimensional,
there is even no big loss in doing so. Therefore the clue to comparing the 
spectra of $\Lp_T$ and $\Lp_{\tilde T}$ is a good bound on
$\cnorm{R(z)-\tilde R(z)}$, which in turn is obtained by approximating both 
transfer operators by suitable randomly perturbed operators.

\begin{prop}\label{prop:determ-perturb}
Given an Anosov map $T:\To^2\to\To^2$ and a constant $K>\|T\|_{\CoT}$, 
a real number $r$ bigger than the constant $\sigma$ associated with $T$ 
in Lemma~\ref{lem:LY} and some $\delta>0$, 
there are constants $\kappa=\kappa(T)$,
$C_i=C_i(T,K,r,\delta)$ and $\ell_0=\ell_0(T,K,r,\delta)$
such that for each map $\tilde T:\To^2\to\To^2$ with
$\dist_{\Co^1}(\tilde T,T)\leq\ell_0$ and $\|\tilde T\|_\CoT\leq K$
\footnote{Note that also $\tilde T$ is an Anosov diffeomorphism if 
$\dist_{\Co^1}(\tilde T,T)$ is small enough.}
\begin{align}
  \tildenorm{\tilde R(z)}&\leq C_0
  \quad\text{ and }\quad\label{eq:ihes3}\\
  \cnorm{\tilde R(z)-R(z)}&\leq C_1\cdot
  \dist_{\Co^1}(\tilde T,T)^{\kappa\eta(\gamma-\beta)}
  \label{eq:ihes4}
\end{align}
provided that $|z|>r$ and $\dist(z,\spectr(\Lp_T))>\delta$ where again
$\eta=1-\frac{\log r}{\log\sigma}$.
\end{prop}
\noindent The proof of this proposition can be found in 
section~\ref{subsec:deterministic-proof} and rests on two observations:
\smallskip\\
1) If $\dist_{\Co^1}(T,\tilde T)$ is bounded by a suitable power of $\ve$, 
then $\|Q_\ve f\|\leq\const\cdot\tildenorm{f}$. This is proved in 
Lemma~\ref{lem:norm-comparison} and leads directly to (\ref{eq:ihes3}).
\smallskip\\
2) In view of (\ref{eq:four-norms}) we have
\begin{equation}\label{eq:three-term-decomp}
\begin{split}
  &\cnorm{\tilde R(z)-R(z)}\\
  &\qquad\leq
  |||\tilde R(z)-\tilde R_\ve(z)|||^\sim+\cnorm{\tilde R_\ve(z)-R_\ve(z)}
 +|||R_\ve(z)-R(z)|||
\end{split}
\end{equation}  
where
$R_\ve(z):=(z-Q_\ve\Lp_{T}Q_\ve)^{-1}$ 
and $\tilde R_\ve(z):=(z-Q_\ve\Lp_{\tilde T}Q_\ve)^{-1}$ 
for a suitable smooth averaging operator 
$Q_\ve$ with a $\Co^2$ kernel.
In section~\ref{subsec:deterministic-proof} we estimate these three terms.
\relax For the third one we use Proposition~\ref{prop:resolvent-estimates},
the second one will turn out to be small if $\dist_{\Co^1}(T,\tilde T)$
is of the order of a certain power of $\ve$, and the first one can again
be estimated by help of Proposition~\ref{prop:resolvent-estimates},
but with the r\^oles of $\Lp$ and $\Lp_\ve$ interchanged, which
requires an estimate on $\|\tilde R_\ve(z)\|$ in terms of quantities that
depend only on the unperturbed map $T$, see
Remark~\ref{rem:resolvent-constants}.
In a different context this idea was already used in \cite{Liverani-paderborn}.
\smallskip

Observing the integral representation (\ref{eq:spectral-projectors})
of the spectral projectors in
terms of the resolvent, the following theorem is an easy corollary:

\begin{Thm}
Given an Anosov map $T:\To^2\to\To^2$, 
for each map $\tilde T$ sufficiently close to
$T$ in $\Co^1$--distance and satisfying uniform $\CoT$-bounds 
the non-essential spectra
$\spectr_r(\Lp_T,\B_T(\manif,\R)))$ and
$\spectr_r(\Lp_{\tilde T},\B_{\tilde T}(\manif,\R))$ are close with 
multiplicities and the $\cnorm{\cdot}$--distance of corresponding spectral
projectors is small.
\end{Thm}

\subsection{Spectral stability -- Ulam finite rank approximation}
\label{subsec:Ulam}

An Ulam approximation is constructed in the following way: Let 
$\Ac=\{A_i\}$ be a partition of $\To^2$ into convex polytopes 
(\eg into squares). Define the operator, in $L^1(\To^2,\R)$,
\begin{equation}\label{eq:partition}
  \Pi_{\Ac}f(x) 
  := 
  \sum\limits_{A\in \Ac} \frac1{m(A)}~ \eins_A(x)\int_A f\,dm
\end{equation}
Ulam's idea is then to look at the finite rank operator
\begin{equation}
\label{eq:ulam}
P_\Ac:=\Pi_\Ac\Lp\Pi_\Ac
\end{equation}
and to prove that, in some sense, its spectral properties are close to
the ones of the original transfer operator; this is usually referred
to as the {\em Ulam conjecture}. 
Since $P_\Ac$ is essentially a finite matrix, its spectral properties
are numerically accessible. Obviously these spectral properties
are the same as those of $\Lp_\Ac:=\Pi_\Ac\Lp$.

Clearly we cannot compare directly the spectra of $\Lp$ and $\Lp_\Ac$,
since the space $\Pi_\Ac(\B(\To^2))$ consists of piecewise
constant functions which are, in particular, discontinuous along 
unstable fibres, in contrast to the functions belonging to our space 
$\B(\To^2)$.
This problem could be overcome by extending the space
$\B(\To^2)$ to a space including the indicator functions 
$\eins_{A_i}$ as we will discuss
briefly in the next section.
It turns out, however, that this would not really help.
At the end of this section we provide examples of linear hyperbolic
torus automorphisms for which straightforward Ulam discretisations $\Lp_\Ac$
produce eigenvalues far away from all possible eigenvalues of $\Lp$.

Another possibility to avoid this problem is to replace
the indicator functions $\eins_{A}(x)$ in (\ref{eq:partition})
by a smooth partition of unity. For example one could
use a convolution kernel 
\[
q_\ve(x,y):=\ve^{-2}\bar q(\ve^{-1}(x-y)); \quad \bar
q\in\Co^1(\R^2,\R);\quad \int \bar q(x,y)m(dy)=1, 
\]
and its associated smooth averaging operator operator $Q_\ve$
and consider the partition of unity $\phi_A:=Q_\ve\eins_{A}$
with $\ve$ of the order of the diameter of the sets $A$.
With this choice of $\ve$, however, it does not seem possible to prove 
spectral stability of the resulting discretisation $Q_\ve\Pi_\Ac\Lp$, 
the obstacles being basically the same as in the case of the original 
Ulam procedure. A much bigger $\ve$, roughly of the order of 
$\diam(\Ac)^{1/3}$, is needed to achieve a spectral stability result 
where the approximation error for the resolvent is again of the order 
$\ve^{\eta(\gamma-\beta)}$ as it was the case for smooth perturbations. 

Since we will use $L^1_m(\To^2)$ as an auxiliary space which contains
the ranges of both $Q_\ve$ and $\Pi_\Ac$ (the price to pay for this is 
the big $\ve$!), the discontinuity
of the indicator functions $\eins_A(x)$ plays absolutely no r\^ole in our
proof, and for technical reasons we prefer to work with a slightly
different discretisation.
Namely, we define the finite rank operators
\begin{equation}
\label{eq:smoothulam}
  P_{\Ac,\ve}:=\Pi_\Ac Q_\ve\Lp\Pi_\Ac
  \quad\text{and}\quad
  \Lp_{\Ac,\ve}:=\Pi_\Ac Q_\ve\Lp\ .
\end{equation}
Both operators have the same spectral properties. So we will
compare the resolvent of $\Lp_{\Ac,\ve}$ with that of $\Lp$, while 
$P_{\Ac,\ve}$ can be used for numerical investigations, 
since it can be considered as a matrix 
acting on the same space as $P_\Ac$. 
So it can serve
as a tool to study the spectral properties of $\Lp$ on the
same footing as the standard Ulam discretisation.
\relax For results on the 
numerical approximation of SRB-measures such an approach was
previously used in \cite{DJ1,DJ2}\footnote{In these papers the 
discretisation of smoothly perturbed Anosov diffeomorphisms serves 
as a theoretical justification for numerical discretisations. The 
effectively implemented procedures, however, do not make use of random 
noise. In this context observe the example at the end of the current 
section.}; see also \cite{Kifer2} for stabilising effects of noise on 
numerical simulations of dynamical systems.

As an intermediate step of our program it turns out to be
necessary to consider $\Pi_\Ac$, $Q_\ve$ and $\Lp$ as
operators acting on the Banach space of functions of bounded
variation $\BV(\To^2,\R)$. 
Observe that $\BV(\To^2,\R)$ consists exactly of those functions 
$f\in L^1(\To^2,\R)$ that can be approximated in $L^1$--norm
by functions $f_n$ 
from the Sobolev space $W^{1,1}$ (integrable functions with 
integrable derivatives) in such a way that $\|f\|_{\BV}=\limn\|f_n\|_{1,1}$.
Recall also that for any sequence $(f_n)$ from $W^{1,1}$ with
$\limn\|f-f_n\|_1=0$ holds: $\|f\|_{\BV}\leq\liminf_{n\to\infty}\|f_n\|_{1,1}$.
Therefore it suffices to prove the following lemma
for $f\in W^{1,1}$ (and hence only for $f\in\Co^1$) 
with the $\|\cdot\|_{1,1}$--norm 
instead of the $\|\cdot\|_{\BV}$--norm. But that is a simple exercise.

\begin{lem}
\label{lem:bvaverage} There exists a constant $K>0$ such that, for
each $\ve\in(0,1)$ and $f\in\BV(\To^2,\R)$, 
\begin{displaymath}
  \begin{split}
    \|Q_\ve f\|_\BV&\leq \|f\|_\BV\\
    \|Q_\ve f\|_\BV&\leq K\ve^{-1}\|f\|_1\\
    \|\Lp f\|_\BV&\leq K\|f\|_\BV\ .
  \end{split}
\end{displaymath}
\end{lem}

This lemma shows in particular that $Q_\ve$ and $\Lp$ are well
defined bounded operator on $\BV(\To^2,\R)$. The next lemma shows that
the same is true for $\Pi_\Ac$.

\begin{lem}[\cite{BuK}]\label{lem:buzzikeller}
If the partition $\Ac$ is made of convex polytopes of size\footnote{The size 
here is the side length of the smallest coordinate cube which contains the 
polytope.} at most $N^{-1}$, then, for each $f\in \BV(\To^2,\R)$,
\begin{displaymath}
  \begin{split}
    \|\Pi_\Ac f\|_\BV&\leq 2\|f\|_\BV\\
    \|f-\Pi_\Ac f\|_1&\leq 2N^{-1}\|f\|_\BV
  \end{split}
\end{displaymath}
\end{lem}

The first inequality of this lemma is proven in Lemma 10 of
\cite{BuK}, the second one in Corollary 3 of the same paper.

The following proposition contains the key estimates of this section.

\begin{prop}\label{prop:ulam-resolvent}
Let $\Ac$ be a partition of $\To^2$ into convex polytopes of size at most
$N^{-1}$.
If $N^{-1}\ve^{-3}$ is smaller than some constant\footnote{This constant
depends on $a$ and $b$ from Proposition~\ref{prop:resolvent-estimates}
and hence
from the distance of $z$ to $\spectr(\Lp)$.},
then $(z-\Lp_{\Ac,\ve})$ is invertible as an operator on $L^1(\To^2,\R)$ and
\begin{equation}\label{eq:ulam-resolvent-bound}
  \|(z-\Lp_{\Ac,\ve})^{-1}f\|_1
  \leq
  \const\,(\ve^{-1}\,\|Q_\ve\Lp f\|+\|f\|_1)
\end{equation}
\relax Furthermore,
\begin{equation}\label{eq:ulam-resolvent-difference}
  \cnorm{(z-\Lp_{\Ac,\ve})^{-1}-(z-Q_\ve\Lp)^{-1}}
  \leq
  \const\,N^{-1}\,\ve^{-3}\ .
\end{equation}
\end{prop}

Combined with Proposition~\ref{prop:resolvent-estimates} this yields at once
that under the conditions of the last proposition
\begin{equation}
  \cnorm{(z-\Lp_{\Ac,\ve})^{-1}-(z-\Lp)^{-1}}
  \leq
  \const\,\left(N^{-1}\,\ve^{-3}+\ve^{\eta(\gamma-\beta)}\right)\ ,
\end{equation}
where the constant depends on $z$ through the quantity $H(\delta,r)$ from
(\ref{eq:H}). Recall from section~\ref{subsec:stability-smooth}
that 
$\eta=1-\frac{\log r}{\log\sigma}$ where $r$ must be chosen such that 
$|z|>r>\sigma$.

The main result of this section now follows again from spectral calculus.

\begin{Thm}\label{thm:ulam}
Let $\Ac$ be a partition of $\To^2$ into convex polytopes of size at most
$N^{-1}$ where
$N=N(\ve)\approx\ve^{-3-\eta(\gamma-\beta)}$. Then the assertions
of Theorem~\ref{thm:spectral-stability} hold for $\Lp_{\Ac,\ve}$ instead 
of $\Lp_\ve$. In particular,
the non-essential spectra
$\spectr_r(\Lp,\B(\To^2))$ and
$\spectr_r(\Lp_{\Ac,\ve},L^1(\To^2,\R))$ are close with multiplicities
and the $\cnorm{\cdot}$--distance of corresponding spectral
projectors is of order $\ve^{\eta(\gamma-\beta)}$.
\end{Thm}

Since $\Lp_{\Ac,\ve}$ is a finite rank operator, hence its
spectrum and eigenvalues can be explicitly computed, the above
theorem provides a constructive tool to study the invariant
measure, the rate of decay of correlations etc. of the
map. Some related numerical implementations 
(using the original Ulam method modified by an adaptive choice
of cell sizes) are 
reviewed in \cite{Froyland2,DJ2}. It seems worth to mention that
our rate of convergence compares favourably to the one
proved in \cite{Junge} for the SRB measure, \ie for the
``eigenfunction'' of the eigenvalue $1$.

Just to have a more precise idea on how the business really works let
us briefly, and not optimally, discuss the determination of the
invariant measure in cases where the system is ergodic, i.e. where $1$ is a
simple eigenvalue of $\Lp$. 
\relax For a more detailed discussion of this type of questions see
\cite{Liverani-paderborn}.

Let $\mu$ be the invariant measure and
$h\in\B(\To^2)$ the corresponding eigenvector of $\Lp$ (that is
$\Lp h=h$ and $\mu(\vf)=\langle h,\vf\rangle$ for all
$\vf\in\D_\gamma$). Let $h_{\ve}, h_{\Ac,\ve,}\in L^1(\To^2,\R)$ be such that
$\Lp_{\Ac,\ve}h_{\Ac,\ve}=h_{\Ac,\ve}$ and
$Q_\ve\Lp h_\ve=h_\ve$, and suppose that 
$N\approx\ve^{-3-\eta(\gamma-\beta)}$. 
Then there exists $\rho>0$ such that
\begin{displaymath}
\begin{split}
  \cnorm{h_\ve-h_{\Ac,\ve}}
  &\leq 
  \int_{\{z\in\C:\;|z-1|=\rho\}}
  \cnorm{(z-\Lp_{\Ac,\ve})^{-1}(1)-(z-Q_\ve\Lp)^{-1}(1)}\\
  &\leq \const\,\ve^{\eta(\gamma-\beta)}\ .
\end{split}
\end{displaymath}
In addition, the results of subsection \ref{subsec:stability-smooth}
together with \cite{KL} imply
\begin{displaymath}
  \|h_\ve-h\|_w\leq \const\, \ve^{\eta(\gamma-\beta)}\ .
\end{displaymath}
Hence, calling $\mu_{\Ac,\ve}(\vf):=\int \vf\,h_{\Ac,\ve}\,dm$,
for each $\vf\in \Co^{1}(\To^2,\R)$ holds
\begin{equation}\label{eq:closemeasure}
|\mu(\vf)-\mu_{\Ac,\ve}(\vf)|\leq  C_2
\ve^{\eta(\gamma-\beta)}\,|\vf|_{\Co^1}\ .
\end{equation}
Since we are focusing on the eigenvalue $1$, the radius $r$ can be chosen
arbitrarily close to $1$ in which way also $\eta$ can be made as close to
$1$ as we wish. A closer look at the proofs of \cite{KL} reveals in fact
that convergence takes place at a rate $\ve^{\gamma-\beta}\log\ve^{-1}$.

Inequality (\ref{eq:closemeasure}) implies, by standard approximation
arguments, that the measures $\mu_{\Ac,\ve}$ converge weakly to
$\mu$, yet it provides much more informations since it tells us exactly
how much the two measures differ on a wide class of observables.
\medskip

\noindent
\textbf{Counterexamples to spectral stability of the pure Ulam method. }
As we already mentioned in the introduction the coexistence of expansion
and contraction in hyperbolic maps is a serious obstacle to the
applicability of the original Ulam scheme. Some results related to this
problem were discussed earlier in \cite{BlKe,Blank}, where in particular
problems with the discretisation of Arnold's cat map were pointed out.
Here we discuss a family of torus automorphisms whose transfer operators 
are all poorly approximated by Ulam discretisations.

Let $T_{a,k}:\To^2\to\To^2$ be the linear torus
automorphisms defined by the following two-parameter family of
$2\times2$ matrices
\begin{displaymath}
  \Amatrix_{a,k} := 
  \left(
    \begin{matrix}
      a&ka-1\cr a+1&k(a+1)-1\cr
    \end{matrix}
  \right),
  \qquad a,k \in\Z_{+}^{1},\ a\geq2 . 
\end{displaymath}
Trivially, the $T_{a,k}$ have stable and unstable foliations of class
$\Co^2$, so $\alpha=\tau-1=1$. Therefore we may choose 
$\gamma=1$ and any $0<\beta<1$ as parameters for our Banach spaces.
Denote by $\lambda_{a,k}>1$ and $\lambda_{a,k}^{-1}$ 
the two eigenvalues of $\Amatrix_{a,k}$. Then 
$\lambda_u=\lambda_{a,k}$ and $\lambda_s=\lambda_{a,k}^{-1}$.
In particular 
$\max\{\lambda_u^{-1},\lambda_s^{\beta}\}=\lambda_{a,k}^{-\beta}$.

Denote by $\Lp_{a,k}$ the transfer operator of $T_{a,k}$. We know from 
Example~\ref{exmp:linear-eigenvalues} that
\begin{displaymath}
  \spectr(\Lp_{a,k})\subseteq
  \left\{z\in\C\,:\;|z|\leq\max\{\lambda_u^{-1},\lambda_s^\beta\}\right\}
  \cup\{1\}
  =
  \left\{z\in\C\,:\;|z|\leq\lambda_{a,k}^{-\beta}\right\}
  \cup\{1\}\ .
\end{displaymath}
Consider Ulam discretisations $P_{\Ac_n}$ of $\Lp_{a,k}$ using
partitions $\Ac_n$ into squares with sides 
of length $n^{-1}$ parallel to the coordinate axes,
see (\ref{eq:ulam}). Mixed numerical and 
theoretical investigations that we line out below show that for many choices 
of $a$ and $k$ the spectra $\spectr(P_{\Ac_n})$ (for arbitrarily large 
even $n$) contain numbers with modulus strictly between $\lambda_{a,k}^{-1}$ 
and $1$ and hence between $\lambda_{a,k}^{-\beta}$ and $1$ when $\beta$ is 
chosen close enough to $1$. Hence the most straightforward Ulam procedure, 
based on squares with equal size and sides parallel to the axes, produces 
fake eigenvalues at arbitrarily fine grid sizes $n^{-1}$.
It is not clear to us whether this effect is an ``arithmetic artifact'' 
of torus \emph{linear} automorphisms 
or if it is a more common phenomenon for hyperbolic diffeomorphisms.

The task to determine (at least some) eigenvalues that occur in $P_{\Ac_n}$ 
for arbitrarily large $n$ is greatly simplified by the following observation:

\begin{prop}\label{prop:str-sp} \cite{BlKe} Let $\Amatrix$ be a $d\times d$
matrix with nonnegative integer entries, and define $T:\To^d\to\To^d$ by
$Tx:=\Amatrix x\mod1$.
Denote by $\Ac_n$ the partition of $\To^d$ into cubes of side length 
$n^{-1}$ with vertices in $n^{-1}\Z^d$. As before let
$P_{\Ac_n}=\Pi_{\Ac_n}\Lp_T\Pi_{\Ac_n}$.
Then
\begin{displaymath}
  \spectr(P_{\Ac_m})\subseteq\spectr(P_{\Ac_{mN}})
\end{displaymath}
for all positive integers $N$ and $m$.
\end{prop}

According to this proposition it suffices
to show that $P_{\Ac_2}$ has an eigenvalue with modulus strictly
between $\lambda_{a,k}^{-1}$ and $1$, since then this number belongs to
$\spectr(P_{\Ac_{2N}})$ for all $N\in\N$.

In case $a=2$ and $k=1$ the matrix representing the operator
$P_{\Ac_2}$ can be figured out by hand. It is
\begin{displaymath}
  \left(
    \begin{matrix}
      1/6&1/3&1/3&1/6\\
      1/3&1/6&1/6&1/3\\
      1/3&1/6&1/6&1/3\\
      1/6&1/3&1/3&1/6
    \end{matrix}
    \right)\ .
\end{displaymath}
This matrix has the
four eigenvalues $1,-1/3,0,0$.
In particular, $-1/3$ is a common eigenvalue of all $P_{\Ac_{2N}}$.
On the other hand, the matrix $\Amatrix_{2,1}$ defining the automorphism 
has the leading eigenvalue $\lambda_{2,1}=2+\sqrt 3$ so that
$\lambda_{2,1}^{-1}=2-\sqrt{3}\approx0.268$. With $\beta=0.9$ we have 
indeed $\lambda_{2,1}^{-\beta}<\frac 13<1$.

This, in fact, is not the worst case and other choices
of the parameters $a,k$ demonstrate even more striking differences
between the eigenvalues of the Ulam approximation by square partitions
and the spectrum of $\Lp_{a,k}$.
We presented the case $a=2$, $k=1$ in some detail, because in
this case everything can be calculated analytically, while for
general values of the parameters only numerical approximations are
available. By means of a program written in 
``Mathematica'' we have calculated eigenvalues for several other
choices of the parameter values.
\begin{center}
{\footnotesize
\begin{tabular}{|r|| r|r|l|l|| r|r|l|l|} \hline
$n$& $a$&$k$&$r_2$&$\lambda_{a,k}^{-1}$& $a$&$k$&$r_2$&$\lambda_{a,k}^{-1}$ \\
\hline
 2 & 20 & 1 & 0.3400 & 0.0250 & 20 & 3 & 0.1115 & 0.0122 \\
 4 & 20 & 1 & 0.3400 & 0.0250 & 20 & 3 & 0.2240 & 0.0122 \\
 6 & 20 & 1 & 0.3400 & 0.0250 & 20 & 3 & 0.1115 & 0.0122 \\
 8 & 20 & 1 & 0.3400 & 0.0250 & 20 & 3 & 0.2240 & 0.0122 \\
\hline\hline
 2 &500 & 1 & 0.5000 & 0.0010 &500 & 3 & 0.2500 & 0.0005 \\
 4 &500 & 1 & 0.5000 & 0.0010 &500 & 3 & 0.2500 & 0.0005 \\
 6 &500 & 1 & 0.5000 & 0.0010 &500 & 3 & 0.2500 & 0.0005 \\
 8 &500 & 1 & 0.5000 & 0.0010 &500 & 3 & 0.2500 & 0.0005 \\
\hline \end{tabular}}
\end{center}
Here $r_{2}$ is the modulus of the second largest eigenvalue
of $P_{\Ac_n}$. It is to be compared to
$\lambda_{a,k}^{-1}$, which we consider to be the ``true'' second eigenvalue.

What is still possible is that the leading eigenfunction -- the SRB
measure -- may be stable for the class of maps we consider (the
above example does not contradict to this -- the SRB measure is
preserved). For piecewise expanding interval maps this is known to be true,
see for example \cite{Blank} and further references therein.
Moreover numerous numerical studies confirm this stability for a
much broader classes of dynamical systems, see \eg \cite{DJ2,Froyland2},
but presently we do not have an
adequate explanation for this numerical effectiveness of the Ulam method.

\relax For Ulam discretisations based on Markov partitions the situation 
is much better: In the case of linear automorphisms of $\To^2$ 
Brini et al.\cite{BSTV} showed that there are arbitrarily 
fine Markov partitions $\Ac$ for which
$\spectr(P_\Ac)\subseteq\{1,\lambda_s,\lambda_s^2,0\}$.
In particular they have no eigenvalue with modulus in
$(\lambda_{a,k}^{-1},1)$.
\relax Froyland \cite{Froyland1} proved for general Anosov maps of $\To^2$
that the eigenvector to the eigenvalue $1$ of Ulam discretisations 
based on suitable Markov partitions approximates the SRB measure 
if the partitions get finer and finer.

\subsection{Remarks on an extension of the space $\B(\manif)$}
\label{subsec:remarks}
\relax Following one of the paths that lead from the Sobolev space
$W^{1,1}$ of differentiable functions with integrable derivative
to the space $\BV$ of functions of bounded variation,
we show in this section how to extend the norm on our space $\B(\manif)$ 
to a larger subspace $\bar\B(\manif)$ of $\B_w(\manif)$ that contains, 
in particular, the space $\BV(\manif)$. Most of this program is abstract 
and works equally well in any dimension. Only to show that $\B(\manif)$ 
is a \emph{closed} subspace of $\bar\B(\manif)$ needs much
more effort, and since we do not really make use of it except in the 
discussion of Ulam discretisations on the two-dimensional torus, we will 
prove that only in the case $\manif=\To^2$.

\relax For $f\in\B_w(\manif)$ let
\begin{equation}
  \newnorm{f}
  :=
  \lim_{\epsilon\to0}\inf\left\{\|g\|\,:\;g\in\B(\manif)
     ,\,\|f-g\|_w\leq\epsilon\right\}\ .
\end{equation}
Observe that one can replace the limit by a supremum over $\ve>0$.
It is easy to see that $\newnorm{\cdot}$ is a norm on
\begin{displaymath}
  \bar\B(\manif):=\left\{f\in\B_w(\manif)\,:\;\newnorm{f}<\infty\right\}\ .
\end{displaymath}
Then $(\bar\B(\manif),\newnorm{\cdot})$ is clearly a normed linear space,
and since $\newnorm{f}\leq\|f\|$ for $f\in\B(\manif)$ by definition, we
have $\B(\manif)\subseteq\bar\B(\manif)$.\footnote{Recall 
Remark~\ref{rem:norm-seminorm} for a discussion of 
the embedding of $\B(\manif)$ into $\B_w(\manif)$.}
We list some simple consequences:
\begin{enumerate}[i)]
\item 
The unit ball $\bar\B_1:=\{f\in\bar\B(\manif)\,:\;\newnorm{f}\leq 1\}$
is the $\|\cdot\|_w$--closure of the unit ball
$\B_1:=\{f\in\B(\manif)\,:\;\|f\|\leq 1\}$.
\item 
$\B_1$ is closed in $\B_w(\manif)$ if and only if the norms 
$\|\cdot\|$ and $\newnorm{\cdot}$ are equivalent on $\B$.
\item
Since $\B_1$ is a relatively compact subset of $\B_w(\manif)$,
$\bar \B_1$ is a compact subset of $\B_w(\manif)$.
\item 
For $f\in\Co^1(\manif,\R)$ we have $\|f\|\leq\|f\|_{1,1}$ and
$\|f\|_w\leq\|f\|_1$. Therefore
\begin{displaymath}
\BV_1=\|\cdot\|_1\text{-closure}((W^{1,1})_1)
\subseteq\|\cdot\|_w\text{-closure}((W^{1,1})_1)
\subseteq\|\cdot\|_w\text{-closure}(\B_1)
=\bar\B_1\ .
\end{displaymath}
(Here the lower index $1$ denotes the unit ball of the respective space.)
In particular, $\bar\B(\manif)$ contains indicator functions of sufficiently 
regular sets.
\item $\Lp$ extends to a bounded operator on $\bar\B(\manif)$.
\item The Lasota Yorke type inequality carries over to $\bar \B(\manif)$:
      $\newnorm{\Lp^n f}\leq3A^2\sigma^n\newnorm{f}+K\|f\|_w$.
\item In particular, $\Lp$ acts on $\bar\B(\manif)$ as a quasicompact operator
      with essential spectral radius bounded by $\sigma$.
\end{enumerate}

What we cannot conclude at this point is that
$\B(\manif)$ embeds as a \emph{closed} linear subspace into $\bar\B(\manif)$.
This is the case if and only if there is some $C>0$ such that
\begin{equation}\label{eq:crucial}
  \|f\|\leq C\cdot\newnorm{f}\quad\text{for $f\in\B(\manif)$,}
\end{equation}
\ie if the two norms are equivalent.\footnote{The ``if'' implication is 
obvious. For the ``only if'' implication observe that if $\B(\manif)$
is closed in $\bar\B(\manif)$, then $(\B(\manif),\newnorm{\cdot})$
is a Banach space. Since $\newnorm{\cdot}\leq\|\cdot\|$, 
the open mapping theorem
implies (\ref{eq:crucial}).}
In section~\ref{subsec:unorm-proof} we will prove the following lemma.
\begin{lem}\label{lem:crucial}
  The estimate (\ref{eq:crucial}) holds for $\manif=\To^2$.
\end{lem}
It turns out that for torus automorphisms (in any dimension, in fact)
the constant $C=1$ so that both norms coincide.

An important consequence of this observation is:
\begin{lem}
\item If $\Lp f=\lambda f$ for some $f\in\bar \B(\To^2)$ 
and $|\lambda|>\sigma$,
then $f\in\B(\To^2)$.
\end{lem} 
\begin{proof}
Since $f\in\bar\B(\To^2)$, there are $g_n\in\B(\To^2)$ 
with $\|f-g_n\|_w\leq\frac 1n$
and $\|g_n\|\leq\newnorm{f}$. Hence
\begin{displaymath}
\begin{split}
  \newnorm{f-\lambda^{-n}\Lp^n g_n}
  &=
  \newnorm{\lambda^{-n}\Lp^n(f-g_n)}
  \leq
  3A^2\left(\frac\sigma{|\lambda|}\right)^n\newnorm{f-g_n}+C\,\|f-g_n\|_w\\
  &\leq
  3A^2\left(\frac\sigma{|\lambda|}\right)^n\,2\newnorm{f}+\frac Cn
  \to 0
\end{split}
\end{displaymath}
so that $f$ belongs to the $\newnorm{\cdot}$--closure of $\B(\To^2)$.
But since $\B(\To^2)$ is closed in $\bar\B(\To^2)$, this implies that 
$f\in\B(\To^2)$.
\end{proof}
With an only slightly more complicated argument one shows that also
if $(\lambda-\Lp)^kf=0$ for some $k>1$ and $f\in\bar\B(\To^2)$ 
and $\lambda$ as above, then $f\in\B(\To^2)$.

\section{Proofs}

In the sequel we will mostly write $\B$ and $\B_w$ instead of $\B(\manif)$
and $\B_w(\manif)$.

\subsection{Proofs: Lasota-Yorke type estimate}
\label{subsec:LYproof}

\begin{proof}[\bf Proof of Lemma \ref{lem:LY}]
Let us start by considering only $f\in \Co^1(\manif,\R)$.

Recall from (\ref{eq:jacobian}) that 
$\Lp^n f(x)=f\circ T^{-n}\cdot g_n$ 
where the Jacobian $g_n$ of $T^{-n}$ is of class $\CoJ$. Then, for
$v\in \V_\beta$, holds
\begin{equation}
\label{eq:differenzia}
  d_x(\Lp^n f)(v(x))=d_xg_n(v(x))\cdot f\circ T^{-n}(x)
  +g_n(x)\cdot d_{T^{-n}(x)}f(d_{x}T^{-n}(v(x))).
\end{equation}
To continue we need to note the following.

\begin{sublem}\label{vectorlem}
For each $n\in\N$ there is $\delta(n)>0$ such that if $0<\delta<\delta(n)$
($\delta$ from the definition of
$H_\beta^s(v)$ in (\ref{eq:vectornorm})), then
\begin{displaymath}
  {R_nv}(x):=A^{-2}\lambda_u^n\,d_{T^nx}T^{-n}(v(T^nx))\in\V_\beta
\end{displaymath}
for each $v\in \V_\beta$.
\end{sublem}
\begin{proof}
Obviously, 
$|{R_nv}|_\infty\leq A^{-2}\lambda_u^n A\lambda_u^{-n}|v|_\infty\leq1$.

To investigate the H\"older norm of ${R_nv}$ it is more convenient
to introduce normal coordinates. Let us be more precise. Given an
arbitrary point $x\in\manif$ let us consider a neighbourhood $x\in
U\subset \manif$ so small to be contained in the domain of
injectivity of the exponential map.\footnote{Recall that the
exponential map is obtained by flowing along the geodesics,
\cite{Boo}.} We assume that $\delta$ is sufficiently small that
the ball of centre $x$ and radius $\delta$ is always contained in
$U$. We consider an isometric identification of $\T_x\manif$ with
$\R^d$ and then consider the chart $\exp_x:V\subset\R^d\to U\subset \manif$
given by the exponential map.

On the torus $\manif=\To^d$ the situation is rather trivial. Here
all tangent spaces $\T_x\manif$ are canonically identified with each other 
and with $\R^d$ and $\exp_x(\eta)=x+\eta\mod\Z^d$. Since there is no danger 
of confusion, we will even skip the ``mod $\Z^d$'' henceforth.
(It suffices to choose $\delta<\frac 12$.)
Then, for each $y\in W_\delta^s(x)$,  we can identify
$d_{T^ny}T^{-n}$ with a $d\times d$ matrix $\Amatrix_y$,
so ${R_nv}(y)=A^{-2}\lambda_u^n \Amatrix_yv(T^ny)$. 
Since $T^n$ is of class $\CoT$
there are $c_n>0$ such that
\begin{displaymath}
  \|\Amatrix_x-\Amatrix_y\|\leq c_n\,d(x,y)\quad\hbox{for $y\in U$.}
\end{displaymath}
It follows that
\begin{equation}
  \begin{split}
    \label{eq:Ax-estimate}
    \|{R_nv}(x)-{R_nv}(y)\|
    &=
    A^{-2}\lambda_u^n\,\|\Amatrix_x v(T^nx)-\Amatrix_y v(T^ny)\|\\
    &\leq
    A^{-2}\lambda_u^n\,\|\Amatrix_x(v(T^nx)-v(T^ny))\|+c_nd(x,y)\ .
  \end{split}
\end{equation}
\relax For manifolds other than $\To^d$ more care has to be taken to derive
an analogous estimate.\footnote{The first fact to notice for general $\manif$
is that, in normal coordinates, the parallel
transport of a tangent vector from $y\in U$ to $x$, along the geodesic,
coincides with the vector obtained by simply identifying the tangent space
by the Cartesian structure of $\R^d$. More precisely:
Let us call $P_{yx}:\T_x\manif\to\T_y\manif$ the above
mentioned parallel transport. Let $V$ be a neighbourhood of the origin in
$\T_x\manif$ which is mapped by the exponential map $\exp_x$ diffeomorphically 
to the neighbourhood $U$ of $x$. For $\xi\in V$ let $y=\exp_x(\xi)$. Then
\begin{displaymath}
  \tilde P_{\xi0}\circ d_x\exp_x^{-1}\circ P_{xy}
  =
  d_y\exp_x^{-1}:\T_y\manif\to\T_\xi(\T_x\manif)
\end{displaymath}
where $\tilde P_{\xi 0}:\T_0(\T_x\manif)\to\T_\xi(\T_x\manif)$ denotes the 
trivial parallel transport by the vector $\xi$.

Next, we introduce normal coordinates in a neighbourhood $U'$ of $Tx$ as
well. (For simplicity of notation we consider only the case $A=1$, $n=1$.)
Let $\exp_{Tx}:V'\to U'$ be diffeomorphic, $V'$ be a neighbourhood of the
origin in
$\T_{Tx}\manif$. Let $\xi'=\exp_{Tx}^{-1}(Ty)$. Then
\[
  \Amatrix_y
  :=
  d_y\exp_x^{-1}\circ d_{Ty}T^{-1}\circ d_{\xi'}\exp_{Tx}
  :\T_{\xi'}(\T_x\manif)\to\T_\xi(\T_x\manif)
\]
can be viewed as a matrix -- due to the canonical identification of the
spaces $\T_{\xi'}(\T_x\manif)$ and $\T_\xi(\T_x\manif)$ with $\R^d$.
Clearly, due to the smoothness properties of $T$
\[
\|\Amatrix_x-\Amatrix_y\|\leq c_1 d(x,y).
\]
With these conventions we have, just as in (\ref{eq:Ax-estimate}),
\begin{displaymath}
  \begin{split}
    \|d_x\exp_x^{-1}(Rv(x))-d_y\exp_x^{-1}(Rv(y))\|
    =&
    \lambda_u
      \|\Amatrix_x\cdot d_{Tx}\exp_{Tx}^{-1}(v(Tx))-\Amatrix_y\cdot
      d_{Ty}\exp_{Tx}^{-1}(v(Ty))\|\\
    \leq&
    \lambda_u
       \|\Amatrix_x\cdot(d_{Tx}\exp_{Tx}^{-1}(v(Tx))-d_{Ty}\exp_{Tx}^{-1}(v(Ty)))\|
    +\lambda_uc_1d(x,y).
  \end{split}
\end{displaymath}
} To continue we need to recall that the vectors $v$ are adapted
to the unstable foliation. Given a vector $w$ in $\R^d$ and a
point $x\in\manif$ we can write $w$ as $w=\xi+\eta$, with $\xi\in
E^u(x)$ and $\eta\in E^s(x)$. Therefore, for each $n\geq0$,
an unstable subspace
$E^u(T^ny)$ can be represented by an operator $\Vmat_y^n:E^u(T^nx)\to
E^s(T^nx)$; the subspace is just $\{\xi+ \Vmat_y^n\xi\}_{\xi\in E^u(T^nx)}$.
Clearly $\Vmat_x^n=0$.\footnote{ On general manifolds $\manif$ we use
$d_x\exp_x^{-1}$ to lift $E^u(y)$ and $E^s(y)$ from $\T_y\manif$ to
$\T_0(\T_x\manif)$, the latter being identified with $\R^d$. Then we
proceed along the same lines.} In terms of the operators $\Vmat_y^n$
$(y\in U)$ the $\tau$-H\"older continuity of the unstable
distribution implies \cite[19.1.6]{KH}: There is a constant
${C}>0$, independent of $x$ and $y$, such that
\[
  \|\Vmat_y^n\|\leq {C} d(T^nx,T^ny)^{{\tau'}};
\]
where ${\tau'}=\min\{\tau,1\}$.
\relax For $y\in W^s_{\delta}(x)$, let $v(T^ny)=\zeta+\Vmat_{y}^n\zeta$,
$\zeta\in E^u(T^nx)$.
Denote $c'_n:=\sup_{x\in\manif}\|\Amatrix_x\|$.
Then, from (\ref{eq:Ax-estimate}) we have
\begin{displaymath}
  \begin{split}
    \|{R_nv}(x)-{R_nv}(y)\|
    &\leq
    A^{-2}\lambda_u^n\,\|\Amatrix_x(v(T^nx)-\zeta)\|
      +A^{-2}\lambda_u^n\,\|\Amatrix_x(\Vmat_{y}^n\zeta)\|
      +c_nd(x,y)\\
    &\leq
    A^{-2}\lambda_u^nA\lambda_u^{-n}\|v(T^nx)-\zeta\|
      +A^{-2}\lambda_u^n c'_n \|\Vmat_{y}^n\zeta\|+c_nd(x,y)\\
    &\leq
    A^{-1}\|v(T^nx)-v(T^ny)\|
    +(A^{-2}\lambda_u^n c'_n+A^{-1}) \|\Vmat_{y}^n\zeta\|
    +c_nd(x,y).
  \end{split}
\end{displaymath}
As 
\begin{displaymath}
  \|\Vmat_{y}^n\zeta\|
  \leq 
  {C}d(T^nx,T^ny)^{{\tau'}}\|\zeta\|
  \leq 
  {C}\lambda_s^{n{\tau'}}d(x,y)^{{\tau'}}\|\zeta\|
  \leq 
  {C}\lambda_s^{n{\tau'}}\delta^{{\tau'}}\|\zeta\|\ ,
\end{displaymath} 
we find for sufficiently small $\delta$
(such that ${C}\lambda_s^{n{\tau'}}\delta^{{\tau'}}\leq\frac 12$)
that $\|\zeta\|\leq 2\|v(T^ny)\|\leq 2$ and hence
$\|\Vmat_{y}^n\zeta\|\leq 2{C} d^s(x,y)^{{\tau'}}|v|_\infty$. So we conclude
\begin{displaymath}
  \|{R_nv}(x)-{R_nv}(y)\|
  \leq
  \big(\lambda_s^{\beta n}
  +2{C}\lambda_s^{{\tau'}n}(A^{-2}\lambda_u^n c'_n+A^{-1})
     \delta^{{{\tau'}}-\beta}
  +c_n\delta^{1-\beta}\big)d^s(x,y)^\beta
  \leq
  d^s(x,y)^\beta
\end{displaymath}
provided $\delta$ is chosen small enough.
This finishes the proof of the sub-lemma.
\end{proof}
It is convenient to introduce the {\em distortion}
\begin{equation}
\label{eq:distorsione}
\Delta^v_n(x)=\frac{d_x g_n(v(x))}{g_n(x)}.
\end{equation}
We can then write
\begin{equation}
\label{eq:scambio}
  \begin{split}
    \int_\manif d(\Lp^n f)(v)\,dm
    &=
    \int_\manif d(f\circ T^{-n})(v)\cdot g_n\,dm
      +\int_\manif f\circ T^{-n}\cdot\Delta^v_n\cdot g_n\,dm\\
    &=
    A^2\lambda_u^{-n}\,\int_\manif\Lp^n(df({R_nv}))\,dm
      +\int_\manif \Lp^n f\cdot \Delta^v_n\,dm\\
    &=
    A^2\lambda_u^{-n}\int_\manif df({R_nv})\,dm
    +\int_\manif f\cdot (\Delta^v_n\circ T^n)\,dm
  \end{split}
\end{equation}
Since the Jacobian $g_n$ is of class $\CoJ$ by assumption and since 
$\inf g_n>0$, the following estimate is immediate:

\begin{sublem}\label{lem:distor} 
For each $n\in\N$ there exists a constant $K_n\in\R^+$ such that,
for each $v\in \V_{\beta}$,
$K_n^{-1}\Delta^v_n\circ T^n\in\D_{\beta}$.
\end{sublem}
Thus
\begin{equation}\label{LYstep1}
  \|\Lp^n f\|_u
  \leq
  A^2\lambda_u^{-n}\|f\|_u+K_n\|f\|_s
\end{equation}

To estimate the stable norm we write
\[
\int_\manif \Lp^n f\cdot \vf\,dm=\int_\manif f\cdot(\vf\circ T^n)\,dm.
\]
But $|\vf\circ T^n|_\infty=|\vf|_\infty$ and 
$H^s_\beta(\vf\circ T^n)\leq
A\lambda_s^{\beta n}H^s_\beta(\vf)$. Applied with
$\gamma$ instead of $\beta$ this yields at once
\begin{equation}\label{eq:LYstep2}
\|\Lp^n f\|_w\leq \max\{1,A\lambda_s^{\beta n}\}\cdot\|f\|_w.
\end{equation}
To obtain strict contraction for the $\|\cdot\|_s$-norm,
we must approximate $\vf\in\D_\beta$ by a suitable function
$\tilde\vf\in\D_\gamma$.

\begin{sublem}
\label{sublem:phi-approx}
There are $\delta_0>0$ and $B_1>0$ which do not depend on the parameter 
$\delta$ from the definition of $H_\beta^s(.)$ and $H_\gamma^s(.)$
such that for any choice of $\delta\in(0,\delta_0]$ and
for each $\vf\in\D_\beta$ there exists $\tilde\vf:\manif\to\R$ such that
\begin{align*}
    |\tilde\vf|_\infty&\leq 1
    &
    |\vf-\tilde\vf|_\infty&\leq 2^{-\beta}\,\delta^\beta\\
    H_\gamma^s(\tilde\vf)&\leq B_1\, \delta^{-\gamma}
    &
    H_\beta^s(\tilde\vf)&\leq1+B_1\,\delta^{1-\beta}\ .
\end{align*}
\end{sublem}
We postpone the proof of the above result to the end of the
section and finish the proof of Lemma~\ref{lem:LY} before.
So we choose $\sigma>\max\{\lambda_u^{-1},\lambda_s^\beta\}$ and fix an
integer $N$ such that 
\begin{equation}\label{eq:fix-N}
(\sigma\,\min\{\lambda_u,\lambda_s^{-\beta}\})^N>9A^2\ .
\end{equation}
Then we choose $\delta\in(0,\delta_0]$ so small that 
$\delta\leq2\lambda_s^N$ and
$B_1\delta^{1-\beta}\leq1$.
The sub-lemma implies now
\begin{displaymath}
  B_1^{-1}\left(\delta^{-1}A\lambda_s^n\right)^{-\gamma}
    \cdot\tilde\vf\circ T^n\in\D_\gamma\quad\hbox{and}\quad
  \frac 1{3A}\lambda_s^{-\beta n}(\vf-\tilde\vf)\circ T^n\in\D_\beta\quad
  \text{for all $n=1,\dots,N$\ .}
\end{displaymath}
Using such a decomposition and denoting 
$K_n':=B_1\left(\delta^{-1}A\lambda_s^n\right)^\gamma$, 
\begin{displaymath}
  \int_\manif\Lp^n f\cdot\vf\,dm
  =
  3A\lambda_s^{\beta n}\int_\manif f\cdot\frac1{3A}\lambda_s^{-\beta n}
    (\vf-\tilde\vf)\circ T^n\,dm
  +K_n'\int_\manif f\cdot({K_n'}^{-1}\cdot\tilde\vf\circ T^n)\,dm 
\end{displaymath}
so that
\begin{displaymath}
  \|\Lp^n f\|_s
  \leq
  3A\lambda_s^{\beta n}\,\|f\|_s+K_n'\,\|f\|_w\quad
  \text{for $n=1,\dots,N$.}
\end{displaymath}
Combined with (\ref{LYstep1}) this yields for 
$\|\cdot\|=\|\cdot\|_u+b\,\|\cdot\|_s$ with a sufficiently large $b=b(N)$
\begin{align}
  \|\Lp^n f\|\label{eq:LYn}
  &\leq
  A^2\cdot\max\{3\lambda_s^{\beta n},\lambda_u^{-n}\}\,\|f\|
  +K_n'\,\|f\|_w\quad
  \text{for $n=1,\dots,N$}\\
\intertext{and, by choice of $N$,}
\label{eq:LYN}
  \|\Lp^N f\|
  &\leq
  \frac 13\sigma^N\,\|f\|+K_N'\|f\|_w\ .
\end{align}
Combining and iterating the last two estimates (and observing
(\ref{eq:LYstep2}))
finally yields
\begin{equation}\label{eq:fullest}
  \|\Lp^n f\|
  \leq
  3A^2\sigma^n\,\|f\|+B\,\|f\|_w\quad\text{for all $n=1,2,\dots$}
\end{equation}
for a suitable constant $B$ that depends on $N$.
To conclude the proof of the Lemma note that (\ref{eq:LYstep2}) and
(\ref{eq:fullest}) extend by continuity to $\B_w(\manif)$ and
$\B(\manif)$ respectively. The boundedness of $\Lp$ is then obvious.
\end{proof}

\begin{proof}[\bf Proof of Sub-lemma \ref{sublem:phi-approx}:]
The basic idea is to define $\tilde \vf(x)$ as an average of $\vf$ on
a small ``ball'', centred at $x$, in $W^s(x)$.

Since only uniform H\"older continuity along stable leaves is
demanded from the approximating function $\tilde\vf$ (except for
global measurability, of course), its construction can be carried
out ``leafwise''. We fix a stable leaf $W=W^s(x_0)$. Since
$W$ is a stable leaf, it is a $d^s$-dimensional $\Co^3$
manifold with the Riemannian structure inherited from $\manif$ (as
smooth as the map $T$). In order to simplify the exposition we
restrict again to the case $\manif=\To^d$ where all tangent spaces
$\T_xW$ can be naturally immersed into $\To^d$, 
yet the general case can be treated in essentially the same way.

\relax For each $x,y\in W$ we will define an isometric identification
$I_{x,y}:\T_xW\to\T_yW$ of the respective tangent spaces.\footnote{This 
can be done in many ways. On general manifolds the natural choice is to 
do it by parallel transport. Here we will make a simpler and more 
na\"\i ve choice that uses explicitly the possibility to identify 
locally $\To^d$ with its tangent space.\label{footnote:I}} In addition, 
for each $x\in W$ we consider a chart
$\Upsilon_x:B_{\rho_*}\subseteq\T_x W\to W$ where
$B_{\rho_*}=\{\xi\in\R^{d_s}:\|\xi\|<\rho_*\}$ with a radius
$\rho_*>0$ that can be chosen the same for all $W$ and all $x\in W$.
We will use $\Upsilon_x$ to define
``balls centred at $x$'' in $W$.\footnote{For general 
manifolds the best choice seems to be to consider the exponential map 
$\Upsilon_x:=\exp^s_x:B_{\rho_*}\subseteq\T_x W\to W$. We do not do this 
in the following since this seems to yield weaker estimates (although 
sufficient for our present needs) and since the proof of the following 
properties (A-C) would need a more heavy machinery from differential 
geometry.\label{footnote:Upsilon}} 
The charts $\Upsilon_x$ enjoy the following useful properties: 
There exists a constant $C_{\rho_*}>0$
(also independent of $W$ and of $x$) such that for 
$0<\rho <\frac{\rho_*}4$ and sufficiently small $\delta>0$ holds:
\begin{enumerate}[A)]
\item
  For all $x\in W$ and all $\xi\in B_\rho$,
  \begin{displaymath}
    (1+C_{\rho_*}\rho)^{-1}\leq J\Upsilon_x(\xi)\leq (1+C_{\rho_*}\rho)
  \end{displaymath}
  where $J\Upsilon_x$ denotes the Jacobian determinant of $\Upsilon_x$. 
\item
\relax For all $x,y\in W$ with $d^s(x,y)<\delta<\frac{\rho_*}2$ and all 
$\xi\in B_\rho$,
\begin{displaymath}
  d^s(\Upsilon_x(\xi),\Upsilon_y (I_{x,y}\xi))\leq (1+C_{\rho_*}\rho)\,d^s(x,y)
\end{displaymath}
and
\begin{displaymath}
  \left|J\Upsilon_x(\xi)-J\Upsilon_y(I_{x,y}\xi)\right|
  \leq 
  C_{\rho_*}\,d^s(x,y)\ .
\end{displaymath}
\item
\relax For all $x,y\in W$ with $d^s(x,y)<\delta<\frac{\rho_*}2$,
  \begin{displaymath}
    \frac{\vol(\Upsilon_x^{-1}\Upsilon_y I_{x,y}B_\rho\setminus B_\rho)}
         {\vol(B_\rho)}
    \leq C_{\rho_*}\,\rho^{-1}\cdot d^s(x,y)\ .
  \end{displaymath}
\end{enumerate}
Before defining the maps $\Upsilon_x$ let us see how such maps allow to 
construct the wanted smooth approximation of $\vf$.

Let $\rho\leq\frac\delta 2$, denote by $m^s$ the Riemannian volume on $W$ 
and define for $\vf\in\D_\beta$ and $x\in W$
\begin{displaymath}
  \tilde\vf(x)
  :=
  \frac{\int_{\Upsilon_x B_\rho}\vf(y)\,m^s(dy)}{m^s(\Upsilon_x B_\rho)}
  =
  \frac{\int_{B_\rho}\vf(\Upsilon_x\xi)\,J\Upsilon_x(\xi)\,d\xi}
  {\int_{B_\rho}J\Upsilon_x(\xi)\,d\xi}
  =
  \frac 1{m(B_\rho)}\int_{B_\rho}\vf(\Upsilon_x\xi)\,K(x,\xi)\,d\xi
\end{displaymath}
where 
$K(x,\xi):=\frac{J\Upsilon_x(\xi)}{m(B_\rho)^{-1}\int_{B_\rho} 
           J\Upsilon_x(\zeta)\,d\zeta}$. 
Then 
\begin{displaymath}
  K(x,\xi)\leq(1+C_{\rho_*}\rho)^2\qquad\text{and}\qquad
  |K(x,\xi)-K(y,I_{x,y}\xi)|\leq \const \cdot\,d^s(x,y)
\end{displaymath}
because $I_{x,y}$ maps the ball $B_\rho\subset\T_xW$ onto
the ball $B_\rho\subset\T_yW$ isometrically. It follows
\begin{displaymath}
  |\tilde\vf|_\infty\leq|\vf|_\infty\leq 1\quad\text{and}\quad
  |\tilde\vf-\vf|_\infty\leq 2^{-\beta}\,\delta^\beta\,H_\beta^s(\vf)\ .
\end{displaymath}
We estimate the H\"older constant of $\tilde\vf$.
To this end consider
$x,y\in W$ with $d^s(x,y)<\delta$. Then
\begin{equation}\label{eq:final}
\begin{split}
  \left|\tilde\vf(y)-\tilde\vf(x)\right|
  &\leq
  \frac 1{m(B_\rho)}
  \left|\int_{B_\rho}\left[\vf(\Upsilon_x\xi)-\vf(\Upsilon_yI_{x,y}\xi)\right]
  \,K(x,\xi)\,d\xi\right|\\
  &\qquad+
  \frac 1{m(B_\rho)}
    \left|\int_{B_\rho}\vf(\Upsilon_yI_{x,y}\xi)\,
    \left[K(x,\xi)-K(y,I_{x,y}\xi)\right]\,d\xi\right|\\
  &\leq
  H_\beta^s(\vf)\,\left(1+C_{\rho_*}\rho\right)^{2+\beta}\,d^s(x,y)^\beta
  +
  \const\cdot d^s(x,y)\\
  &\leq
  \left(1+C_{\rho_*}\rho\right)^3\,d^s(x,y)^\beta
  +
  \const\cdot\delta^{1-\beta}\,d^s(x,y)^{\beta}\ .
\end{split}
\end{equation}
Choose now $\rho=\frac\delta 2$. Then
this is the desired estimate on $H_\beta^s(\tilde\vf)$.

In order to estimate $H_\gamma^s(\tilde\vf)$ we proceed differently.
To this end denote
$\tilde B_\rho=\Upsilon_x^{-1}\Upsilon_y I_{x,y}B_\rho$. Then
\begin{displaymath}
  \begin{split}
    &\left|\int_{B_\rho}\vf(\Upsilon_x\xi)\,J\Upsilon_x(\xi)\,d\xi
      -\int_{B_\rho}\vf(\Upsilon_yI_{x,y}\xi)\,J\Upsilon_y(I_{x,y}\xi)
        \,d\xi\right|\\
    &\hspace{2cm}\leq
    \int_{B_\rho\setminus\tilde B_\rho}
        |\vf(\Upsilon_x\xi)|\,J\Upsilon_x(\xi)\,d\xi
    \leq
    (1+C_{\rho_*}\rho)\,\vol(B_\rho\setminus\tilde B_\rho)
  \end{split}
\end{displaymath}
Hence
\begin{align}
    |\tilde\vf(y)-\tilde\vf(x)|
    &\leq
     \left|\int_{B_\rho}J\Upsilon_x(\xi)\,d\xi
        -\int_{B_\rho}J\Upsilon_y(I_{x,y}\xi)\,d\xi\right|
    \cdot
    \left|\frac{\int_{B_\rho}\vf(\Upsilon_yI_{x,y}\xi)\,
          J\Upsilon_y(I_{x,y}\xi)\,d\xi}
      {\int_{B_\rho}J\Upsilon_y(I_{x,y}\xi)
        \,d\xi\cdot\int_{B_\rho}J\Upsilon_x(\xi)\,d\xi}
    \right|\nn \\
    &\hspace{0.3cm}+
    \frac 1{\int_{B_\rho}J\Upsilon_x(\xi)\,d\xi}\cdot
    \left|\int_{B_\rho}\vf(\Upsilon_yI_{x,y}\xi)\,J\Upsilon_y(I_{x,y}\xi)\,d\xi
      -\int_{B_\rho}\vf(\Upsilon_x\xi)\,J\Upsilon_x(\xi)\,d\xi
    \right|\nn \\
\label{eq:Hesse}
    &\leq
    \const\cdot|\vf|_\infty\,
    \frac{\vol(B_\rho\setminus \tilde B_\rho)}{\vol(B_\rho)}
    \leq
    \const\cdot\rho^{-1}\,d^s(x,y)\\
    &\leq
    \const\cdot\delta^{-\gamma}\,d^s(x,y)^\gamma\nn
\end{align}
with a constant that depends only on $C_{\rho^*}$,
because $\rho=\frac\delta 2$ and $d^s(x,y)\leq\delta$. It follows that
$H_\gamma^s(\tilde\vf)\leq\const\,\delta^{-\gamma}$.

To conclude we need to finally define the charts $\Upsilon_x$ 
and the identifications $I_{x,y}$ and prove properties (A-C). 
Throughout the rest of this proof various constants $C$ will appear
that we do not specify precisely, but that all depend at most
on the first and second derivatives of the stable manifolds $W$
which in turn are uniformly bounded, see the appendix for a reference.
In addition, they depend on a suitably small choice of the radius $\rho_*$.

To begin with, we define
for each pair $x,y\in W$ a linear map
 $U_{x,y}:E^s(x)\to E^s(x)^\perp$ by
$\xi+U_{x,y}\xi\in E^s(y)$ for each $\xi\in E^s(x)$.
Clearly
\begin{equation}
\label{eq:basta1}
  \|U_{x,y}\|\leq C\,d(x,y),
\end{equation}
where $C$ depends only on the second derivative (curvature) of $W$.
In addition, locally we can view $W$ as a graph over $E^s(x)$. 
Let $F_x:E^s(x)\to E^s(x)^\perp$ such that $x+\zeta+F_x(\zeta)\in W$ for 
each $\zeta\in B_{\rho_*}(0)\subset E^s(x)$, and define
$\Upsilon_x:E^s(x)\to W$, \footnote{Remark that here,
and in the following, we are using explicitly the (in $\To^d$) trivial 
identification of tangent vectors and vectors in the space.}
\begin{equation}
\label{eq:basta2}
  \Upsilon_x(\zeta):=x+\zeta+F_x(\zeta)\ .
\end{equation}
It is easy to see that $d_\zeta F_x=U_{x,\Upsilon_x(\zeta)}$
and hence $d_\zeta\Upsilon_x=\Id+U_{x,\Upsilon_x(\zeta)}$ so that
$\|F_x(\zeta)\|\leq C\,\|\zeta\|$, see also Figure~\ref{fig:basta}. 
\begin{figure}[ht]
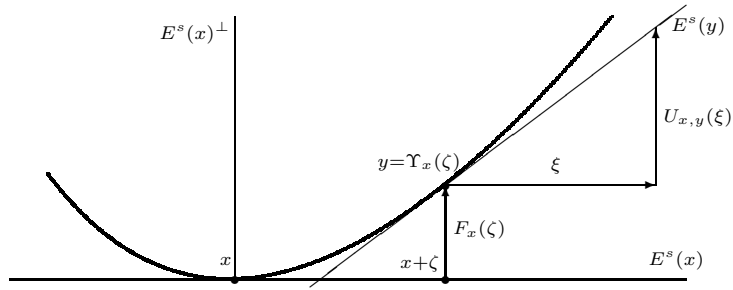
\ 
\put(-50,0){\line(1,0){90}}         \put(35,2){$\scriptstyle E^s(x)$}
\put(-20,0){\line(0,1){35}}       \put(-30,32){$\scriptstyle E^s(x)^\perp$}
\put(8,0){\vector(0,1){12.5}}        \put(9,6){$\scriptstyle F_x(\zeta)$}
\put(8,12.5){\vector(1,0){28}}   \put(22,14.5){$\scriptstyle \xi$} 
\put(36,12.5){\vector(0,1){21}}    \put(37,21){$\scriptstyle U_{x,y}(\xi)$}
{\linethickness{0.3mm}\qbezier(-45,14)(-18,-22)(30,35)}  
\put(-20,0){\circle*{1}}           \put(-22,2){$\scriptstyle x$}
\put(8,0){\circle*{1}}             \put(2,1.5){$\scriptstyle x+\zeta$}
\put(8,12.5){\circle*{1}}        \put(-1,15){$\scriptstyle y=\Upsilon_x(\zeta)$}
\put(10,14){\line(-4,-3){20}}  
\put(10,14){\line(4,3){30}}        \put(38,33){$\scriptstyle E^s(y)$}
\caption{The chart $\Upsilon_x$ and related quantities}
\label{fig:basta}
\end{figure}
Similarly,
$d(x,\Upsilon_x(\zeta))\leq C\,\|\zeta\|$.
This implies in particular that $\Upsilon_x$ is a diffeomorphism close to 
the identity.
Note also that
\begin{equation}
\label{eq:basta3}
  J_\zeta \Upsilon_x
  :=
  \sqrt{\det(\Id+U_{x,\Upsilon_x(\zeta)}^*U_{x,\Upsilon_x(\zeta)})}.
\end{equation}
The last ingredient needed is a way to identify the tangent space of
$W$ at different points. Choose a basis $\{e_i\}$ of $E^s(x)$, then
$\{e_i+U_{x,y}e_i\}=:\{\tilde b_i\}$ is a basis of $E^s(y)$,
unfortunately it may not be orthonormal. Yet, 
\[
\langle \tilde
b_i,\tilde b_j\rangle=\delta_{ij}+\langle
e_i,U_{x,y}^*U_{x,y}e_j\rangle =: (\Eins + B)_{ij}.
\]
This means that setting $\Lambda:=(\Id + B)^{-\frac 12}$  
the new basis $b_i:=\sum_j\Lambda_{ij}\tilde b_j$ is orthonormal.  
We then choose the isometric identification $I_{x,y}e_i:=b_i$.
We are now in the position to prove properties (A-C).
\medskip\\
A)\;This is nothing else than (\ref{eq:basta1}) and
(\ref{eq:basta3}).
\smallskip\\
B)\;Consider $x,y\in W$ and let $\zeta\in E^s(x)$ such that
$\Upsilon_x(\zeta)=y$. 
Consider the map $\Xi_{x,y}:E^s(x)\to E^s(x)$ defined by
\begin{equation}\label{eq:basta7}
\Upsilon_y\circ I_{x,y}(\xi)= \Upsilon_x(\zeta+\Xi_{x,y}(\xi)).
\end{equation}
We can apply the orthogonal projection $\Pi_x$ on $E^s(x)$ to both
sides of
the above equation to obtain
\begin{equation}
\label{eq:basta4}
\Xi_{x,y}(\xi)=\Pi_x I_{x,y}\xi+\Pi_x F_y(I_{x,y}\xi).
\end{equation}
Note that if we consider $I_{x,y}$ as a map in the ambient space, as
above, then for $\xi=\sum_i\xi_ie_i$ we have 
\[
I_{x,y}\xi=\sum_i\xi_i b_i=\sum_{ij}\Lambda_{ij}\xi_i\tilde b_j
\]
and thus it immediately follows 
$\|I_{x,y}\xi-\xi\|\leq C\,d(x,y)\|\xi\|$, 
which implies 
\begin{displaymath}
  \|\Xi_{x,y}(\xi)-\xi\|
  \leq
  \|\Pi_x(I_{x,y}\xi-\xi)\|+\|(\Pi_x-\Pi_y)(F_yI_{x,y}\xi)\|
  \leq
  C\,\|\xi\|d(x,y)
\end{displaymath}
because $x\mapsto\Pi_x$ is a $\Co^1$ function.
By using the above facts and recalling the definition of
$\Xi_{x,y}$ in equation (\ref{eq:basta7}) we can compute
\begin{displaymath}
\begin{split}
  \|\Upsilon_y\circ I_{x,y}(\xi)-\Upsilon_x(\xi)\|
  &=
  \|\Upsilon_x(\zeta+\Xi_{x,y}(\xi))-\Upsilon_x(\xi)\|\\
  &\leq
  \sup_{\xi\in B_\rho}\|d_\xi\Upsilon_x\|\cdot\|\zeta+\Xi_{x,y}(\xi)-\xi\|\\
  &\leq
  (1+C\,\rho)\left(\|\zeta\|+\|\Xi_{x,y}(\xi)-\xi\|\right)
\leq (1+C\,\rho)d(x,y),
\end{split}
\end{displaymath}
where all the constants $C$ depend only on the curvature of $W$.

To obtain the estimate on the Jacobian we need to differentiate
(\ref{eq:basta7}) which yields as before (observing that $JI_{x,y}=1$)
\begin{displaymath}
\begin{split}
  |J\Upsilon_y(I_{x,y}\xi)-J\Upsilon_x(\xi)|
  &=
  |J\Upsilon_x(\zeta+\Xi_{x,y}(\xi))J\Xi_{x,y}(\xi)-J\Upsilon_x(\xi)|\\
  &\leq
  (1+C\,\rho)\,d(x,y)+C\,|1-J\Xi_{x,y}(\xi)|\ .
\end{split}
\end{displaymath}
The wanted estimate follows now from
$d_\xi\Xi_{x,y}=\Pi_x
I_{x,y}+\Pi_xU_{y,\Upsilon_y(I_{x,y}\xi)}I_{x,y}$.
\smallskip\\
C)\;Again we use (\ref{eq:basta7}) to have
\[
\Upsilon_x^{-1}\circ\Upsilon_y(\xi)= \zeta+\Xi_{x,y}(I_{x,y}^{-1}\xi)
\]
this means that
\[
\big|\|\Upsilon_x^{-1}\circ\Upsilon_y(\xi)\|-\|\xi\|\big|\leq C\,d(x,y),
\]
from which C) immediately follows.
\footnote{In the case of general manifolds, using for 
$I_{x,y}$ and $\Upsilon_x$
the choices specified in Footnotes~\ref{footnote:I} and
\ref{footnote:Upsilon}, the same results can be obtained,
\ie $d(\Upsilon_x\circ I_{x,y})$ can be computed, as usual,
via Jacobi fields, \cite[Chapter 5]{DoCa}.
Doing so the constants would depend also on the derivative of the 
curvature, which would make no difference, however, since in view of our
$\CoT$ hypothesis on $T$ and of the results in the appendix, the curvature 
of the stable and unstable manifolds is uniformly $\Co^1$.}
\end{proof}

\begin{proof}[\bf Proof of Lemma \ref{lem:inverseLp}]
Since $\Lp^{-n}$ is the transfer operator of $T^{-n}$, we have
$\int\Lp^{-n}f\,\vf\,dm=\int f\,(\vf\circ T^{-n})\,dm$ 
for each $\vf\in\D_\beta$ and $|\vf\circ T^{-n}|_\infty=|\vf|_\infty\leq1$.
In order to estimate $H_\beta^s(\vf\circ T^{-n})$ let $x,y\in\manif$ with
$y\in W^s(x)$ and $d^s(x,y)<\delta$. Since $T^{-n}$ expands $W^s(x)$,
$d^s(T^{-n}x,T^{-n}y)$ may be as large as $A\mu_s^{-1}\delta>\delta$. 
Therefore let $\ell=\ell(n)$ be the smallest integer such that
$\ell d^s(x,y)> d^s(T^{-n}x,T^{-n}y)$ and fix a chain of points
$z_0=T^{-n}x,z_1,\dots,z_\ell=T^{-n}y$ 
such that $d^s(z_{i-1},z_i)\leq d^s(x,y)$ for all $i$. Since 
$d^s(T^{-n}x,T^{-n}y)\leq A\mu_s^{-n}d^s(x,y)$, it follows that
$\ell\leq A\mu_s^{-n}+1$ and hence
\begin{displaymath}
  \frac{|\vf(T^{-n}x)-\vf(T^{-n}y)|}{d^s(x,y)^\beta}
  \leq
  \sum_{i=1}^\ell\frac{|\vf(z_{i-1})-\vf(z_i)|}{d^s(z_{i-1},z_i)^\beta}
  \cdot\frac{d^s(z_{i-1},z_i)^\beta}{d^s(x,y)^\beta}
  \leq
  (A\mu_s^{-n}+1)\cdot H_\beta^s(\vf)\ .
\end{displaymath}
It follows that $\|\Lp^{-n}f\|_s\leq(A\mu_s^{-n}+1)\|f\|_s$.
Exactly the same estimate holds also for the $\|\cdot\|_w$--norm.

\relax For the unstable norm one proceeds as
in (\ref{eq:scambio}) and in Sub-lemma~\ref{vectorlem}, taking the same
care when estimating the H\"older constant 
of $v\circ T^{-n}$ along stable fibres that we took
above in the case of $\vf\circ T^{-n}$. For sufficiently small
$\delta>0$ this leads to the estimate
\begin{displaymath}
  \|\Lp^{-n}f\|_u
  \leq
  A\mu_u^n\,(A\mu_s^{-n}+1)\|f\|_u+K_n\|f\|_s
\end{displaymath}
and for large enough $b$ we obtain with
arguments analogous to those leading to
(\ref{eq:fullest}) that
\begin{displaymath}
  \|\Lp^{-n}f\|
  \leq
  A\mu_u^n\,(A\mu_s^{-n}+2)\|f\|+K_n'\|f\|_w\ .
\end{displaymath}
Iterating this estimate for fixed $n$ and recalling that
$\|\Lp^{-n}f\|_w\leq(A\mu_s^{-n}+1)\|f\|_w$ leads to the estimate
$\rho_{\spectr}(\Lp^{-n})=\limsup_{k\to\infty}\sqrt[kn]{\|\Lp^{-kn}\|}
\leq\sqrt[n]{A\mu_u^n\,(A\mu_s^{-n}+2)}$. Since we can choose $n\in\N$
as large as we like, 
$\rho_{\spectr}(\Lp^{-1})\leq\mu_u\mu_s^{-1}$.
\end{proof}

\subsection{Proofs: Peripheral spectrum and SRB measures}
\label{subsec:peripheral}

\begin{proof}[\bf \bf Proof of Proposition \ref{prop:SRB-existence}]
Let us start with $f\in\Co^1(\manif,\R)$. For
$\vf\in \Co^0(\manif,\R)\cap \D_\gamma$, and $|\lambda|=1$,
  \begin{equation}
    \label{eq:Plambda}
    \langle \Pi_\lambda f,\vf\rangle
    =
    \limn\frac 1n\sum_{k=0}^{n-1}\lambda^{-k}\langle\Lp^kf,\vf\rangle
    =
    \limn\frac 1n\sum_{k=0}^{n-1}\lambda^{-k}\int\vf\circ T^k\cdot f\,dm\ .
  \end{equation}
Hence $\langle \Pi_\lambda f,\vf\rangle\leq|\vf|_\infty\cdot\int|f|\,dm$, so
that $\Pi_\lambda f$ extends to a continuous linear functional on
$\Co^0(\manif,\R)$, and by Riesz' theorem there exists $\mu_{\Pi_\lambda
f}$ such that $\langle \Pi_\lambda
f,\vf\rangle=\int\vf\,d\mu_{\Pi_\lambda f}$ for each $\vf\in
\Co^0(\manif,\R)\cap \D_\gamma $.

The equality for all the functions in $\D_\gamma$ follows by an
approximation argument. The idea is to
approximate a function in $\D_\gamma$ by a smooth function but keeping
under control the regularity along the stable direction. To this end
it is convenient to introduce the following singular convolution:
\begin{equation}\label{eq:stableaverage}
\A_\epsilon(\vf)(x):=\frac1{Z_\epsilon(x)}\int_{W^s_\epsilon(x)}\vf(y)m_s(dy),
\end{equation}
where $W^s_\epsilon(x)$ is the intersection of the local stable manifold
with the ball of radius $\epsilon$ centred at $x$, $m_s$ is the
Riemannian measure restricted to the stable manifold, and
$Z_\epsilon(x)$ is chosen so that $\A_\epsilon 1=1$.
Obviously, 
\begin{equation}\label{eq:ihes7}
  \|\A_\epsilon\vf-\vf\|_\infty\leq\epsilon^\gamma\quad\text{ for } 
  \vf\in\D_\gamma\ .
\end{equation}

\begin{sublem}\label{lem:stableaverage}
There exist constants $c,\epsilon_*>0$ such that for each
$\epsilon\leq \epsilon_*$ we have 
$c\epsilon\A_\epsilon:\Co^0(\manif,\R)_1\to
\Co^0(\manif,\R)\cap \D_\gamma$.
\end{sublem}
\begin{proof}
Clearly $\A_\epsilon$ is bounded with respect to the sup norm. If
$y\in W^s_\delta(x)$ then, by the uniform smoothness of the
stable manifold it follows that there exists $c_0>0$ such that,
for $\epsilon$ small enough
\[
\left|\int_{W^s_\epsilon(x)\setminus W^s_\epsilon(y)}\vf(z)m_s(dz)\right|
\leq
|\vf|_\infty m_s\left(\{z\in W^s_{2\epsilon}(x)\;:\; |\epsilon-d(z,x)|
\leq c_0d(x,y)\}\right).
\]
\relax From the above equation, remembering that the stable manifolds have
uniformly bounded curvature, it follows that there exists
$c>0$ such that $c\epsilon\A_\epsilon: L^\infty(\manif,\R)_1\to \D_\gamma$.
The continuity of $c\epsilon A_\epsilon(\vf)$ 
is an  immediate consequence of 
the continuity of the
stable foliation.
\end{proof}

Notice that
$\mu_{\lambda,f,\epsilon}(\vf):=\mu_{\Pi_\lambda f}(\A_\epsilon\vf)$ and
$m_\epsilon(\vf):=m(\A_\epsilon \vf)$ are measures.
By Lusin's theorem, each $\vf\in\D_\gamma$ can be approximated  
$(m+\mu_{\lambda,f,\epsilon})$-almost everywhere by
a sequence $\{\vf_n\}\subset \Co^{0}(\manif,\R)$, $|\vf_n|_\infty\leq 1$,
and in view of Lebesgue's dominated convergence theorem also the
$m$-- and $\mu_{\lambda,f,\epsilon}$--integrals of $|\vf-\vf_n|$
tend to zero.

Now fix $\vf\in\D_\gamma$ and $\epsilon>0$.
Because of (\ref{eq:P_lambda})
there is $N=N(\epsilon)\in\N$ such that
\begin{displaymath}
  \left\|\frac 1N\sum_{k=0}^{N-1}\lambda^{-k}\Lp^kf-\Pi_{\lambda}f\right\|_w
  \leq\epsilon^2\ .
\end{displaymath}
Hence, recalling (\ref{eq:ihes7}) and Sub-lemma \ref{lem:stableaverage}, 
we have
\begin{displaymath}
\begin{split}
  \left|\mu_{\Pi_\lambda f}(\vf)-\mu_{\Pi_\lambda f}(\A_\epsilon\vf)\right|
  &\leq
  \epsilon^\gamma\\
  \left|\mu_{\Pi_\lambda f}(\A_\epsilon\vf)
    -\langle\Pi_\lambda f,\A_\epsilon\vf_n\rangle\right|
  =
  \left|\mu_{\Pi_\lambda f}(\A_\epsilon\vf)
    -\mu_{\Pi_\lambda f}(\A_\epsilon\vf_n)\right|
  &\leq
  \int|\vf-\vf_n|\,d\mu_{\lambda,f,\epsilon}\\
  \left|\langle \Pi_\lambda f, \A_\epsilon \vf_n\rangle
    -\frac 1N\sum_{k=0}^{N-1}\lambda^{-k}\langle
  \Lp^kf,\A_\epsilon\vf_n\rangle\right|
  &\leq 
  (c\epsilon)^{-1}\epsilon^2=c^{-1}\epsilon\\
  \left|\frac 1N\sum_{k=0}^{N-1}\lambda^{-k}
    \langle\Lp^kf,\A_\epsilon(\vf_n-\vf)\rangle\right|
  &\leq
  \max_{k=0,\dots,N-1}\|\Lp^kf\|_\infty \int|\vf_n-\vf|\,dm_\epsilon \\
  \left|\frac 1N\sum_{k=0}^{N-1}\lambda^{-k}
    \langle\Lp^kf,\A_\epsilon\vf-\vf\rangle\right|
  &\leq \|f\|_1\,\epsilon^\gamma\\
  \left|\frac 1N\sum_{k=0}^{N-1}\lambda^{-k}
    \langle\Lp^kf,\vf\rangle-\langle\Pi_\lambda f,\vf\rangle\right|
  &\leq\epsilon^2
\end{split}
\end{displaymath}
The second and the fourth expression tend to $0$ as $n\to\infty$ for fixed 
$\epsilon$ and $N$. Hence
\begin{displaymath}
  \left|\mu_{\Pi_\lambda f}(\vf)-\langle\Pi_\lambda f,\vf\rangle\right|
  \leq
  \epsilon^{\gamma}(1+\|f\|_1)+c^{-1}\epsilon+\epsilon^2\ .
\end{displaymath}
Hence $\langle\Pi_\lambda f,\vf\rangle=\mu_{\Pi_\lambda f}(\vf)$ 
for all $f\in\Co^1(\manif,\R)$ and all $\vf\in\D_\gamma$.

  If $\lambda=1$ and $f,\vf\geq0$,
  then $\int\vf\,d\mu_{\Pi_1f}=\langle \Pi_1f,\vf\rangle\geq0$
  and $\langle \Pi_1f,1\rangle=\int f\,dm$ by (\ref{eq:Plambda}).
  It follows that $\mu_{\Pi_11}$ is a positive measure and
  $\lambda=1$ is an eigenvalue of $\Lp$. Finally equation
  (\ref{eq:Plambda}) implies that for $\vf\geq0$
  \begin{displaymath}
    |\mu_{\Pi_j f}(\vf)|
    =
    |\langle \Pi_jf,\vf\rangle|
    \leq
    \limn\frac 1n\sum_{k=0}^{n-1}\int\vf\circ T^k\cdot |f|\,dm
    \leq
    \|f\|_\infty\,\langle \Pi_11,\vf\rangle
    =
    \|f\|_\infty\,\int\vf\,d\mu\ .
  \end{displaymath}
  It follows that $\mu_{\Pi_jf}$ is absolutely continuous with respect to
  $\mu$ and $\left|\frac{d\mu_{\Pi_j f}}{d\mu}\right|\leq\|f\|_\infty$ 
  $\mu$-almost surely.
  Observe also that, because of (\ref{eq:same-ranks}),
  each $h\in\Pi_j\B$ can be represented as $h=\Pi_jf$ for some 
  $f\in\Co^1(\manif,\R)$.
  
Next, if $h=\Pi_1f$, then $T^*\mu_h=\mu_h$ follows since
\begin{displaymath}
  \int\vf\,d(T^*\mu_h)=\int\vf\circ T\,d\mu_h=\limn\frac 1n\sum_{k=0}^{n-1}
  \int\vf\circ T^{k+1}\,f\,dm=\int\vf\,d\mu_h
\end{displaymath}
for all $\vf\in\Co^1(\manif,\R)$. The last statement follows immediately 
from the observation $\langle \Pi_1f,1\rangle=\langle f,1\rangle$, see
(\ref{eq:Plambda}).
\end{proof}

\begin{proof}[\bf Proof of Proposition \ref{prop:SRB-properties}]
a)\; For each $\vf\in\Co^1(\manif,\R)$,
  \begin{displaymath}
    \limn\int\vf\,d\bigg(\frac 1n\sum_{k=0}^{n-1}T^{*k}m\bigg)
    =
    \limn\frac 1n\sum_{k=0}^{n-1}\int\vf\circ T^k\,dm
    =
    \langle \Pi_11,\vf\rangle
    =
    \int\vf\,d\mu\ .
  \end{displaymath}
\smallskip
b)\;
Let $|\lambda|=1$. Recall from (\ref{eq:PiBw-contraction}) 
and (\ref{eq:same-ranks}) that
$\Pi_1:\B_w\to\B$ is a bounded linear operator with 
$\Pi_1(\B_w)=\Pi_1(\B)$.

Introduce the map $\Phi:\Pi_1\B\to L^\infty_{\mu,\text{inv}} :=
        \{h\in L^\infty_\mu(\manif,\R): h\circ T=h\}$ by the relation
\begin{displaymath}
    \Phi(f)=\frac{d\mu_f}{d\mu}
\end{displaymath}
with $\mu_f$ and $\mu$ as in Proposition~\ref{prop:SRB-existence}.\footnote{
$\Phi(f)\circ T=\Phi(f)$ because both measures, 
$\mu_f$ and $\mu$, are $T$-invariant.} Then
  \begin{itemize}
  \item $\Phi$ is linear by inspection of the definition
  \item $\Phi$ is injective, because 
  \begin{center}
    $\Phi(f)=0$ $\Rightarrow$
    $\mu_f=0$ $\Rightarrow$
    $\langle f,\phi\rangle=0\,\forall\,\phi\in\D_\gamma$ $\Rightarrow$
    $\|f\|_w=0$,
  \end{center}
  and since $f\in\Pi_1\B$, \ie $f=\Pi_1f$, it follows from
    (\ref{eq:PiBw-contraction}) that also $\|f\|=0$ and hence $f=0$.
  \item $\Phi$ is surjective, as we show below.
  \end{itemize}
Hence $\dim(L^\infty_{\mu,\text{inv}})=\dim(\Pi_1\B)$ is the number
of ergodic components of $\mu$.

\emph{Proof of the surjectivity of $\Phi$:} 
Let $h\in L^\infty_{\mu,\text{inv}}$ and choose 
$h_n\in \Co^1(\manif,\C)\subseteq \D_\gamma$ \footnote{We complexify 
$\D_\gamma$ tacitly.} such that $\lim_{n\to\infty}\int|h-h_n|\,d\mu=0$. 
Then, for each $\vf\in\D_\gamma$,
\begin{displaymath}
    \begin{split}
      \langle h_n\cdot\Pi_11-h_k\cdot\Pi_11,\vf\rangle
      =
      \langle\Pi_11,(h_n-h_k)\cdot\vf\rangle
      =
      \int(h_n-h_k)\cdot\vf\,d\mu\\
      \leq
      \int|h_n-h_k|\,d\mu
    \end{split}
\end{displaymath}
so that
$\|h_n\cdot\Pi_11-h_k\cdot\Pi_11\|_w\leq\int|h_n-h_k|\,d\mu\to0$
when $n,k\to\infty$. It follows that
\begin{displaymath}
    f:=\|\cdot\|_w\text{--}\lim_{n\to\infty}h_n\cdot\Pi_11\in\B_w
\end{displaymath}
exists.

To show that $\Lp f=f$, consider
\begin{displaymath}
    \begin{split}
      \|\Lp f-f\|_w
      &=
\lim_{n\to\infty}\left\|\Lp(h_n\cdot\Pi_11)-h_n\cdot\Pi_11\right\|_w\\
      &=
      \lim_{n\to\infty}\sup_{\vf\in\D_\gamma}
      \left(\langle\Lp(h_n\cdot\Pi_11),\vf\rangle
        -\langle h_n\cdot\Pi_11,\vf\rangle\right)\\
      &=
      \lim_{n\to\infty}\sup_{\vf\in\D_\gamma}
      \left(\langle \Pi_11,h_n\cdot(\vf\circ T)\rangle
        -\langle\Pi_11,h_n\cdot\vf\rangle\right)\\
      &=
      \lim_{n\to\infty}\sup_{\vf\in\D_\gamma}
      \int h_n\cdot(\vf\circ T-\vf)\,d\mu\\
      &\leq
      \sup_{\vf\in\D_\gamma}\int h\cdot(\vf\circ T-\vf)\,d\mu
      +\lim_{n\to\infty}
      \underbrace{\sup_{\vf\in\D_\gamma}
        \int (h_n-h)\cdot(\vf\circ T-\vf)\,d\mu}_
      {\leq2\,\int|h_n-h|\,d\mu}\\
      &=
      \sup_{\vf\in\D_\gamma}\left(\int(h\cdot\vf)\circ T\,d\mu -
        \int h\cdot\vf\,d\mu\right)\quad(\text{since }h\circ T=h)\\
      &=
      0 .
    \end{split}
\end{displaymath}
It follows that $f=\Pi_1f\in\Pi_1\B_w$. Hence
\begin{displaymath}
  \int \vf\cdot h\,d\mu
  =
  \lim_{n\to\infty}\int \vf\cdot h_n\,d\mu
  =
  \lim_{n\to\infty}\langle h_n\cdot\Pi_11,\vf\rangle
  =
  \langle f,\vf\rangle
  =
  \int\vf\,d\mu_f
\end{displaymath}
for all $\vf\in \Co^1(\manif,\R)$, so that $h=\frac{d\mu_f}{d\mu}=\Phi(f)$.

Since $\dim(L^\infty_{\mu,\text{inv}})<\infty$, this space has
a basis consisting of indicator functions of pairwise disjoint sets
$A_i$. Let $f_i:=\Phi^{-1}\eins_{A_i}$ and $\mu_i:=\mu_{f_i}$.
Then $\frac{d\mu_i}{d\mu}=\eins_{A_i}$.
\smallskip\\
c)
The third statement follows from Birkhoff's Theorem. Indeed, if $\vf\in
\Co^{0}(\manif,\R)$ there exists a measurable $T$-invariant set
$\Omega_\vf\subset \manif$, $\mu(\Omega_\vf)=1$, such that
\[
  \vf^+(x)
  =
  \lim_{n\to\infty}\frac
  1n\sum_{i=0}^{n-1}\vf\circ T^i(x) \text{ exists } \;\forall x\in\Omega_\vf.
\]
The characteristic function $\eins_{\Omega_\vf}$ of $\Omega_\vf$ is 
a measurable function, moreover, if $x\in \Omega_\vf$ and $y\in W^s(x)$,
$d^s(x,y)\leq \delta$, then
\[
\lim_{n\to\infty}\frac
1n\sum_{i=0}^{n-1}\vf\circ T^i(x)=
\lim_{n\to\infty}\frac
1n\sum_{i=0}^{n-1}\vf\circ T^i(y)
\]
hence $y\in\Omega_\vf$. Accordingly,
$\eins_{\Omega_\vf}\in\D_\gamma$. Consequently
\[
m(\Omega_\vf)=m(\eins_{\Omega_\vf}\circ T^k)=\lim_{n\to\infty}\frac
1n\sum_{k=0}^{n-1}
T^{*k}m(\eins_{\Omega_\vf})=\langle\Pi_11,\eins_{\Omega_\vf}\rangle=
\mu(\Omega_\vf)=1,
\]
where we have used Proposition~\ref{prop:SRB-existence}. We can then 
define, for each $t\in\R$, the sets 
$\Omega_{\vf,t}=\{x\in\Omega_\vf\,:\;\vf^+(x)\leq t\}$. Clearly
these are invariant sets as well and their characteristic functions
belong to $\D_\gamma$, hence we have
$\mu(\Omega_{\vf,t})=m(\Omega_{\vf,t})$. But, in view of what we just 
proved in b), there can be only finitely many invariant sets of positive 
$\mu$ measure. If $\mu$ is ergodic it follows that 
$m(\Omega_{\vf,t})\in\{0,1\}$ for each $t\in\R$ so that 
$\vf^+(x)=\int_\manif\vf\,d\mu$ for $m$-almost all $x$.
\smallskip\\
d)
The proof of this part depends heavily on Lemma~\ref{lem:weak-unstable} 
and its proof. So we postpone it to section~\ref{subsec:unorm-proof}.
\end{proof}

\subsection{Proofs: Spectral stability -- Smooth random perturbations} 
\label{subsec:smooth-random-proof}

Recall that we are assuming now that $d=\dim\manif=2$, hence
the stable and unstable foliations are $\Co^{\tau}$ with, at least, 
$1<\tau=1+\alpha\leq 2$ and $0<\beta<\gamma\leq\alpha$.
\relax Furthermore we give a detailed proof only for the case $\manif=\To^2$.

What makes the $\Co^{1+\alpha}$ setting so special, is the following lemma.
\begin{lem}
\label{lem:regularity-lemma}
Each point $x\in\manif$ has a neighbourhood $U$
together with a $\Co^{1+\alpha}$-diffeomorphism $\Gamma$ from $U$ into
$\R^2$ which straightens both foliations simultaneously in the
following sense:
  \begin{enumerate}[1)]
  \item $\Gamma(x)=0$,
  \item $\Gamma([y,y'])=(\Gamma_1(y),\Gamma_2(y'))$ for $y,y'\in U$,
    where $[y,y']$ denotes the unique point in $W^{u}(y)\cap
W^{s}(y')$.
  \item Let $\Gamma(y)=(\xi,\eta)$. Then $\{\Gamma^{-1}(t,\eta)\}$ and 
  $\{\Gamma^{-1}(\xi,t)\}$ are the stable
  and unstable manifold of $y$, respectively.
  \item $\{\Gamma^{-1}(t,0)\}$ and $\{\Gamma^{-1}(0,t)\}$ are the stable
and unstable manifold of $x$, respectively, parametrised by arc-length.
  \end{enumerate}
\end{lem}
\begin{proof}
Since the stable and unstable distributions are $\Co^{1+\alpha}$ (see
Remark \ref{rem:bunching} in
the appendix) the regularity statement follows like Theorem 6.1 in
\cite{PSW} from the main result of \cite{Journe} which says in our
context that since $\Gamma^{-1}$ is $\Co^{1+\alpha}$ in each of its
two real variables  separately, it is jointly $\Co^{1+\alpha}$
in both variables.\footnote{G.K. thanks Boris Hasselblatt for pointing
out to him the relevance of the argumentation in \cite{PSW} for the
present setting. Note that the above argument holds in any dimension
provided the distributions are $\Co^{1+\alpha}$.}
The third assertion is a restatement of (2) which implies 
that $\{\Gamma^{-1}(t,\eta)\}$ and
$\{\Gamma^{-1}(\xi,t)\}$ are the stable and unstable manifold of 
$\Gamma^{-1}(\xi,\eta)$. In addition, it
is always possible to reparametrise the two coordinate axis so
that (4) is satisfied. 
\end{proof}

As in the previous section we start by studying the action of smooth
averaging operators on test functions. So let 
$\tilde q_\ve:\To^2\times\R^2\to\R^k$ be $\ve^{-1}$-dominated
by $\tilde p_{\ve}$ as in Definition \ref{def:Lipschitz-domination}. 
In analogy to the definition of $Q_\ve^*$ in 
section~\ref{subsec:stability-smooth} we define the operator $Q_\ve^*$ here by
\begin{displaymath}
  Q_\ve^*\vf(x)
  =
  \int_{\R^2} \tilde q_\ve(x,\xi)\,\vf(x+\xi)\,d\xi\ .
\end{displaymath}
Here $\tilde q_\ve$ is a $\R^k$-valued function (where $k>1$ is
possible), $\vf$ is $\R$-valued. Therefore $Q_\ve^*\vf$ is also
$\R^k$-valued.

\begin{lem}\label{lem:preliminary}
  Let $q_\ve:\To^2\times\R^2\to\R^k$ be such that
  $\tilde q_\ve$ is $\ve^{-1}$-dominated by the stochastic kernel
  $\tilde p_{\ve}$ in the sense of Definition \ref{def:Lipschitz-domination}.
  Then there exist $K>0$ and $\ve_0>0$ such that for $\ve\in(0,\ve_0]$ 
  and for each smooth function $\vf$ holds
  \[
  H_\beta^s(Q_\ve^* \vf)
  \leq
  (1+K\ve^\alpha)^\beta H_\beta^s(\vf)
  +K\delta^{\alpha-\beta}|\vf|_\infty.
  \]
  The same holds with $\gamma$ instead of $\beta$.
\end{lem}
\begin{proof}
Given $x\in\manif$ let us consider a neighbourhood $U$ that is foliated on 
the one hand by the leaves $W^s_\loc(y)$, $y\in U$, and, on the other hand,
by the leaves $W^u_\loc(y)$, $y\in U$.
\relax For each $z\in W^s_\loc(x)$, define the two functions
$\Sigma_z,\Psi_z: U\subset\manif\to\manif$ by\footnote{In general 
Riemannian manifolds the same can be done by simply defining
\[
\Sigma_z(y):=\exp_x(\exp_x^{-1}(y)+\exp_x^{-1}(z))
\].
}
\[
\begin{array}l
\Sigma_z(y):=z+y-x\\[6pt]
\Psi_z(y)=[\Sigma_z(y),\,y].
\end{array}
\]
where $[y,y']$ denotes the unique point in $W^u_\loc(y)\cap W^s_\loc(y')$,
see Figure \ref{fig:figure1}.
\Bfig(300,150)
{\footnotesize{
       \bline(10,15)(1,0)(250)
       \put(114,79){\circle*{3}} \put(118,78){$\Sigma_z(y)=z+\xi$}
       \put(100,40){$W^u(\Sigma_z(y))$}
       \bline(114,79)(1,2)(20)
       \put(125,102){\circle*{3}} \put(130,102){$\Psi_z(y)$}
       \bline(114,79)(-1,-2)(40)
       \dashline{3}(18,15)(108,75)
       \put(18,15){\circle*{3}} \put(20,7){$z$}
       \put(220,20){$W^s(x)$}
       \put(190,95){$W^s(y)$}
       \bline(264,79)(-6,1)(240)
       \bline(264,79)(6,-1)(20)
       \put(264,79){\circle*{3}} \put(260,82){$y=x+\xi$}
       \dashline{3}(168,15)(258,75)
       \put(168,15){\circle*{3}} \put(170,7){$x$}
      }}
{\label{fig:figure1}}

\begin{sublem}
  \label{sublem:Psi-estimates}
  The map $\Psi_z$ is injective and there is a constant $C>0$, which 
  can be chosen
  the same for all $x$ and $z$, such that for all $y\in U$
  \begin{equation}
    \label{eq:Psi-distance}
    d(\Sigma_z(y),\Psi_z(y))\leq C\,\min\left\{d(x,z)\,d(x,y)^\alpha,
      \,d(x,y)\, d(x,z)^\alpha\right\}\ .
  \end{equation}
In addition $\Psi_z$ is absolutely continuous and its Jacobian
$J\Psi_z$ satisfies\footnote{We note that, with some more effort
and using Lemma~\ref{lem:regularity} from the appendix,
it is possible to replace the exponent $\alpha$ by $1$
in this and in the previous estimate.}

  \begin{equation}
    \label{eq:JPsi-estimate}
    \sup_{y\in U}|J\Psi_z(y)-1|\leq C\,d(x,z)^\alpha\ .
  \end{equation}
\end{sublem}
\begin{proof}
The injectivity of $\Psi_z$ is a consequence of the transversality of 
the defining foliations $\W^u$ and $\W^s$. The uniformity of all 
constants is due to the compactness of $\manif=\To^2$.

Because of Lemma \ref{lem:regularity-lemma}, in a suitable neighbourhood 
$U$ of $x$, both foliations can be straightened simultaneously by some
$\Co^{1+\alpha}$-diffeomorphism $\Gamma$ from $U$ onto
$\Gamma(U)\subseteq\R^2$. This means in particular that for any $y,z\in U$
\begin{align*}
  y\in W^u(z)\ &\Rightarrow\ \Gamma_1(y)=\Gamma_1(z)\\
  y\in W^s(z)\ &\Rightarrow\ \Gamma_2(y)=\Gamma_2(z)
\end{align*}
Without loss of generality we can assume that there is a constant $C_1>0$
such that $e^{-C_1}\leq\|D\Gamma_{|y}\|\leq e^{C_1}$ for all $y\in U$.

Now let $z\in W^s(x)$ and $y\in U$. Then
\begin{equation}\label{eq:implicit}
  \Gamma_1(\Psi_zy)
  =
  \Gamma_1(\Sigma_zy),\quad\Gamma_2(\Psi_zy)=\Gamma_2(y)\ .
\end{equation}
It follows that
\begin{displaymath}
  \begin{split}
    d(\Psi_zy,\Sigma_zy)
    &=
    d\left(\Gamma^{-1}\left(\Gamma(\Psi_zy)\right),
      \Gamma^{-1}\left(\Gamma(\Sigma_zy)\right)\right)
    \leq
    e^{C_1}\,\left\|\Gamma(\Psi_zy)-\Gamma(\Sigma_zy)\right\|\\
    &=
    e^{C_1}\,\left|\Gamma_2(y)-\Gamma_2(\Sigma_zy)\right|\ .
  \end{split}
\end{displaymath}
Let $\xi:=y-x$. There are two ways to estimate
$\left|\Gamma_2(y)-\Gamma_2(\Sigma_zy)\right|$, and we will need them both.
Since $\Gamma_2(x)=\Gamma_2(z)$ we have
\begin{equation}\label{eq:trick1}
  \begin{split}
    \left|\Gamma_2(y)-\Gamma_2(\Sigma_zy)\right|
    &=
    \left|\left(\Gamma_2(x+\xi)-\Gamma_2(z+\xi)\right)-
      \left(\Gamma_2(x)-\Gamma_2(z)\right)\right|    \\
    &\leq
    \sup_{t\in[0,1]}
    \left\|{D\Gamma_2}|_{z+t\xi}-{D\Gamma_2}|_{x+t\xi}\right\|\cdot\|\xi\|\\
    &\leq
    C_2\,d(x,z)^\alpha\,d(x,y),
  \end{split}
\end{equation}
because $\Gamma$ is of class $\Co^{1+\alpha}$. Similarly,
\begin{equation}\label{eq:trick2}
  \begin{split}
    \left|\Gamma_2(y)-\Gamma_2(\Sigma_zy)\right|
    &=
    \left|\left(\Gamma_2(x+\xi)-\Gamma_2(x)\right)-
      \left(\Gamma_2(z+\xi)-\Gamma_2(z)\right)\right|    \\
    &\leq
    \sup_{t\in[0,1]}
    \left\|{D\Gamma_2}|_{x+\xi+t(z-x)}-{D\Gamma_2}|_{x+t(z-x)}\right\|\cdot\|z-x\|\\
    &\leq
    C_2\,\|\xi\|^\alpha\,d(x,z)
    =
    C_2\,d(x,y)^\alpha\,d(x,z)\ .
  \end{split}
\end{equation}
This yields (\ref{eq:Psi-distance}).

We turn to the estimate for the Jacobian $J\Psi_z$.
\relax First note that $\Psi_z$ is continuously differentiable.
Indeed, the implicit function theorem gives in view of 
(\ref{eq:implicit}) for all $y\in U$
\begin{displaymath}
  D_y\Psi_z
  =
  \left(D_{\Psi_z}\Gamma\right)^{-1}
  \left(
    \begin{array}{c}
      D_{\Sigma_zy}\Gamma_1\\
      D_y\Gamma_2
    \end{array}
  \right)
  =
  1+\left(D_{\Psi_z}\Gamma\right)^{-1}
  \left(
    \begin{array}{c}
      D_{\Sigma_z y}\Gamma_1-D_{\Psi_z y}\Gamma_1\\
      D_y\Gamma_2-D_{\Psi_z y}\Gamma_2
    \end{array}
  \right)
\end{displaymath}
from which it follows that
\begin{displaymath}
\begin{split}
  \left|J\Psi_z(y)-1\right|
  &\leq
  \const\left(d(\Psi_z y,\Sigma_z y)^\alpha+d(\Psi_z y,y)^\alpha\right)
  \leq
  \const\left(d(\Psi_z y,\Sigma_z y)^\alpha+d(\Sigma_z y,y)^\alpha\right)\\
  &\leq
  \const\,d(x,z)^\alpha
\end{split}
\end{displaymath}
by the previous estimate and since $\Sigma_zy=(z-x)+y$.
\end{proof}

We return to the proof of Lemma \ref{lem:preliminary}.
In our estimates, in order to compare $Q^*_\ve\vf(x)$ and
$Q^*_\ve\vf(z)$, we will have to compare $\vf(y)$ to $\vf(\Sigma_z(y))$. 
But since we have control over $\vf$ only along stable manifolds, we will
compare $\vf(y)$ to $\vf(\Psi_z(y))$ and care for the remainder by
estimating the difference of $q_\ve(z,\Sigma_z(y))$ and $q_\ve(z,\Psi_z(y))$.

We start by noticing that
\begin{displaymath}
  \begin{split}
    |\tilde q_\ve(x,y-x)-\tilde q_\ve(z,\Sigma_z(y)-z)|
    &=
    |\tilde q_\ve(x,y-x)-\tilde q_\ve(z,y-x)|\\
    &\leq
    M\,d(x,z)^\alpha\,\tilde p_{a\ve}(x,y-x)\\
    &=
    M\,d(x,z)^\alpha\,p_{a\ve}(x,y),\\
  \end{split}
\end{displaymath}
in view of (\ref{eq:Lipschitz-domination-b}) so that
\begin{displaymath}
    \int_\manif|\tilde q_\ve(x,y-x)-\tilde q_\ve(z,\Sigma_z(y)-z)|\,m(dy)
    \leq
    M\,d(x,z)^\alpha\ .
\end{displaymath}
Similarly, denoting $B_{2\ve}(x)=\{y\in\manif:d(x,y)\leq 2\ve\}$ we have
\begin{displaymath}
  \begin{split}
    &\hspace*{-1cm}\int_{B_{2\ve}(x)}
    |\tilde q_\ve(z,\Sigma_z(y)-z)-\tilde q_\ve(z,\Psi_z(y)-z)|\,m(dy)\\
    &\leq
    M\,
    \int_{B_{2\ve}(x)} \ve^{-1}\,d(\Psi_z(y),\Sigma_z(y))\,
    \left(p_{a\ve}(z,\Sigma_z(y))+p_{a\ve}(z,\Psi_z(y))\right)\,m(dy)\\
    &\leq
    (1+\|J\Psi_z^{-1}\|_\infty)\,2MC\,d(x,z)^\alpha
    \leq
    6MC\,d(x,z)^\alpha
  \end{split}
\end{displaymath}
where we used (\ref{eq:Lipschitz-domination-c}),
the second estimate in (\ref{eq:Psi-distance}) and the fact that 
$J\Sigma_z=1$. Accordingly,
\begin{equation}
  \label{eq:combined-estimate}
  \int_{B_{2\ve}(x)}
  |\tilde q_\ve(x,y-x)-\tilde q_\ve(z,\Psi_z(y)-z)|\,m(dy)
  \leq
  (M+6MC)\,d(x,z)^\alpha\ .
\end{equation}

Now we can estimate the H\"older constant of $Q_\ve^*\vf$:
\begin{align*}
    &|Q_\ve^*\vf(x)-Q_\ve^*\vf(z)|
  =\left|\int_{\R^2}\left(\tilde q_\ve(x,\xi)\vf(x+\xi)
      -\tilde q_\ve(z,\xi)\vf(z+\xi)\right)
      \,d\xi\right|\\
    &=
    \left|\int_{\manif}
      \left(\tilde q_\ve(x,y-x)\vf(y)-\tilde q_\ve(z,y-z)\vf(y)
      \right)\,m(dy)\right|\\
    &\leq
    \int_{\manif} |\tilde q_\ve(x,y-x)|\,|\vf(y)-\vf(\Psi_zy)|\,m(dy)\\
    &\hspace{2cm}+
    \int_{\manif}|\tilde q_\ve(x,y-x)-\tilde q_\ve(z,\Psi_zy-z)J\Psi_z|\,
       |\vf(\Psi_zy)|
    \,m(dy)\\
    &\leq
    M\,\int_\manif p_{a\ve}(x,y)\,m(dy)\,H_\beta^s(\vf)\,
    \sup_{|y-x|\leq\ve}d(y,\Psi_z(y))^\beta\\
    &\hspace{2cm}+
    \int_{B_{2\ve}(x)}|\tilde q_\ve(x,y-x)-\tilde q_\ve(z,\Psi_zy-z)|
    \,|\vf(\Psi_zy)|\,m(dy)\\
    &\hspace{2cm}+
    M\,\int_\manif p_{a\ve}(x,y)\,m(dy)\,
    |\vf|_\infty\,\sup_{y\in U}|J\Psi_z^{-1}(y)-1|
\end{align*}
where we used in the last step that 
$|\Psi_z y-z|\leq|\Psi_z y-\Sigma_z y|+|\Sigma_z y-z|
\leq d(x,y)(C\delta^\alpha+1)\leq 2\ve$ if $\delta$ is small enough.
Therefore
\begin{align*}
    |Q_\ve^*\vf(x)-Q_\ve^*\vf(z)|
    &\leq
    H_\beta^s(\vf)(1+C\ve^\alpha)^\beta\,d(x,z)^\beta
    +(M+6MC)\,d(x,z)^\alpha\,|\vf|_\infty\\
    &\hspace{2cm}+M\, |\vf|_\infty\,C\,d(x,z)^\alpha\\
    &\leq
    d(x,z)^\beta\left[H_\beta^s(\vf)(1+C\ve^\alpha)^\beta
      +|\vf|_\infty\big(M+6MC+MC\big)\delta^{\alpha-\beta}\right]
\end{align*}
where we used the first estimate in (\ref{eq:Psi-distance}), estimate
(\ref{eq:combined-estimate}) and also (\ref{eq:JPsi-estimate}).
This proves the lemma.
\end{proof}

\begin{proof}[\bf Proof of Lemma \ref{lem:properties}]
Let $(q_\ve)_{0<\ve\leq1}$ be an admissible stochastic kernel family.
We draw first conclusions for this family
from Lemma~\ref{lem:preliminary}. It is immediately seen that
$\sup_{0<\ve\leq\ve_0}\|Q_\ve\|_w<\infty$; this is the first assertion of
Lemma~\ref{lem:properties}.
 
Combining Lemma~\ref{lem:preliminary} with Sub-lemma~\ref{sublem:phi-approx} 
we see that there is a constant $B_2\geq1$ such that for each 
$\vf\in\D_\beta$ there is $\tilde\vf$ with
\begin{displaymath}
  H_\gamma^s(Q_\ve^*\tilde\vf)\leq B_2\ ,
  \quad|Q_\ve^*\tilde\vf|_\infty\leq 1
\end{displaymath}
and
\begin{displaymath}
  H_\beta^s(Q_\ve^*(\vf-\tilde\vf))
  \leq
  (1+1 + B_1\delta^{1-\beta})(1+K\ve^\alpha)^\beta
  +K2^{-\beta}\delta^{\alpha}
  \ ,\quad
  |Q_\ve^*(\vf-\tilde\vf)|_\infty\leq2^{-\beta}\delta^\beta\ .
\end{displaymath}

Our first conclusion is that for sufficiently small $\ve$ and $\delta$,
\begin{displaymath}
  \frac 25 Q_\ve^*(\vf-\tilde\vf)\in\D_\beta\ .
\end{displaymath}
Since
\begin{displaymath}
  \int_\manif Q_\ve f\cdot\vf\,dm
  =
  \frac 52\int_\manif f\cdot
  \frac 25Q_\ve^*(\vf-\tilde\vf)\,dm
  +B_2 \int_\manif f\cdot B_2^{-1}Q_\ve^*\tilde\vf\,dm
\end{displaymath}
it follows that
\begin{equation}
\label{eq:Qeps-estimate1}
  \|Q_\ve f\|_s
  \leq
  \frac 52\|f\|_s+B_2\|f\|_w\ .
\end{equation}

We turn to the estimate for $\|Q_\ve f\|_u$. Remember that, in dimension 
two, the stable and unstable distributions are of
class $\Co^{1+\alpha}$, (see Remark \ref{rem:bunching} in the
appendix). Therefore they can be described by $\Co^{1+\alpha}$ fields
of unit tangent vectors $v^s(x)$ and $v^u(x)$. Each vector field
$v\in\V_\beta$ can be written as $v=\vf v^u $ where $\vf$ is
measurable, $|\vf|_\infty\leq 1$ and
$H_\beta^s(\vf)\leq2H_\beta^s(v)$. One just has to make sure 
by choosing $\delta$ sufficiently small that the angle between $v(x)$ 
and $v(y)$ is small enough if $d(x,y)<\delta$.

So let $v=\vf v^u\in\V_\beta$. Then
\begin{displaymath}
  \begin{split}
    \int_\manif d(Q_\ve f)(v)\,dm
    &=
    \int_\manif\int_\manif 
    f(x)\,\frac{\partial}{\partial z}\tilde q_\ve(x,z-x)\cdot v^u(z)\,\vf(z)\,
    m(dx)\,m(dz)\\
    &=
    \int_\manif f(x)\left(\Phi_1(x)+\Phi_2(x)\right)\,m(dx)
  \end{split}
\end{displaymath}
where
\begin{displaymath}
  \begin{split}
    \Phi_1(x)
    &:=
    \int_\manif\frac{\partial}{\partial z}
    \tilde q_\ve(x,z-x)\cdot(v^u(z)-v^u(x))\,\vf(z)\,m(dz)\ ,\\
    \Phi_2(x)
    &:=
    \int_\manif\frac{\partial}{\partial z}\tilde q_\ve(x,z-x)\cdot v^u(x)\,
    \vf(z)\,m(dz)\ .
  \end{split}
\end{displaymath}
Observe that we interpret $\frac{\partial}{\partial z}\tilde q_\ve$ as a
vector so that the dot-products under the integral signs are inner
products. Let $w(x,\xi):=v^u(x+\xi)-v^u(x)$. Then
$|w(x,\xi)|\leq\const\,|\xi|$ and
\begin{align*}
  \Phi_1(x)
  =&
  \int_\manif\partial_2\tilde q_\ve(x,z-x)\cdot w(x,z-x)\ \vf(z)\,m(dz)\\
  \leq& 
  M|\vf|_\infty \ve^{-1}\int_{\manif}\tilde p_{a\ve}(x,z-x)|w(x,z-x)|m(dz) .
\end{align*}
Since this integral extends only over points $z\in B_\ve(x)$,
we have
\begin{displaymath}
  |\Phi_1|_\infty\leq\const |\vf|_\infty
\end{displaymath}
for sufficiently small $\ve>0$.

In order to estimate $H_\beta^s(\Phi_1)$ we apply Lemma
\ref{lem:preliminary} to the $\R^2$-valued function
$\partial_2\tilde q_\ve\cdot w$.
Since $w$ is $\Co^{1+\alpha}$ in both coordinates and since $q_\ve$ is
admissible, $\partial_2\tilde q_\ve\cdot w$ is $\ve^{-1}$-dominated by 
the same $\tilde p_{\ve}$ that dominates $\tilde q_\ve$.\footnote
{The proof is a direct computation that uses the
same ideas employed in (\ref{eq:trick1}), and (\ref{eq:trick2}). The
details are left to the reader.} Hence

\begin{equation}
\label{eq:Phi1}
  H_\beta^s(\Phi_1)
  \leq
  (1+K\ve^\alpha)^\beta H_\beta^s(\vf)
  +K\delta^{\alpha-\beta}|\vf|_\infty
  \leq
  (1+K\ve^\alpha)^\beta
  +K\delta^{\alpha-\beta}
  \leq
  2
\end{equation}
for sufficiently small $\ve$ and $\delta$.
So there exists a constant $M_1>0$ such that $\frac 1{M_1}\Phi_1\in\D_\beta$.

We turn to the estimate of $\Phi_2$.
\begin{displaymath}
  \begin{split}
    \Phi_2(x)
    &=
    -\int_\manif\frac{\partial}{\partial x}\tilde q_\ve(x,z-x)\cdot v^u(x)\
\vf(z)\,dm(z)
    +\int_\manif\partial_1\tilde q_\ve(x,z-x)\cdot v^u(x)\ \vf(z)\,dm(z)\\
    &=
    -\divg\left(v^u\,Q_\ve^*\vf\right)(x)
    +(\divg v^u)(x)\,Q_\ve^*\vf(x)
    +v^u(x)\cdot\int_\manif\partial_1\tilde q_\ve(x,z-x)\,\vf(z)\,m(dz)\\
    &=:
    -\divg\left(v^u\,Q_\ve^*\vf\right)(x)
    +\Phi_{2,1}(x)+\Phi_{2,2}(x).
  \end{split}
\end{displaymath}
Since $\frac 12\vf\in\D_\beta$, it follows from
Lemma~\ref{lem:preliminary} that $B_3^{-1}\Phi_{2,1}\in\D_\beta$ 
for a suitable constant $B_3>0$
that depends on the $\Co^{1+\alpha}$-norm of $v^u(x)$. 
Since $\partial_1\tilde q_\ve$ is $\ve^{-1}$-dominated by 
by $\tilde p_{\ve}$ (see the definition
of admissible kernels in Definition~\ref{def:kernel}), 
we see as above that $\frac 1{M_1}\Phi_{2,2}\in\D_\beta$.

In order to collect the last estimates we write
\begin{equation}
\label{eq:conclusion}
  \begin{split}
    \int_\manif d(Q_\ve f)(v)\,dm
    &=
    \int_\manif f\cdot
     \left(\Phi_1+\Phi_{2,1}+\Phi_{2,2}-\divg(v^u Q_\ve^*\vf)\right)\,dm\\
    &=
    \int_\manif f\cdot(\Phi_{1}+\Phi_{2,1}+\Phi_{2,2})\,dm
    +\int_\manif df(v^u\,Q_\ve^*\vf)\,dm\ .
  \end{split}
\end{equation}
Observe that $H_\beta^s(v^u\,Q_\ve^*\vf)\leq H_\beta^s(Q_\ve^*\vf)+1$
when $\delta$ is small enough
because $|v^u|_\infty=1$ and $|Dv^u|_\infty<\infty$.
We invoke Lemma \ref{lem:preliminary} once more and conclude
that $H_\beta^s(v^u\,Q_\ve^*\vf)\leq3$
when $\delta$ and $\ve$ are sufficiently small.
Note also that $|v^uQ_\ve^*\vf|_\infty\leq|\vf|_\infty\leq 1$.
Hence $\frac 13v^uQ_\ve^*\vf\in\V_\beta$ so that
\begin{displaymath}
  \|Q_\ve f\|_u
  \leq
  3\|f\|_u+(2M_1+B_3)\|f\|_s\ .
\end{displaymath}
Combined with the estimate (\ref{eq:Qeps-estimate1}) for the stable norm
this yields for sufficiently large $b$ 
\begin{displaymath}
  \|Q_\ve f\|
  \leq
  3\|f\|+bB_2\|f\|_w
\end{displaymath}
which finishes the proof of Lemma \ref{lem:properties}.
\end{proof}

\begin{proof}[\bf Proof of Lemma \ref{lem:propbound}]
We start by proving the lemma not for $\Lp_\ve=Q_\ve\Lp$ but for
$Q_\ve\Lp^N$ where $N$ is the iterate from (\ref{eq:fix-N}) for which 
$(\sigma\,\min\{\lambda_u,\lambda_s^{-\beta}\})^N>9A^2$.
The reduction to the case $N=1$ is then accomplished in
Lemmas~\ref{lem:Qswitch} and \ref{lem:Qmult}.

Let $\vf\in\D_\gamma$.
Since
$(Q_\ve\Lp^N)^*\vf=(\Lp^N)^*(Q_\ve^*\vf)=(Q_\ve^*\vf)\circ T^N$,
it follows from Lemma \ref{lem:preliminary} that
\begin{displaymath}
  H_\gamma^s\left((Q_\ve\Lp^N)^*\vf\right)
  \leq
  A\lambda_s^{\gamma N} H_\gamma^s(Q_\ve^*\vf)
  \leq
  A\lambda_s^{\gamma N} (1+K\ve^\alpha)^\gamma H_\gamma^s(\vf)
  +A\lambda_s^{\gamma N} K\delta^{\alpha-\gamma}|\vf|_\infty
\end{displaymath}
whence, for sufficiently small $\ve$,
\begin{displaymath}
  H_\gamma^s\left((Q_\ve\Lp^N)^{*}\vf\right)
  \leq
  \sigma^N
  H_\gamma^s(\vf)+K\delta^{\alpha-\gamma}|\vf|_\infty\ .
\end{displaymath}
By induction, it follows
\[
  H_\gamma\left((Q_\ve\Lp^N)^{*k}\vf\right)
  \leq
  \sigma^{kN} H_\gamma^s(\vf)+\frac
  K{1-\sigma^N}\delta^{\alpha-\gamma}|\vf|_\infty\ .
\]
(Observe that $|(Q_\ve\Lp^N)^*\vf|_\infty\leq|\vf|_\infty$.)
Hence there is $K_2>0$ such that
$K_2^{-1}(Q_\ve\Lp^N)^{*k}\vf\in\D_\gamma$ for all $k\in\N$ so that 
\begin{equation}\label{eq:recover1}
  \|(Q_\ve\Lp^N)^kf\|_w
  \leq 
  K_2\|f\|_w\qquad\text{for all $k$.}
\end{equation}

 We turn to the second estimate from Lemma~\ref{lem:propbound}.
 Again we consider first $Q_\ve\Lp^N$ instead of $\Lp_\ve=Q_\ve\Lp$.
 For each $n=1,\dots,N$ we have
 \begin{equation}\label{eq:recover2}
 \begin{split}
    &\|(Q_\ve\Lp^n)^kf\|\\
    &\leq
    3\|\Lp^n(Q_\ve\Lp^n)^{k-1}f\|+K\|\Lp^n(Q_\ve\Lp^n)^{k-1}f\|_w\\
    &\leq
    3A^2\,\max\{3\lambda_s^{\beta n},\lambda_u^{-n}\}
      \|(Q_\ve\Lp^n)^{k-1}f\|+3B\|(Q_\ve\Lp^n)^{k-1}f\|_w
      +KA\|(Q_\ve\Lp^n)^{k-1}f\|_w\\
    &\leq
    3A^2\,\max\{3\lambda_s^{\beta n},\lambda_u^{-n}\}
      \|(Q_\ve\Lp^n)^{k-1}f\|+K_2\big(3B+K\big)\|f\|_w
  \end{split}
  \end{equation}
where we used Lemma \ref{lem:properties} for the first,
equations (\ref{eq:LYstep2}) and (\ref{eq:LYn}) for the second, and
equation (\ref{eq:recover1}) for the third inequality. For $n=N$ this 
leads, by induction on $k$ and observing the choice of $N$, to
\begin{displaymath}
    \|(Q_\ve\Lp^N)^kf\|
    \leq
    \sigma^{kN}\|f\|+\frac{K_2(3B+K)}{1-\sigma^N}\|f\|_w\ .
\end{displaymath}

It remains to carry over these results for $Q_\ve\Lp^N$ to 
$\Lp_\ve^N=(Q_\ve\Lp)^N$. For this we use repeatedly the next two 
lemmas to show that there exists a smooth
averaging operator $Q_{\ve_N}^{(N)}$ such that 
$\Lp_\ve^N=(Q_\ve\Lp)^N=Q_{\ve_N}^{(N)}\Lp^N$
with $\ve_N\leq C_N\ve$ for some $C_N>0$.
Applying (\ref{eq:recover1}) and 
the last estimate to this operator instead of $Q_\ve$ and using
estimate (\ref{eq:recover2}) for $n=1$ at most $N-1$ times then finishes
the proof of Lemma~\ref{lem:propbound}.
\end{proof}

\begin{lem}\label{lem:Qswitch}
Denote by $\Lambda^{-1}$ the Lipschitz constant of $T^{-1}$.
For each smooth averaging operator $Q_\ve$ there exists 
another smooth averaging operator $Q_\ve'$ such that 
$\Lp Q_\ve=Q_{\Lambda^{-1}\ve}'\Lp$.
\end{lem}

\begin{lem}\label{lem:Qmult}
If $Q_{\ve}$ and $Q_{\ve}'$ are smooth averaging operators, then
also $Q_\ve'':=Q_{(1-p)\ve}'Q_{p\ve}$ are smooth averaging operators
for each $p\in(0,1)$.
\end{lem}

\begin{proof}[Proof of Lemma~\ref{lem:Qswitch}]
We define $Q_\ve':=\Lp Q_{\Lambda\ve}\Lp^{-1}$. Then 
$Q_\ve'f(y)=\int_{\manif} f(x)q_{\Lambda\ve}'(x,y)\,m(dx)$ with
\begin{displaymath}
\begin{split}
  q_\ve'(x,y)&=q_{\Lambda\ve}(T^{-1}x,T^{-1}y)\cdot g(y)
  \qquad\text{and hence}\\
  \tilde q_\ve'(x,\xi)&=\tilde q_{\Lambda\ve}(G(x,\xi))\cdot g(x+\xi)
  \quad\text{where}\quad G(x,\xi):=(T^{-1}x,T^{-1}(x+\xi)-T^{-1}x)\ .
\end{split}
\end{displaymath}
It follows immediately that $q_\ve'(x,y)=0$ if $d(x,y)>\ve$,
thus $q_\ve'$ is a kernel family. Clearly,
$\int_{\To^2}q_\ve'(x,y)\,m(dy)=1$, so $q_\ve'$ is a
stochastic kernel family. 
Moreover,
\begin{displaymath}
\begin{split}
  \partial_1\tilde q_\ve'(x,\xi)
  &=
  \underbrace{\partial_1\tilde q_{\Lambda\ve}(G(x,\xi))
     \cdot DT^{-1}_{|x}\cdot g(x+\xi)}_{=:A_1}\\
  &\qquad+\underbrace{\Lambda\ve\,\partial_2\tilde q_{\Lambda\ve}(G(x,\xi))
    \cdot(\Lambda\ve)^{-1}\left(DT^{-1}_{|x+\xi}-DT^{-1}_{|x}\right)
    \cdot g(x+\xi)}_{=:A_2}
  +\underbrace{\tilde q_{\Lambda\ve}(G(x,\xi))\cdot Dg_{|x+\xi}}_{=:A_3}\\
  \ve\,\partial_2\tilde q_{\ve}'(x,\xi)
  &=
  \underbrace{\ve\,\partial_2\tilde q_{\Lambda\ve}(G(x,\xi))
    \cdot DT^{-1}_{|x+\xi}\cdot g(x+\xi)}_{=:A_4}
  +\underbrace{\ve\,\tilde q_{\Lambda\ve}(G(x,\xi))\cdot Dg_{|x+\xi}}_{=:A_5}
\end{split}
\end{displaymath}
Observe that
\begin{itemize}
\item[-] $G$ is Lipschitz continuous,
\item[-] $DT^{-1}$, $g$ and $Dg$ are bounded and Lipschitz continuous, 
  because $T$ is of class $\CoT$.
\end{itemize}
Therefore the verification of (\ref{eq:Lipschitz-domination-a}) --
(\ref{eq:Lipschitz-domination-c}) for the kernel $q_\ve'$ 
and for the terms $A_i$, $i\neq2$, is straightforward.\footnote{They
are all $\ve^{-1}$-dominated by the same stochastic kernel family
that dominates $q_\ve$.}
The term $A_2$ needs some more care. First note that
\begin{displaymath}
  \left\|DT^{-1}_{|x+\xi}-DT^{-1}_{|x}\right\|
  \leq
  \const\cdot\|\xi\|\leq\const\cdot\ve\ .
\end{displaymath}
This controls the $A_2$-contribution to
(\ref{eq:Lipschitz-domination-a}) and reduces the H\"older estimates for 
$A_2$ in (\ref{eq:Lipschitz-domination-b}) and 
(\ref{eq:Lipschitz-domination-c}) to the corresponding estimates for
$\ve^{-1}\,(DT^{-1}_{|x+\xi}-DT^{-1}_{|x})$. But
\begin{displaymath}
\begin{split}
  &\hspace*{-2cm}\ve^{-1}\left(DT^{-1}_{|x+\xi}-DT^{-1}_{|x}\right)-
    \ve^{-1}\left(DT^{-1}_{|z+\xi}-DT^{-1}_{|z}\right)\\
  &=
  \ve^{-1}\left(DT^{-1}_{|x+\xi}-DT^{-1}_{|z+\xi}\right)
    -\ve^{-1}\left(DT^{-1}_{|x}-DT^{-1}_{|z}\right)\\
  &=
  \frac d{dt}\left(DT^{-1}_{|x+t\xi}-DT^{-1}_{|z+t\xi}\right)_{t=t_0}
    \cdot\ve^{-1}\xi
\end{split}
\end{displaymath}
for a suitable $t_0\in(0,1)$ so that this term is bounded by
$\const\cdot d(x+t_0\xi,z+t_0\xi)=\const\cdot d(x,z)$, and finally
\begin{displaymath}
  \ve^{-1}\left(DT^{-1}_{|x+\xi}-DT^{-1}_{|x}\right)
    -\ve^{-1}\left(DT^{-1}_{|x+\zeta}-DT^{-1}_{|x}\right)
  =
  \ve^{-1}\left(DT^{-1}_{|x+\xi}-DT^{-1}_{|x+\zeta}\right)
\end{displaymath}
so that this term is bounded by $\const\cdot\ve^{-1}\,d(x+\xi,x+\zeta)$.
Hence $q_\ve'$ is an admissible kernel family.
\end{proof}

\begin{proof}[Proof of Lemma~\ref{lem:Qmult}]
Let $q_\ve''(x,y):=\int_{\manif}q_{p\ve}(x,z)q_{(1-p)\ve}'(z,y)\,m(dz)$.
Obviously $q_\ve''(x,y)=0$ if $d(x,y)>\ve$.
The associated kernel $\tilde q_\ve''(x,\xi)$ is
\begin{displaymath}
  \tilde q_\ve''(x,\xi)
  =
  \int_{\R^2}\tilde q_{p\ve}(x,\zeta)\tilde q_{(1-p)\ve}(x+\zeta,\xi-\zeta)
    \,d\zeta
\end{displaymath}
It follows that
\begin{displaymath}
\begin{split}
  \partial_1\tilde q_\ve''(x,\xi)
  &=
  \int_{\R^2}\partial_1\tilde q_{p\ve}(x,\zeta)
       \tilde q_{(1-p)\ve}(x+\zeta,\xi-\zeta)\,d\zeta\\
  &\qquad+
  \int_{\R^2}\tilde q_{p\ve}(x,\zeta)
       \partial_1\tilde q_{(1-p)\ve}(x+\zeta,\xi-\zeta)\,d\zeta
  \\
  \ve\,\partial_2\tilde q_\ve''(x,\xi)
  &=
  \int\tilde q_{p\ve}(x,\zeta)\cdot\ve\,
      \partial_2\tilde q_{(1-p)\ve}(x+\zeta,\xi-\zeta)\,d\zeta\ ,
\end{split}
\end{displaymath}
and (\ref{eq:Lipschitz-domination-a}) -- (\ref{eq:Lipschitz-domination-c})
for $\tilde q_\ve''$, $\partial_1\tilde q_\ve''$ and 
$\partial_2\tilde q_\ve''$ can be deduced easily from the corresponding 
properties for $\tilde q_\ve$, $\partial_1\tilde q_\ve$ and 
$\partial_2\tilde q_\ve$.
\end{proof}

\begin{proof}[\bf Proof of Lemma \ref{lem:three-norm}]
The last step in the discussion of smooth random perturbations
consists in showing that there exists a constant $K>0$ such that
$|||Q_\ve-\Id|||\leq K\ve^{\gamma-\beta}$. To this end consider
$f\in\Co^1(\manif,\R)$.

Let $\vf\in\D_\gamma$. Denoting as before $[x,y]:=W^u(x)\cap W^s(y)$ we have
\begin{equation}
  \begin{split}
    \label{eq:I1I2-def}
    &\int_\manif (Q_\ve f-f)\, \vf \,dm \\
    &= \int_\manif\int_{\R^2} f(x) \tilde q_{\ve}(x,\zeta)
\vf(x+\zeta)\,d\zeta\,m(dx)
    - \int_\manif f(x)\,\vf(x)\,m(dx)\\
    &=
    \int_\manif\int_{\R^2} (f(x) - f([x,x+\zeta])) \tilde q_{\ve}(x,\zeta)
    \vf(x+\zeta)\,d\zeta\,m(dx)       \\
    &\hspace{0.4cm}
    +\int_\manif\left(\int_{\R^2}
    \tilde q_{\ve}(x,\zeta)f([x,x+\zeta])
\vf(x+\zeta)\,d\zeta-f(x)\vf(x)\right)
    \,m(dx)\\
    &=:I_1+I_2\ .
  \end{split}
\end{equation}
Our goal is to show that
$|I_1|,|I_2|\leq\const\cdot\ve^{\gamma-\beta}\cdot\|f\|$.

We denote by $U_t$ the flow at unit speed along the unstable manifolds.
Let $r(x,\zeta)$ be the signed distance between $x$ and $[x,x+\zeta]$
within $W^u(x)$. The direction is chosen such that
$U_{r(x,\zeta)}(x)=[x,x+\zeta]$
(see Fig.~\ref{int-fibers-2}). We first look at $I_1$, a term which
involves a difference of $f$-values along an unstable fibre. It is natural
to try to estimate this difference in terms of $d_xf(v^u(x))$ where 
$v^u(x)$, as in the preceding proof, denotes the unit tangent field in 
unstable direction.
\Bfig(300,150)
      {\footnotesize{
       \bline(20,20)(1,0)(250)
       \bline(30,20)(1,2)(45)  \bline(128,57)(-6,1)(100)
       \put(30,20){\circle*{3}} \put(25,10){$x$}
       \dashline{3}(30,20)(111,60)
       \put(122,22){$W^{s}(x)$}
       \put(55,69){\circle*{3}} \put(52,58){$[x,x+\zeta]$}
       \put(111,60){\circle*{3}} \put(115,60){$x+\zeta$}
       \put(5,77){$W^{s}(x+\zeta)$} \put(68,110){$W^{u}(x)$}
       \bline(180,20)(1,2)(35)  \bline(260,60)(-6,1)(240)
       \put(180,20){\circle*{3}} \put(175,10){$y$}
       \put(272,20){$W^{s}(y)$}
       \dashline{3}(180,20)(261,60)
       \put(205,69){\circle*{3}} \put(202,58){$[y,y+\zeta]$}
       \put(261,60){\circle*{3}} \put(265,58){$y+\zeta$}
       \put(155,77){$W^{s}(y+\zeta)$} \put(218,90){$W^{u}(y)$}
       \put(66,92){\circle*{3}}   \put(71,93){$[x,y+\zeta]$}
      }}
{      \label{int-fibers-2}}

\begin{displaymath}
  I_1
  =
  -\int_\manif\int_{\R^2} \tilde q_{\ve}(x,\zeta)
  \int_0^{r(x,\zeta)} d_{U_tx}f(v^u(U_t(x))) \,
  \vf(x+\zeta)\,dt\,d\zeta\,m(dx)
\end{displaymath}
Changing variables to $\hat x=U_{t}(x)$ 
and setting
\begin{displaymath}
  \Omega(x,\zeta):=\left\{t\in\R\,:\;\sign(t)=\sign(r(U_{-t}x,\zeta))\,;\, 
                  |t|\leq|r(U_{-t}x,\zeta)\right\}
\end{displaymath}
we get
\begin{equation}
\label{eq:I1-def}
  \begin{split}
    I_1
    &=
    \int_\manif\int_{\R^2}\int_\R \tilde q_{\ve}(U_{-t}\hat x,\zeta)\,
    1_{\Omega(\hat x,\zeta)}(t)
    \, d_{\hat x}f(v^u(\hat x)) \, \vf(U_{-t}\hat x+\zeta)
    \,JU_{-t}(\hat x)
    \,dt\,d\zeta\,m(d\hat x) \\
    &=\int_\manif d_{\hat x}f(\Phi(\hat x)\, v^u(\hat x))\,m(d\hat x)
  \end{split}
\end{equation}
where $JU_{-t}=JU_{t}^{-1}$ denotes the Jacobian of $U_{-t}$,
$\log JU_t(x)=\int_0^t(\divg v^u)(U_sx)\,ds$,
and
\begin{displaymath}
  \Phi(x)
  :=
  \int_{\R^2}\int_\R \tilde q_{\ve}(U_{-t}x,\zeta)\,
  1_{\Omega(x,\zeta)}(t)
  \, \vf(U_{-t}x+\zeta) \,JU_{-t}(x) \,dt\,d\zeta .
\end{displaymath}
To proceed we will show that $C^{-1}\ve^{-1}\Phi\,v^u$ is a test vector
field in $\V_\beta$ for a suitable $C>0$. To this end it suffices to 
check that $C^{-1}\ve^{-1}\Phi\in\D_\beta$ for some $C>0$.

It is easy to estimate $|\Phi|_\infty$. Because of the uniform
transversality of the stable and unstable foliation, 
$r(U_{-t}x,\zeta)\leq\const\,\|\zeta\|$ so that,
for each $\zeta$ with $\tilde q_\ve(U_{-t}x,\zeta)\neq0$,
the set $\{t\in\R:0\le t \le r(U_{-t}x,\zeta)\}$
is contained in the interval $[0,\const\,\ve]$. Hence
$|\Phi|_{\infty}\le C\ve |\vf|_{\infty}\leq C\ve$
with a constant $C>0$ which depends only
on the stable and unstable foliation.

It remains to estimate $H_\beta^s(\Phi)$. Various points and corresponding 
stable and unstable fibres which we use in the following computation are
shown schematically in Figure~\ref{int-fibers-2-new}.
Here are some estimates on distances between related points.

We abbreviate $U_{-t}x$ as $x_t$ \etc Then, if $\ve$ is sufficiently small,
\begin{equation}
  \label{eq:distances}
  \begin{split}
    d(x_t+\zeta,y_t+\zeta)&=d(x_t,y_t)\leq 2d(x,y)\\
    d(x_t+\zeta,[x_t+\zeta,y_t+\zeta])&\leq \const\,d(x_t+\zeta,y_t+\zeta)
    \leq\const\,d(x,y)\\
    d(y_t+\zeta,[x_t+\zeta,y_t+\zeta])&\leq \const\,d(x_t+\zeta,y_t+\zeta)
    \leq\const\,d(x,y)\\
  \end{split}
\end{equation}
for suitable constants that depend on the lower bound of the angles of
intersection
of the stable and unstable foliation and other quantities associated with
them.\footnote{On general manifolds a term of higher order in $d(x,y)$ must
be added on the r.h.s. of these estimates, but this has no consequence
if $\delta$ is chosen small enough.}

\Bfig(300,150)
      {\footnotesize{
       \bline(20,40)(1,0)(250)
       \bline(30,40)(1,2)(45)  
       \put(30,40){\circle*{3}} \put(25,42){$x$}
       \put(114,79){\circle*{3}} \put(118,78){$x_t+\zeta$}
       \put(125,120){$W^u(x_t+\zeta)$}
       \bline(114,79)(1,2)(20)  \put(125,102){\circle*{3}}
       \put(68,130){$W^{u}(x)$}
       \bline(30,40)(-1,-2)(15)
       \dashline{3}(18,15)(108,75) \put(85,50){$\zeta$}
       \dashline{3}(18,15)(120,100) \put(60,60){$\hat\zeta$}
       \put(18,15){\circle*{3}} \put(20,7){$x_t:=U_{-t}(x)$}
       \put(130,102){$[x_t+\zeta,y_t+\zeta]$}
       \bline(180,40)(1,2)(35)   \bline(264,79)(-6,1)(240)
       \put(180,40){\circle*{3}} \put(175,43){$y$}
       \put(237,42){$W^{s}(x)=W^{s}(y)$}
       \put(205,89){\circle*{3}} \put(202,78){$[y,y_t+\zeta]$}
       \put(264,79){\circle*{3}} \put(267,78){$y_t+\zeta$}
       \put(5,125){$W^{s}(y_t+\zeta)$} \put(218,110){$W^{u}(y)$}
       \bline(180,40)(-1,-2)(15)
       \dashline{3}(168,15)(258,75) \put(215,55){$\zeta$}
       \put(168,15){\circle*{3}} \put(170,7){$y_t:=U_{-t}(y)$}
      }}
{\label{int-fibers-2-new}}

We turn to the H\"older estimate for $\Phi$.
\begin{equation}
\label{eq:hoelder-phi-estimate}
  \begin{split}
    &|\Phi(x) - \Phi(y)|\\
    &\leq
    \left|\int_{\R^2}\int_\R \tilde q_{\ve}(x_t,\zeta)\,
    1_{\Omega(x,\zeta)}(t) \,
    \big(\vf(x_t+\zeta) - \vf([x_t+\zeta,y_t+\zeta])\big)
    \,JU_{-t}(x)\,dt\,d\zeta\right|\\
    &\hspace{3mm}+
    \left|\int_{\R^2}\int_\R \tilde q_{\ve}(x_t,\zeta)
    \big\{1_{\Omega(x,\zeta)}(t) \,\vf([x_t+\zeta,y_t+\zeta])
    \,JU_{-t}(x)\right.\\
    &\hspace{3cm}
    - 1_{\Omega(y,\zeta)}(t) \, \vf(y_t+\zeta)
    \,JU_{-t}(y)\big\}\,dt\,d\zeta \bigg| \\
    &\phantom{\leq}+
    \left|\int_{\R^2}\int_\R (\tilde q_{\ve}(x_t,\zeta)-\tilde
q_{\ve}(y_t,\zeta))\,
    1_{\Omega(y,\zeta)}(t) \, \vf(y_t+\zeta)
    \, JU_{-t}(y)\,dt\,d\zeta\right|   \\
    &=:A_1+A_2+A_3
  \end{split}
\end{equation}
Notice that according to our assumptions $JU_{-t}$ is
$\Co^{\alpha}$-close to 1.

The $A_3$-part in (\ref{eq:hoelder-phi-estimate}) is most easily estimated,
since 
\begin{displaymath}
  |\tilde q_\ve(x_t,\zeta)-\tilde q_\ve(y_t,\zeta)|\leq 
  M\,2^\alpha\,d(x,y)^\alpha\,
  \tilde p_{a\ve}(x_t,\zeta)
\end{displaymath} 
by Definition~\ref{def:kernel} and (\ref{eq:distances}), and
since $0\leq |t|\leq |r(y_t,\zeta)|\leq\const \|\zeta\|$. This implies
\begin{equation}
  \label{eq:A3-estimate}
  A_3
  \leq
  \const\,\ve\,|\vf|_\infty\,d(x,y)^\alpha
  \leq
  \const\,\ve\,d(x,y)^\beta\delta^{\gamma-\beta}\ .
\end{equation}

Denote the expression in curly brackets in $A_2$ by $G(t,\zeta)$.
\relax For each fixed $\zeta$, the Lebesgue measure of the set of the points 
$t\in\R$ that belong to exactly one of the two indicator sets is of order
\begin{equation}
\label{eq:r-difference}
  \begin{split}
    &|r(x_t,\zeta)-r(y_t,\zeta)|\\
    &\leq
\left|(\Gamma_2(x_t+\zeta)-\Gamma_2(x_t))-(\Gamma_2(y_t+\zeta)-\Gamma_2(y_t))
      \right|
    +C\,d(x_t,y_t)^\alpha\,\left|\Gamma_2(y_t+\zeta)-\Gamma_2(y_t)\right|\\
    &=
\left|(\Gamma_2(x_t+\zeta)-\Gamma_2(y_t+\zeta))-(\Gamma_2(x_t)-\Gamma_2(y_t))
    \right|
    +C\,d(x_t,y_t)^\alpha\,\left|\Gamma_2(y_t+\zeta)-\Gamma_2(y_t)\right|\\
    &\leq
    \|\zeta\|\,\|D\Gamma_2(x_t+q\zeta)-D\Gamma_2(y_t+q\zeta)\|
    +C\,d(x_t,y_t)^\alpha\,\|\zeta\|\,\|D\Gamma_2(y_t+\tilde q\zeta)\|\\
    &\leq
    \|\zeta\|\,C\,d(x_t,y_t)^\alpha
    \leq
    \|\zeta\|\,C\,d(x,y)^\beta\delta^{\alpha-\beta}
  \end{split}
\end{equation}
where $\Gamma$ is a local $\Co^{1+\alpha}$ diffeomorphisms that straightens
the stable and unstable foliation simultaneously, see
Lemma~\ref{lem:regularity-lemma}.
Therefore the corresponding part of the integral can be estimated as 
$C\,\ve\,d(x,y)^\beta\delta^{\alpha-\beta}
|\vf|_\infty\leq C\,\ve\,d(x,y)^\beta\delta^{\gamma-\beta}$.

Now consider for fixed $\zeta$ the set $Y_\zeta$ of those $t\in\R$ which 
belong to both indicator sets. Then $Y_\zeta\subseteq[0,\const\,\ve]$, and 
for $t\in Y_\zeta$ we have
\begin{displaymath}
  \begin{split}
    G(t,\zeta)
    &=\phantom{:}  \big(\vf([x_t+\zeta,y_t+\zeta]) -
\vf(y_t+\zeta)\big)\,JU_{-t}(y)
    +\vf([x_t+\zeta,y_t+\zeta])\,\big(JU_{-t}(x)-JU_{-t}(y)\big)\\
    &=:
    A_{2,1}(t,\zeta)+A_{2,2}(t,\zeta)\ .
  \end{split}
\end{displaymath}

$A_{2,1}$ and $A_{2,2}$ are easy to estimate:
\begin{align*}
  |A_{2,1}(t,\zeta)|
  &\leq
H_\gamma^s(\vf)\,d([x_t+\zeta,y_t+\zeta],y_t+\zeta)^\gamma\,|JU_{-t}|_\infty\\
  &\leq
  \const\,H_\gamma^s(\vf)\,d(x,y)^{\gamma}
  \leq
  \const\,d(x,y)^{\beta}\delta^{\gamma-\beta}\\
  |A_{2,2}(t,\zeta)|
  &\leq
  \const\,|\vf|_\infty\,d(x,y)^\alpha
  \leq
  \const\,d(x,y)^\beta\delta^{\gamma-\beta},
\end{align*}
where we have used $\vf\in\D_\gamma$. It follows that
\begin{equation}
\label{eq:A2-estimate}
  A_2
  \leq
  \const\,\delta^{\gamma-\beta}\,\ve\,d(x,y)^\beta\ .
\end{equation}

To estimate $A_1$ we shift the difference in $A_{1}$
to the kernel $\tilde q_{\ve}$ using the following change of variables:
$\Psi_t:\{\zeta\in\R^2:\|\zeta\|<\ve\}\to\manif$,
$\zeta\mapsto\hat\zeta:=[x_t+\zeta,y_t+\zeta]-x_t$.

\begin{sublem}
  The map $\Psi_t$ is injective and there is a constant $C>0$, which can 
  be chosen the same for all $x$, $y$ and $t\leq r(U_{-t}x,\zeta)$, such that
  \begin{equation}
    \label{eq:Psit-distance}
    \sup_{\|\zeta\|<\ve}\|\Psi_t(\zeta)-\zeta\|
    \leq C\,\min\{\ve\,d(x,y)^\alpha,d(x,y)\}\ .
  \end{equation}
  In addition $\Psi_t$ is absolutely continuous and its Jacobian
  $J\Psi_t$ satisfies
  \begin{equation}
    \label{eq:JPsit-estimate}
    \sup_{\|\zeta\|<\ve}|J\Psi_t(\zeta)-1|\leq C\,d(x,y)^\alpha\ .
  \end{equation}
\end{sublem}
\begin{proof}
  The proof is very similar to that of Sub-lemma~\ref{sublem:Psi-estimates}.

  The r\^ole of the points $z$ and $x$ in that Sub-lemma is played by
  the points $x_t$ and $y_t$ here, that of  
  $\Sigma_z(y)$ and $\Psi_z(y)$ by $x_t+\zeta$ and $[x_t+\zeta,y_t+\zeta]$.
  The only difference is that
  $\Gamma_2(z)=\Gamma_2(x)$ in that proof, while in the present situation
  $|\Gamma_2(x_t)-\Gamma_2(y_t)|
  \leq C\,\min\{d(x_t,y_t),t\,d(x_t,y_t)^\alpha\}$.
  Therefore the estimate (\ref{eq:trick1}) does not change at all
  (except for the constant) whereas the estimate in
  (\ref{eq:trick2}) is replaced by the weaker one 
  $\ldots\leq\const\,d(x_t,y_t)$. This leads to
\begin{displaymath}
  \|\Psi_t(\zeta)-\zeta\|
  =
  \left\|[x_t+\zeta,y_t+\zeta]-(x_t+\zeta)\right\|
  \leq
  C\,\min\{\ve\,d(x,y)^\alpha,d(x,y)\}\ .
\end{displaymath}
\end{proof}

Armed with these estimates we continue to estimate $A_1$ from
(\ref{eq:hoelder-phi-estimate}). Observe first that
\begin{displaymath}
  \begin{split}
    &\int_{\R^2}\tilde q_\ve(x,\zeta)\,1_{\Omega(x,\zeta)}(t)\,
    \vf([x_t+\zeta,y_t+\zeta])\,d\zeta\\
    &=
    \int_{\R^2}\tilde q_\ve(x,\Psi_t^{-1}\hat\zeta)\,
    1_{\Omega(x,\Psi_t^{-1}\hat\zeta)}(t)\,
    \vf(x_t+\hat\zeta)\,
    J\Psi_t^{-1}(\hat\zeta)\,d\hat\zeta
  \end{split}
\end{displaymath}
and that, as in (\ref{eq:r-difference}),
\begin{displaymath}
  \left|r(x_t,\zeta)-r(x_t,\Psi_t^{-1}\zeta)\right|
  \leq
  \|\zeta\|\,C\,d(x,y)^\alpha\ .
\end{displaymath}
This implies
\begin{displaymath}
  \begin{split}
    A_1
    &\leq
    \int_{\R^2}\int_\R|\vf(x_t+\zeta)|
    \big|\tilde q_\ve(x,\zeta)-\tilde q_\ve(x,\Psi_t^{-1}\zeta)\big|JU_{-t}(x)
    \,1_{\Omega(x,\zeta)}(t)
    \,dt\,d\zeta\\
    &\hspace{0.2cm}
    +\int_{\R^2}\int_\R|\vf(x_t+\zeta)|\tilde q_\ve(x,\Psi_t^{-1}\zeta)
    \left|1_{\Omega(x,\zeta)}(t)-
      1_{\Omega(x,\Psi_t^{-1}\hat\zeta)}(t)\right|
    JU_{-t}(x)\,dt\,d\zeta\\
    &\hspace{0.2cm}
    +\int_{\R^2}\int_\R|\vf(x_t+\zeta)|
    \tilde q_\ve(x,\Psi_t^{-1}\zeta)\,
    1_{\Omega(x,\Psi_t^{-1}\hat\zeta)}(t)\,
    \big|J\Psi_t^{-1}(\zeta)-1\big|\,JU_{-t}(x)
    dt\,d\zeta    \\
    &\leq
    \const\,|\vf|_\infty \left(\ve\,M
      \sup_{0\leq t\leq r(x_t,\zeta)}
        \ve^{-1}\|\zeta-\Psi_t^{-1}(\zeta)\|
    + 2\cdot C\,\ve\,d(x,y)^\alpha\right)\\
    &\leq
    \const\,\ve\,d(x,y)^\beta\delta^{\gamma-\beta}
  \end{split}
\end{displaymath}
where we used (\ref{eq:Psit-distance}) for the last step.

Combining this with (\ref{eq:hoelder-phi-estimate}), (\ref{eq:A3-estimate})
and (\ref{eq:A2-estimate}) we find that, for $\vf\in\D_\gamma$,
\begin{displaymath}
  \left|\Phi(x)-\Phi(y)\right|
  \leq
  \const\,\ve\,\delta^{\gamma-\beta}\,d(x,y)^\beta
\end{displaymath}
so that $\ve^{-1}\Phi\in\D_\beta$ for sufficiently small $\delta$.
It follows that $\ve^{-1}\Phi\, v^u\in\V_\beta$. Therefore,
\begin{equation}
  \label{eq:I1-est}
  |I_1|\leq \ve\,\|f\|_u\ ,
\end{equation}
see (\ref{eq:I1I2-def}) and (\ref{eq:I1-def}) to recall the meaning of $I_1$.

It remains to estimate the integral
\begin{displaymath}
  I_2=\int_{\manif}\left(\int_{\R^2}
    \tilde q_{\ve}(x,\zeta)f([x,x+\zeta])  \vf(x+\zeta)\,d\zeta
    -f(x)\vf(x)\right)\,m(dx)\ ,
\end{displaymath}
see (\ref{eq:I1I2-def}). The strategy is similar to that for the estimate 
of $I_1$. For each $\zeta$, we use the coordinate change 
$\Psi_\zeta(x):=[x,x+\zeta]$.

\begin{sublem}
  \label{sublem:Psi-zeta-lemma}
  The map $\Psi_\zeta$ is injective and
  there is a constant $C>0$, which can be chosen
  the same for all $x$ and $\zeta$ and $y\in W^s(x)$, such that
  \begin{equation}
    \label{eq:Psizeta-distance-1}
    \|\Psi_\zeta^{-1}(x)-x\|\leq C\,\|\zeta\|
  \end{equation}
  \begin{equation}
    \label{eq:Psizeta-distance-2}
    \|\Psi_\zeta^{-1}(x)-\Psi_\zeta^{-1}(y)\|\leq C\,d(x,y)\ .
  \end{equation}
  In addition $\Psi_\zeta$ is continuously differentiable and its Jacobian
  $J\Psi_\zeta$ satisfies
  \begin{equation}
    \label{eq:JPsizeta-estimate-1}
    |J\Psi_\zeta(x)-1|\leq C\,\|\zeta\|^\alpha
  \end{equation}
  \begin{equation}
    \label{eq:JPsizeta-estimate-2}
    |J\Psi_\zeta(x)-J\Psi_\zeta(y)|\leq C\,d(x,y)^\alpha\ .
  \end{equation}
\end{sublem}
We postpone the proof of this Sub-lemma to the end of this section.
\medskip\par
Using the coordinate change $\Psi_\zeta$,
\begin{displaymath}
  \begin{split}
    I_2
    &=\phantom{:}
    \int_\manif\left(\int_{\R^2} \left(\tilde q_{\ve}(\Psi_\zeta^{-1}x,\zeta)
    \vf(\Psi_\zeta^{-1}x+\zeta)J\Psi_\zeta^{-1}(x)
    -\tilde q_{\ve}(x,\zeta)\vf(x)\right)
    \,d\zeta\right)\,f(x)\,m(dx)\\
    &=\phantom{:}
    \int_\manif\left(\int_{\R^2} \tilde q_{\ve}(\Psi_\zeta^{-1}x,\zeta)
    \big(\vf(\Psi_\zeta^{-1}x+\zeta)J\Psi_\zeta^{-1}(x)-\vf(x)\big)
    \,d\zeta\right)\,f(x)\,m(dx)\\
    &\hspace{2cm}
    +\int_\manif\left(\int_{\R^2}\left(\tilde q_\ve(\Psi_\zeta^{-1}x,\zeta)
      -\tilde q_\ve(x,\zeta)\right)
    \vf(x)\,d\zeta\right)\,f(x)\,m(dx)\\
    &=:\int_\manif\Phi_1(x)f(x)\,m(dx)+\int_\manif\Phi_2(x)f(x)\,m(dx)\ ,
  \end{split}
\end{displaymath}
and we have to estimate $|\Phi_i|_\infty$ and $H_\beta^s(\Phi_i)$ in order
to bound $I_2$ in terms of $\|f\|_s$.

We start with the estimate for $\Phi_1$.
In view of Remark~\ref{rem:Lipschitz-domination}
and of (\ref{eq:Psizeta-distance-1}) we have 
$\tilde q_\ve(\Psi_\zeta^{-1}x,\zeta)\leq\const\,\tilde p_{a\ve}(x,\zeta)$
So it suffices to show that, for each fixed $\zeta$ with
$\|\zeta\|<\ve$, the supremum and the $\beta$--H\"older constant of
\begin{displaymath}
\Phi_{1,1}(x,\zeta):=\vf(\Psi_\zeta^{-1}x+\zeta)-\vf(x)\quad\hbox{and}\quad
  \Phi_{1,2}(x,\zeta):=\vf(\Psi_\zeta^{-1}x+\zeta)(J\Psi_\zeta^{-1}(x)-1)
\end{displaymath}
are bounded (uniformly in $\zeta$) by $\const\,\ve^{\gamma-\beta}$.
The relevant points are sketched in Figure \ref{fig:figure4}.

\Bfig(300,150)
      {\footnotesize{
       \bline(30,20)(1,2)(45)
       \dashline{3}(42,41)(123,81)
       \put(77,110){$W^{u}(x)$}
       \put(124,82.5){\circle*{3}} \put(122,72){$\Psi_\zeta^{-1}x+\zeta$}
       \put(40,40){\circle*{3}} \put(45,35){$\Psi_\zeta^{-1}x$}
       \put(66,92){\circle*{3}}   \put(70,85){$x$}
       \bline(180,20)(1,2)(35)  \bline(260,60)(-6,1)(240)
       \put(180,20){\circle*{3}} \put(185,15){$\Psi_\zeta^{-1}y$}
       \dashline{3}(180,20)(261,60)
       \put(205,69){\circle*{3}} \put(202,58){$y$}
       \put(261,60){\circle*{3}} \put(265,58){$\Psi_\zeta^{-1}y+\zeta$}
       \put(0,105){$W^{s}(x)=W^s(y)$} \put(218,90){$W^{u}(y)$}
      }}
{\label{fig:figure4}}

Recall that $\vf\in\D_\gamma$. Then
\begin{displaymath}
  |\Phi_{1,1}|_\infty\leq((C+1)\ve)^\gamma H_\gamma^s(\vf)
  \leq
  \const\,\ve^{\gamma-\beta}
\end{displaymath}
because of (\ref{eq:Psizeta-distance-1}).
Similarly, invoking (\ref{eq:JPsizeta-estimate-1}),
\begin{displaymath}
  |\Phi_{1,2}|_\infty
  \leq
  |\vf|_\infty\,C\,\ve^\alpha
  \leq
  \const\,\ve^{\gamma-\beta}\ .
\end{displaymath}

To analyse the H\"older constant of $\Phi_{1,1}$ we consider two
different cases which may happen for points $x$ and $y\in W^{s}(x)$.
\relax First, if $d(x,y)\le\ve$, we use (\ref{eq:Psizeta-distance-2}) 
to estimate
\begin{displaymath}
  \begin{split}
    |\Phi_{1,1}(x,\zeta) - \Phi_{1,1}(y,\zeta)|
    &\le
    |\vf(\Psi_\zeta^{-1}x+\zeta) - \vf(\Psi_\zeta^{-1}y+\zeta)|
    +|\vf(x) - \vf(y)|\\
    &\le
    \const\,d(x,y)^\gamma\,H^s_\gamma(\vf)\\
    &\leq
    \const\,\ve^{\gamma-\beta}\,d(x,y)^\beta
  \end{split}
\end{displaymath}
(Observe that
$\Psi_\zeta^{-1}(x)+\zeta \in W^s(x) = W^s(y) \ni
\Psi_\zeta^{-1}(y)+\zeta$.)

Otherwise, if $d(x,y)>\ve$ we proceed as follows:
\begin{displaymath}
  |\Phi_{1,1}(x,\zeta) - \Phi_{1,1}(y,\zeta)|
  \leq
  2|\Phi_{1,1}|_\infty
  \leq
  \const\,\ve^{\gamma}
  \leq
  \const\,\ve^{\gamma-\beta}\,d(x,y)^\beta\ .
\end{displaymath}
In any case,
\begin{displaymath}
  H^s_\beta(\Phi_{1,1}(\,.\,,\zeta))\leq\const\,\ve^{\gamma-\beta}\ .
\end{displaymath}

The H\"older constant of $\Phi_{1,2}$ is estimated along the same lines
using (\ref{eq:JPsizeta-estimate-2}) -- the details are
left to the reader.

We turn to the estimates for $\Phi_2$. Because of
(\ref{eq:Lipschitz-domination-b}) and (\ref{eq:Psizeta-distance-1}) we have
\begin{displaymath}
  |\Phi_2|_\infty
  \leq
  |\vf|_\infty\,M(C\|\zeta\|)^\alpha
  \leq
  \const\,\ve^\alpha
  \leq
  \const\,\ve^{\gamma-\beta}
\end{displaymath}
In order to estimate the H\"older constant of $\Phi_2$ we
consider the cases $d(x,y)\leq\ve$ and $d(x,y)>\ve$ separately as
in the estimate for $H_\beta^s(\Phi_{1,1})$. This yields
\begin{displaymath}
  \begin{split}
    H_\beta^s(\Phi_2)
    &\leq
    \max\left\{\const\,H_\gamma^s(\vf)\,\ve^{\gamma-\beta}
      +\const\,\ve^{\alpha-\beta},\const\,\ve^{\alpha-\beta}\right\}\\
    &\leq
    \const\,\ve^{\gamma-\beta}\ .
  \end{split}
\end{displaymath}
\end{proof}

\begin{proof}[\bf Proof of Sub-lemma \ref{sublem:Psi-zeta-lemma}]
  Let $\Gamma$ be a $\Co^{1+\alpha}$-diffeomorphism as in the proof of
  Sub-lemma~\ref{sublem:Psi-estimates}. This means that
  \begin{displaymath}
    \Gamma_1(\Psi_\zeta^{-1}x)=\Gamma_1(x)\ ,\quad
    \Gamma_2(\Psi_\zeta^{-1}x+\zeta)=\Gamma_2(x)\ .
  \end{displaymath}
  Hence
  \begin{displaymath}
    \begin{split}
      \left\|\Gamma(\Psi_\zeta^{-1}x)-\Gamma(x)\right\|
      &=
      \left|\Gamma_2(\Psi_\zeta^{-1}x)-\Gamma_2(x)\right|
      =
\left|\Gamma_2(\Psi_\zeta^{-1}x)-\Gamma_2(\Psi_\zeta^{-1}x+\zeta)\right|\\
      &\leq
\left\|D{\Gamma_2}_{|\Psi_\zeta^{-1}x+t\zeta}\right\|\cdot\left\|\zeta\right\|
    \end{split}
  \end{displaymath}
  for a suitable $t\in(0,1)$. This implies (\ref{eq:Psizeta-distance-1}).

  To proceed further, we need the estimate
  \begin{equation}
    \label{eq:auxiliary}
    \left\|D(\Gamma_2\circ\Psi_\zeta^{-1}-{\Gamma_2})_{|x}\right\|
    \leq
    \const\,\|\zeta\|^\alpha\ .
  \end{equation}
  This is proved as follows:
  Since $\Gamma_2(x)=\Gamma_2(\Psi_\zeta^{-1}x+\zeta)$,
  we have
  \begin{displaymath}
    \begin{split}
      \left\|D(\Gamma_2\circ\Psi_\zeta^{-1})_{|x}-D{\Gamma_2}_{|x}\right\|
      &=
      \left\|D{\Gamma_2}_{|\Psi_\zeta^{-1}x}\cdot D{\Psi_\zeta^{-1}}_{|x}
        -D(\Gamma_2\circ(\Psi_\zeta^{-1}+\zeta))_{|x}\right\|\\
      &=
      \left\|\left(D{\Gamma_2}_{|\Psi_\zeta^{-1}x}
          -D{\Gamma_2}_{|\Psi_\zeta^{-1}x+\zeta}\right)\cdot
D{\Psi_\zeta^{-1}}_{|x}
      \right\|\\
      &\leq
\const\,\|\zeta\|^\alpha\,\sup_{x,\zeta}\left\|D{\Psi_\zeta^{-1}}_{|x}\right\|
    \end{split}
  \end{displaymath}
  and
  \begin{equation}
    \label{eq:impl-funct-theo}
    \sup_{x,\zeta}\left\|D{\Psi_\zeta^{-1}}_{|x}\right\|<\infty
  \end{equation}
  since the manifold $\manif$ is compact and since $\Psi_\zeta$
  is determined by the implicit function theorem.

  Now we apply (\ref{eq:auxiliary}) to prove
  estimate~(\ref{eq:Psizeta-distance-2}). Since
  $\Gamma_1(\Psi_\zeta^{-1}y)-\Gamma_1(\Psi_\zeta^{-1}x)
  =\Gamma_1(y)-\Gamma_1(x)$,
  it suffices to estimate
  $\Gamma_2(\Psi_\zeta^{-1}y)-\Gamma_2(\Psi_\zeta^{-1}x)$.
  To this end let $t_0:=\Gamma_1(y)-\Gamma_1(x)$ and assume without loss of
  generality that $t_0>0$. For $0\leq t\leq t_0$ denote
  $x_t:=\Gamma^{-1}(\Gamma x+t\,{\bf e}_1)$. Then $x_0=x$, $x_{t_0}=y$, and
  \begin{displaymath}
    \begin{split}
      \left\|\Gamma_2(\Psi_\zeta^{-1}y)-\Gamma_2(\Psi_\zeta^{-1}x)\right\|
      &=
      \left\|\left(\Gamma_2(\Psi_\zeta^{-1}x_{t_0})-\Gamma_2x_{t_0}\right)
        -\left(\Gamma_2(\Psi_\zeta^{-1}x_0)-\Gamma_2x_0\right)\right\|\\
      &\leq
|t_0|\,\left\|D(\Gamma_2\circ\Psi_\zeta^{-1}-\Gamma_2)_{|x_t}\right\|\\
      &\leq
      \const\,\|\zeta\|^\alpha\,d(x,y)\ .
    \end{split}
  \end{displaymath}

  We turn to the proof of (\ref{eq:JPsizeta-estimate-1}). Obviously it
  suffices to estimate $\|D\Psi_\zeta^{-1}(x)-\Eins\|$ where 
  $\Eins$ denotes the identity matrix.
  \begin{displaymath}
    \begin{split}
      \left\|D{\Psi_\zeta^{-1}}_{|x}-\Eins\right\|
      &=
      \left\|D{\Gamma^{-1}}_{|\Gamma(\Psi_\zeta^{-1}x)}\cdot
        D(\Gamma\circ\Psi_\zeta^{-1})_{|x}
        -D{\Gamma^{-1}}_{|\Gamma x}\cdot D{\Gamma}_{|x}\right\|\\
      &\leq
      \left\|D{\Gamma^{-1}}_{|\Gamma(\Psi_\zeta^{-1}x)}\right\|\,
      \left\|D(\Gamma\circ\Psi_\zeta^{-1})_{|x}-D\Gamma_{|x}\right\|\\
      &\hspace{1cm}+\left\|D{\Gamma^{-1}}_{|\Gamma(\Psi_\zeta^{-1}x)}
        -D{\Gamma^{-1}}_{|\Gamma x}\right\|\,\left\|
       D{\Gamma}_{|x}\right\|\\
      &\leq
  \const\,\left\|D(\Gamma_2\circ\Psi_\zeta^{-1})_{|x}-D{\Gamma_2}_{|x}\right\|
      +\const\,d(\Psi_\zeta^{-1}x,x)^\alpha\\
      &\leq
      \const\,\|\zeta\|^\alpha
    \end{split}
  \end{displaymath}
  where we used (\ref{eq:auxiliary}) and (\ref{eq:Psizeta-distance-1}) for
  the last step.

  The proof of estimate~(\ref{eq:JPsizeta-estimate-2})
  is more subtle. To simplify the notation denote by $G$ and $H$ the maps
  $x\mapsto\Psi_\zeta^{-1}(x)$ and $x\mapsto\Psi_\zeta^{-1}(x)+\zeta$,
  respectively. Then $DG_{|x}=DH_{|x}$ and hence
  \begin{displaymath}
    D{\Gamma^{-1}}_{|\Gamma(Gx)}\, D(\Gamma\circ G)_{|x}
    =
    D{\Gamma^{-1}}_{|\Gamma(Hx)}\, D(\Gamma\circ H)_{|x}
  \end{displaymath}
  so that
  \begin{displaymath}
    D(\Gamma\circ G)_{|x}
    =
    \left(D\Gamma_{|Gx}\, (D{\Gamma}_{|Hx})^{-1}\right)\,
        D(\Gamma\circ H)_{|x}
    =:
    (\Eins+\Delta_{\zeta,x})\, D(\Gamma\circ H)_{|x}\ .
  \end{displaymath}
  Since $d(Gx,Hx)=\|\zeta\|$, it follows that
  $\|\Delta_{\zeta,x}\|\leq\const\,\|\zeta\|^\alpha$.

  Now consider again $x$ and $y$ with $\Gamma_2(x)=\Gamma_2(y)$. For such
  points
  \begin{displaymath}
    \begin{split}
      D(\Gamma\circ G)_{|y}-D(\Gamma\circ G)_{|x}
      =
      D(\Gamma\circ H)_{|y}-D(\Gamma\circ H)_{|x}
      +\Delta_{\zeta,y}\cdot D(\Gamma\circ H)_{|y}
      -\Delta_{\zeta,x}\cdot D(\Gamma\circ H)_{|x}
    \end{split}
  \end{displaymath}
  Multiplying this equation with the row vector ${\bf e}_2'=(0\;1)$ 
  from the left we obtain
  \begin{displaymath}
    \begin{split}
      D(\Gamma_2\circ G)_{|y}-D(\Gamma_2\circ G)_{|x}
      =
      D{\Gamma_2}_{|y}-D{\Gamma_2}_{|x}
      +{\bf e}_2'\cdot\left(\Delta_{\zeta,y}\cdot D(\Gamma\circ H)_{|y}
      -\Delta_{\zeta,x}\cdot D(\Gamma\circ H)_{|x}\right)
    \end{split}
  \end{displaymath}
  because $\Gamma_2\circ H=\Gamma_2$. Observing now that also
  $\Gamma_1\circ G=\Gamma_1$ we can conclude that
  \begin{displaymath}
    \begin{split}
      &\hspace*{-5mm}\left\|D(\Gamma\circ G)_{|y}-D(\Gamma\circ G)_{|x}
                     \right\|\\
      &\leq
      \left\|D{\Gamma}_{|y}-D{\Gamma}_{|x}\right\|
      +\left\|\Delta_{\zeta,y}\cdot D(\Gamma\circ H)_{|y}
        -\Delta_{\zeta,x}\cdot D(\Gamma\circ H)_{|x}\right\|\\
      &\leq
      \const\,d(x,y)^\alpha
      +\left\|\Delta_{\zeta,y}-\Delta_{\zeta,x}\right\|\,
      \left\|D(\Gamma\circ H)_{|y}\right\|
      +\left\|\Delta_{\zeta,x}\right\|\,
      \left\|D(\Gamma\circ H)_{|y}-D(\Gamma\circ H)_{|x}\right\|\\
      &\leq
      C_1\,d(x,y)^\alpha
      +C_2\,\ve^\alpha\,
      \left\|D(\Gamma\circ H)_{|y}-D(\Gamma\circ H)_{|x}\right\|
    \end{split}
  \end{displaymath}
  for suitable $C_1,C_2>0$,
  where we used the boundedness of $\left\|D(\Gamma\circ H)_{|y}\right\|$,
  the estimate
  $\left\|\Delta_{\zeta,x}\right\|\leq\const\,\|\zeta\|^\alpha$ and
  the fact that
  $\left\|\Delta_{\zeta,y}-\Delta_{\zeta,x}\right\|\leq\const\,d(x,y)^\alpha$ 
  in the last step.

  So we proved
  \begin{displaymath}
    \left\|D(\Gamma\circ G)_{|y}-D(\Gamma\circ G)_{|x}\right\|
    -C_2\,\ve^\alpha\,
    \left\|D(\Gamma\circ H)_{|y}-D(\Gamma\circ H)_{|x}\right\|
    \leq
    C_1\,d(x,y)^\alpha\ .
  \end{displaymath}
  Interchanging the r\^oles of $G$ and $H$ we obtain in the same way
  \begin{displaymath}
    \left\|D(\Gamma\circ H)_{|y}-D(\Gamma\circ H)_{|x}\right\|
    -C_2\,\ve^\alpha\,
    \left\|D(\Gamma\circ G)_{|y}-D(\Gamma\circ G)_{|x}\right\|
    \leq
    C_1\,d(x,y)^\alpha
  \end{displaymath}
  with the same constants. Therefore we can add both inequalities and arrive
at
  \begin{displaymath}
    \left\|D(\Gamma\circ G)_{|y}-D(\Gamma\circ G)_{|x}\right\|
    +\left\|D(\Gamma\circ H)_{|y}-D(\Gamma\circ H)_{|x}\right\|
    \leq
    \frac{2C_1\,d(x,y)^\alpha}{1-C_2\,\ve^\alpha}
  \end{displaymath}
  provided $\ve>0$ is sufficiently small. Hence
  \begin{displaymath}
    \begin{split}
      &\left\|D{\Psi_\zeta^{-1}}_{|y}-D{\Psi_\zeta^{-1}}_{|x}\right\|\\
      &=
      \left\|DG_{|y}-DG_{|x}\right\|\\
      &=
      \left\|D\Gamma^{-1}_{|\Gamma(Gy)}\cdot D(\Gamma\circ G)_{|y}
        -D\Gamma^{-1}_{|\Gamma(Gx)}\cdot D(\Gamma\circ G)_{|x}\right\|\\
      &\leq
      \left\|D\Gamma^{-1}_{|\Gamma(Gy)}-D\Gamma^{-1}_{|\Gamma(Gx)}\right\|\,
      \left\|D(\Gamma\circ G)_{|y}\right\|
      +\left\|D\Gamma^{-1}_{|\Gamma(Gx)}\right\|\,
      \left\|D(\Gamma\circ G)_{|y}-D(\Gamma\circ G)_{|x}\right\|\\
      &\leq
      \const\,d(\Psi_\zeta^{-1}y,\Psi_\zeta^{-1}x)^\alpha\,
      \left\|D{\Psi_\zeta^{-1}}_{|y}\right\|
      +\const\,d(x,y)^\alpha\\
      &\leq
      \const\,d(x,y)^\alpha
    \end{split}
  \end{displaymath}
  where we used (\ref{eq:Psizeta-distance-2}) and
  (\ref{eq:impl-funct-theo}) in the last step.
\relax  From this estimate (\ref{eq:JPsizeta-estimate-2}) follows immediately.
\end{proof}

\subsection{Proofs: Spectral stability -- Deterministic Perturbations}
\label{subsec:deterministic-proof}
\quad\\
Let $T$ be a $\CoT$ Anosov diffeomorphism of the 2-dimensional torus $\To^2$.
Our goal is to prove that the spectral properties of the associated 
transfer operator do not change much under $\Co^1$--small
perturbations as long as the $\CoT$ norm 
of the perturbed maps remains smaller than some constant $K>\|T\|_\CoT$.
\relax From now on we will use $\tilde T$ to denote any Anosov map such that
$d_{\Co^1}(T,\tilde T)=:\ell<<1$ and $\|\tilde T\|_\CoT\leq K$.

Since we are working now with spaces associated to $T$ and also with 
spaces associated to $\tilde T$, we will denote the space of
test functions related
to $\tilde T$ by $\tilde \D_\beta$, the resulting norm by
$\tildenorm{\cdot}$ \etc 

The essential geometric fact that makes it possible 
to compare $\Lp_T:\B\to\B$ and $\Lp_{\tilde T}:\tilde\B\to\tilde\B$
is contained in the following Lemma. 
\begin{lem}\label{lem:norm-geometry}
There exist constants $C,\rho>0$ such that, for each $f\in\Co^2$,
\[
  \|f\|\leq C\ell^{\rho}\|f\|_{2,1}+C\tildenorm{f}
\]
where $\|\cdot\|_{k,1}$ denotes the $L^1$-Sobolev norms. 
\end{lem}
\begin{proof}
By using a smooth partition of
unity we can always restrict ourselves to a situation in which
the functions are all supported in small balls of radius
$r$.\footnote{If in doubt look at Lemma
\ref{lem:multiplication}.} In such a ball one can consider two
changes of variables $\Gamma$ and $\tilde\Gamma$ that straighten the
foliations of the maps $T$ and $\tilde T$, respectively. As seen in
the previous section these are $\Co^{1+\alpha}$.

Accordingly, setting $F:=\Gamma^{-1}\circ\tilde\Gamma$, $F$ is
$\Co^{1+\alpha}$. More can be said.
\begin{sublem}\label{eq:closechart}
There exist constants $C>0$ and $\rho>0$ such that
\[
  \|F-\Id\|_\infty\leq C\ell^\rho
  \quad\text{and}\quad
  \|dF-\Eins\|_\infty\leq C\ell^{\rho}.
\]
\end{sublem}

Since the above result is hardly surprising and its proof is a bit
technical we postpone it to the end of the present section.

\relax For each $\vf\in\D_\beta$ and $f\in\Co^1$ holds
\begin{displaymath}
\begin{split}
  \left|\int_{\To^2} f\vf\,dm\right|
  &\leq
  \left|\int_{\To^2} (f-f\circ F^{-1})\cdot\vf\,dm\right|
  +\left|\int_{\To^2} f\cdot(\vf\circ F) JF\,dm\right|\\
  &\leq
  \|f\|_{1,1}\|\Id-F^{-1}\|_\infty
    +\left|\int_{\To^2} f\,(\vf\circ F)\,JF\,dm\right|\ .
\end{split}
\end{displaymath}
Clearly there exists $C>0$ such that $C^{-1}\vf\circ F\cdot JF\in\tilde
\D_\beta$, hence Sub-Lemma \ref{eq:closechart} implies
\[
\|f\|_s\leq C\|f\|_{1,1} \ell^\rho+C\tildenorm{f}_s.
\]
Analogously, 
\[
  \int_{\To^2} df(\vf v^u)\,dm
  =
  \int_{\To^2} d_Ff[(v^u\circ F)(\vf\circ F)]JF\,dm
  =
  \int_{\To^2} d(f\circ F)[((DF^{-1} v^u) \circ F)(\vf\circ F)]JF\,dm\ .
\]
Since $F$ sends the stable and unstable manifolds of $\tilde T$ to the
corresponding ones of  $T$ and is $\Co^{1+\alpha}$ it follows that
there exists an $\alpha$-H\"older function $\omega$ such that 
$(DF^{-1} v^u)\circ F=\omega\tilde v^u$. Accordingly, 
$\tilde v:=C^{-1} (DF^{-1}v^u) \circ F\cdot\vf\circ F\cdot JF
\in\tilde\V_\beta$, 
for some constant $C>0$. Therefore the following estimate finishes
the proof:
\begin{align*}
  \int_{\To^2}|d(f\circ F)({\tilde v})-df({\tilde v})|\,dm
  &=
  \int_{\To^2}|d_Ff(dF({\tilde v})-{\tilde v})|\,dm
  +\int_{\To^2}|(d_Ff-df)({\tilde v})|\,dm\\
  &\leq 
  \|f\|_{1,1}\|dF-\Eins\|_\infty+\|f\|_{2,1}\|F-\Id\|_\infty\\
  &\leq
  C\,\ell^\rho\,\|f\|_{2,1}
\end{align*}
where the last inequality follows from Sub-Lemma \ref{eq:closechart}.
\end{proof}

Some further estimates are collected in the next lemma.

\begin{lem}\label{lem:elementary-estimates}
Under the above assumptions on $q_\ve$, $T$ and $\tilde T$, we have for all
$f\in\Co^1(\To^2,\R)$
\begin{align}
  \|Q_\ve f\|^{(\sim)}&\leq\const\,\ve^{-1}\,\|f\|_1\label{eq:ihes5}\\
  \|Q_\ve f\|_{1}&\leq\const\,\ve^{-1}\,\|f\|^{(\sim)}\label{eq:ihes6}\\
  \|Q_\ve f\|_{1,1}&\leq\const\,\ve^{-2}\,\|f\|^{(\sim)}\label{eq:ihes8}\\
  \|Q_\ve f\|_{2,1}&\leq\const\,\ve^{-3}\,\|f\|^{(\sim)}\label{eq:ihes8a}
\end{align}
Furthermore,
\begin{equation}\label{eq:ihes9}
  \|(\Lp_T-\Lp_{{\tilde T}})f\|_1\leq\const\cdot
    \dist_{\Co^1}(T,{\tilde T})\,\|f\|_{1,1}\ .
\end{equation}
\end{lem}

\begin{proof}
For $\vf\in\D_\beta$, 
$\int_{\To^2} Q_\ve f\cdot\vf\,dm\leq\|Q_\ve f\|_1\leq\|f\|_1$, so $\|Q_\ve f\|_s\leq\|f\|_1$.
In order to estimate $\|Q_\ve f\|_u$, let $v\in\V_\beta$. Observing that
$q_\ve$ is a convolution kernel, one verifies easily that
for $f\in\Co^1(\To^2,\R)$
\begin{displaymath}
\begin{split}
  \int_{\To^2}d(Q_\ve f)(v)\,dm
  &=
  \int_{\To^2}D(Q_\ve f)\cdot v\,dm
  =
  \int_{\To^2}Q_\ve(Df)\cdot v\,dm
  =
  \int_{\To^2}Df\cdot Q_\ve v\,dm\\
  &=
  -\int_{\To^2}f\,\divg(Q_\ve v)\,dm
  \leq
  \|f\|_1\cdot\|\divg(Q_\ve v)\|_\infty\ ,
\end{split}
\end{displaymath}
and $|\divg(Q_\ve v)|\leq\const\,\ve^{-1}\,|v|_\infty\leq\const\,\ve^{-1}$.
Hence $\|Q_\ve f\|\leq\const\,\ve^{-1}\,\|f\|_1$, and since the same 
arguments apply to the $\tildenorm{\cdot}$ norm, this finishes
the proof of (\ref{eq:ihes5}).

Next we estimate $\|Q_\ve f\|_1$. To this end consider 
$\vf\in\Co^1(\To^2,\R)$ with $|\vf|_\infty\leq1$.
Then
\begin{displaymath}
  \int_{\To^2}\vf\,Q_\ve f\,dm
  =
  \int_{\To^2}Q_\ve^*\vf\,f\,dm
  \leq
  \max\{|Q_\ve^*\vf|_\infty,H_\beta^s(Q_\ve^*\vf)\}\cdot\|f\|_s
  \leq
  \const\,\ve^{-1}\,\|f\|\ .
\end{displaymath}
As $\|Q_\ve f\|_1$ is the supremum of all such integrals,
this proves (\ref{eq:ihes6}).
Again the same estimate applies to the $\tildenorm{\cdot}$ norm.

The proof of (\ref{eq:ihes8}) is similar to that of (\ref{eq:ihes6}).
Consider $v\in\Co^1(\To^2,\R^2)$ with $|v|_\infty\leq1$. Then
\begin{displaymath}
\begin{split}
  \int_{\To^2}D(Q_\ve f)\cdot v\,dm
  &=
  -\int_{\To^2}Q_\ve f\,\divg(v)\,dm
  =
  -\int_{\To^2}f\,\divg(Q_\ve^* v)\,dm\\
  &\leq
  \max\{|\divg(Q_\ve^*v)|_\infty,H_\beta^s(\divg(Q_\ve^*v))\}\cdot\|f\|_s
  \leq
  \const\,\ve^{-2}\,\|f\|\ ,
\end{split}
\end{displaymath}
and the argument is concluded as before. Estimate (\ref{eq:ihes8a})
is proved in the same way.

We turn to (\ref{eq:ihes9}). 
To this end we consider $T^{-1}$ and ${{\tilde T}}^{-1}$
as maps from $[0,1]^2$ to $\R^2$ and denote 
$w(x):={{\tilde T}}^{-1}(x)-T^{-1}(x)$.
\relax For $t\in[0,1]$ define
\begin{displaymath}
  S_t(x):=(1-t)\,T^{-1}(x)+t\,{{\tilde T}}^{-1}(x)\quad\text{and}\quad
  g_t(x):=JS_t(x)\ ,
\end{displaymath}
so $S_0=T^{-1}$ and $S_1={{\tilde T}}^{-1}$ and $g_0$, $g_1$ are the 
Jacobians of $T^{-1}$ and ${{\tilde T}}^{-1}$, respectively. Therefore, 
not writing the variable $x$ explicitly and extending $f$ periodically 
to the covering $\R^2$ of $\To^2$, we have 
\begin{displaymath}
\begin{split}
  \left|\Lp_{{\tilde T}}f-\Lp_T f\right|
  &=
  \left|f\circ S_1\cdot g_1-f\circ S_0\cdot g_0\right|\\
  &\leq
  \int_0^1\left|\frac{\partial}{\partial t}
    \left(f\circ S_t\cdot g_t\right)\right|\,dt\\
  &\leq
  \int_0^1\left|Df_{|S_t}\cdot w\cdot g_t\right|\,dt
  +\int_0^1\left|f\circ S_t\right|
     \cdot\left|\frac{\partial g_t}{\partial t}\right|\,dt\\
  &\leq
  \|w\|_\infty\cdot\int_0^1|Df|\circ S_t\cdot JS_t\,dt
  +\left\|\frac{\partial}{\partial t}\ln g_t\right\|_\infty
  \cdot\int_0^1|f|\circ S_t\cdot JS_t\,dt
\end{split}
\end{displaymath}
Fix $a>0$ such that $K_t:=S_t([0,1]^2)\subset[-a,a]^2$ for all $t\in[0,1]$.
Then
\begin{displaymath}
\begin{split}
  \left\|\Lp_{{\tilde T}}f-\Lp_Tf\right\|_1
  &\leq
  \|w\|_\infty\cdot\int_0^1\left(\int_{K_t}|Df|\,dm\right)\,dt
  +\sup_t\left\|\frac{\partial}{\partial t}\ln g_t\right\|_\infty\cdot
   \int_0^1\left(\int_{K_t}|f|\,dm\right)\,dt\\
  &\leq
  a^2\left(\|w\|_\infty
  +\sup_t\left\|\frac{\partial}{\partial t}\ln g_t\right\|_\infty\right)\cdot
  \|f\|_{1,1}\ .
\end{split}
\end{displaymath}
Clearly, $\|w\|_\infty\leq\const\cdot\dist_{\Co^0}({\tilde T},T)$, and a 
routine calculation shows that
\begin{displaymath}
\begin{split}
  \left|\frac{\partial}{\partial t}\ln g_t\right|
  &=
  \left|\sum_{k=1}^\infty(-t)^{k-1}\tr\left((DS_0)^{-1}DS_1-1\right)^k\right|\\
  &\leq\frac{2\left|(DS_0)^{-1}DS_1-1\right|}
            {1-\left|(DS_0)^{-1}DS_1-1\right|}
       \leq\const\cdot\dist_{\Co^1}({\tilde T},T)
\end{split}
\end{displaymath}
provided this distance is small enough to make the denominator
of the fraction not smaller than $\frac 12$, say.
\end{proof}

Introducing smooth averaging operators $Q_\ve$ we can get rid
of the Sobolev norms in Lemma~\ref{lem:norm-geometry}.
\begin{lem}\label{lem:norm-comparison}
Suppose that $T$ and $\tilde T$ are $\CoT$--Anosov diffeomorphisms of $\To^2$,
$\|T\|_\CoT,\|\tilde T\|_\CoT\leq K$.
Let $q_\ve(x,y)=\ve^{-2}\bar q(\ve^{-1}(y-x))$ 
be $\Co^2$. Then there is a constant $C>0$ that depends
only on $\bar q$ and on $T$ such that for sufficiently small $\ve>0$ holds:
If $\dist_{\Co^1}(T,\tilde T)\leq\ve^{\frac {3}{\rho}}$,
then 
\begin{displaymath}
  \|Q_\ve f\|\leq C\,\tildenorm{f}\quad\text{ and }\quad\tildenorm{Q_\ve f}
  \leq C\,\|f\|\
\end{displaymath}
for any $f\in\Co^1(\To^2,\R)$.
\end{lem}
\begin{proof}
  From Lemmas~\ref{lem:norm-geometry} and \ref{lem:elementary-estimates} 
follows
\[
  \|Q_\ve f\|
  \leq 
  C\ell^{\rho}\|Q_\ve f\|_{2,2}+C\tildenorm{Q_\ve f}
  \leq
  C\ell^{\rho}\ve^{-3}\|f\|+C\tildenorm{Q_\ve f}\ ,
\]
and the first inequality follows by Lemma
\ref{lem:properties}. The second inequality is proven in the same way
since the r\^oles of $T$ and $\tilde T$ are perfectly symmetric.
\end{proof}

\begin{proof}[\bf Proof of Proposition \ref{prop:determ-perturb}]
We fix $r>\sigma$ and $\delta>0$ and consider only $z$ with $r<|z|\leq2$ 
and $\dist(z,\spectr(\Lp))>\delta$.\footnote{Since 
$\rho_{\spectr}(\tilde\Lp)\leq 1$,
estimate (\ref{eq:ihes3}) is trivially verified for $|z|>2$.} 
Several constants introduced in the sequel will
depend on $r$, $\delta$ and $R(z)$ 
in various ways, but we will not precisely keep track of them in this 
section. So ``$\const$'' denotes an arbitrary constant 
that depends in particular on $z$, but not on $\ve$.

Recall from Remark~\ref{rem:double-perturbation} 
that Proposition~\ref{prop:resolvent-estimates}
applies also to $Q_\ve\Lp Q_\ve$ and hence gives estimates on
$R_\ve(z)=(z-Q_\ve\Lp Q_\ve)^{-1}$. 
In particular,
$\|R_\ve(z)\|\leq\const$. Therefore
\begin{displaymath}
\begin{split}
  \|R_\ve(z)Q_\ve(\Lp_T-\Lp_{{\tilde T}})Q_\ve f\|
  &\leq
  \const\,\|Q_\ve(\Lp_T-\Lp_{{\tilde T}})Q_\ve f\|\\
  &\leq
  \const\,\ve^{-1}\,\|(\Lp_T-\Lp_{{\tilde T}})Q_\ve f\|_1\\
  &\leq
  \const\,\ve^{-1}\,\dist_{\Co^1}(T,{\tilde T})\,\|Q_\ve f\|_{1,1}\\
  &\leq
  \const\,\ve^{-3}\,\dist_{\Co^1}(T,{\tilde T})\,\|f\|
\end{split}
\end{displaymath}
so that 
\begin{displaymath}
  \left\|(1+R_\ve(z)Q_\ve(\Lp_T-\Lp_{{\tilde T}})Q_\ve)^{-1}\right\|\leq 2
\end{displaymath}
provided that $\dist_{\Co^1}(T,{\tilde T})\leq\frac 12\,\const\,\ve^{3}$.
Since 
$\tilde R_\ve(z)=
\left[1+R_\ve(z)Q_\ve(\Lp_T-\Lp_{{\tilde T}})Q_\ve\right]^{-1}R_\ve(z)$, it
follows that 
\begin{equation}\label{eq:ihes2-}
  \|\tilde R_\ve(z)\|\leq 2\|R_\ve(z)\|\ .
\end{equation}
Observe next that
\begin{displaymath}
  \tilde R_\ve(z)
  =
  z^{-1}+z^{-2}Q_\ve\Lp_{\tilde T}Q_\ve
  +z^{-2}Q_\ve\Lp_{\tilde T}Q_\ve\tilde R_\ve(z)Q_\ve\Lp_{\tilde T}Q_\ve\ .
\end{displaymath}
Then it follows from (\ref{eq:ihes2-}) and from 
Lemma~\ref{lem:norm-comparison} that
\begin{equation}\label{eq:ihes2}
  \tildenorm{\tilde R_\ve(z)}
  \leq
  \const\,\left(1+\|R_\ve(z)\|\right)\ .
\end{equation}

The next step is to apply Proposition~\ref{prop:resolvent-estimates} 
to $\Lp_{{\tilde T}}$ and $Q_\ve\Lp_{{\tilde T}}Q_\ve$ 
with the r\^oles of these two operators interchanged,
\ie interpreting $\Lp_{{\tilde T}}$ as a perturbation of 
$Q_\ve\Lp_{\tilde T}Q_\ve$. 
This is possible, because of the following two observations:
\smallskip\\
1) The operators $Q_\ve\Lp_{\tilde T}Q_\ve$ satisfy the 
Lasota-Yorke type inequality from Lemma~\ref{lem:propbound}
with uniform constants that are close to those for $\Lp_{T}$ and 
$Q_\ve\Lp_{T}Q_\ve$, \cf Remark~\ref{rem:double-perturbation}.
(The dependence of these constants on the map $T$
is continuous under $\Co^2$ changes of $T$.)
\smallskip\\
2) In view of (\ref{eq:ihes2}) and our convention on the use of
``$\const$'', $\tildenorm{\tilde R_\ve(z)}$ is bounded in terms
of quantities depending only on $T$ and on the kernel $\bar q(x,y)$. 
Therefore Remark~\ref{rem:resolvent-constants} guarantees that
\begin{align}
    \tildenorm{\tilde R(z)}\label{eq:tilde-resolvent}
    &\leq
    \const\qquad\text{and}\\
    |||\tilde R_\ve(z)-\tilde R(z)|||^\sim\label{eq:term3}
    &\leq
    \const\ve^{\eta(\gamma-\beta)}
\end{align}
with constants that do not depend on ${\tilde T}$ and on $\ve$ if $\ve$ and
$\dist_{\Co^1}(T,{\tilde T})$ are small enough.

Estimate (\ref{eq:tilde-resolvent}) is exactly assertion (\ref{eq:ihes3}) 
of Proposition \ref{prop:determ-perturb}.

\relax For assertion (\ref{eq:ihes4}) of this proposition --
as discussed in section~\ref{subsec:deterministic-perturbations} --
we must bound the three terms
\begin{equation}\label{eq:three-terms}
  |||R(z)-R_\ve(z)|||,\quad\cnorm{\tilde R_\ve(z)-R_\ve(z)},\quad
  |||\tilde R_\ve(z)-\tilde R(z)|||^\sim
\end{equation}
where $R(z)$, $\tilde R(z)$, $R_\ve(z)$ and $\tilde R_\ve(z)$ 
denote the resolvents
of $\Lp_T$, $\Lp_{{\tilde T}}$, $Q_\ve\Lp_TQ_\ve$ and 
$Q_\ve\Lp_{{\tilde T}}Q_\ve$, respectively.
Because of (\ref{eq:term3}),
the first and the third term are bounded by $\const\,\ve^{\eta(\gamma-\beta)}$.
It remains to estimate second one.
Observing
the resolvent estimate (\ref{eq:ihes2}), 
the norm comparisons (\ref{eq:four-norms}) and 
Lemma~\ref{lem:elementary-estimates}, we have
\begin{displaymath}
\begin{split}
  \left\|\left(\tilde R_\ve(z)-R_\ve(z)\right)f\right\|_{(\Co^1)*}
  &\leq
  \left\|\tilde R_\ve(z)Q_\ve(\Lp_{{\tilde T}}
       -\Lp_T)Q_\ve R_\ve(z)f\right\|^\sim\\
  &\leq
  \const\cdot\left\|Q_\ve(\Lp_{{\tilde T}}
       -\Lp_T)Q_\ve R_\ve(z)f\right\|^\sim\\
  &\leq
  \const\cdot\ve^{-1}\,\left\|(\Lp_{{\tilde T}}
       -\Lp_T)Q_\ve R_\ve(z)f\right\|_1\\
  &\leq
  \const\cdot\ve^{-1}\cdot\dist_{\Co^1}(T,{\tilde T})\,
       \left\|Q_\ve R_\ve(z)f\right\|_{1,1}\\
  &\leq
  \const\cdot\ve^{-3}\cdot\dist_{\Co^1}(T,{\tilde T})\,
       \left\|R_\ve(z)f\right\|\\
  &\leq
  \const\cdot\ve^{-3}\cdot\dist_{\Co^1}(T,{\tilde T})\,
       \left\|f\right\|_{\Co^1}\ .
\end{split}
\end{displaymath}  
Collecting the estimates for the three terms in (\ref{eq:three-terms})
and setting $\ve=\dist_{\Co^1}(T,{\tilde T})^{\rho/(3+\eta(\gamma-\beta))}$
we conclude that
\begin{displaymath}
  \cnorm{\tilde R(z)-R(z)}\leq \const\cdot
  \dist_{\Co^1}({\tilde T},T)^{\frac
       {\rho\eta(\gamma-\beta)}{3+\eta(\gamma-\beta)}}\ .
\end{displaymath}  
This is (\ref{eq:ihes4}).
\end{proof}

\begin{proof}[\bf Proof of Sub-Lemma \ref{eq:closechart}]
If $v^u(x)$ and $\tilde v^u(x)$ are the unitary vectors fields in the 
unstable directions for the two maps, it is clear that there exists 
$\rho_0>0$ such that\footnote{This follows from the standard construction 
of the unstable distribution by cone field contraction. Given a strictly
invariant continuous cone field $\Co(x)$ for the map $T$, it will be
strictly invariant for $\tilde T$ as well, provided the two maps are
sufficiently close in the $\Co^1$ topology. Accordingly, for each
$x\in\manif$ the cones $\Co_{n}(x):=D_{T^{-n}x}T^{n}\Co(T^{-n}x)$
and the analogously defined $\tilde\Co_n(x)$
shrink at an exponential rate towards the respective unstable directions
$v^u(x)$ and $\tilde v^u(x)$. 
Hence there exist $\nu\in (0,1)$ and $C>0$ such that for each vector 
$v\in\Co(T^{-n}x)$ holds $\|v^u(x) - D_{T^{-n}x}T^{n}v\|\leq C \nu^n$ 
and the same for $\tilde T$. On the other hand
\[
\|D_{T^{-n}x}T^n-D_{{\tilde T}^{-n}x}{\tilde T}^n\|\leq \Lambda^n\ell
\]
for some $\Lambda>1$, from which, by choosing $n$ appropriately, the
statement follows.}
\begin{equation}\label{eq:diststab}
\|v^u(x)-\tilde v^u(x)\|\leq C\ell^{\rho_0}.
\end{equation}
The estimate $\|F-\Id\|_\infty\leq C\ell^{\rho_0}$ 
is an immediate consequence of
(\ref{eq:diststab}) and the definition of $F$. The second is more
subtle and requires some work.

We start by studying the derivative of the map $\Gamma: U\to\R^2$. 
\relax From the definition of $\Gamma$ is follows immediately that
\begin{align*}
\frac{\partial}{\partial\xi}\Gamma^{-1}(\xi,\eta)=
v^s(\Gamma^{-1}(\xi,\eta))J^u\circ \Gamma^{-1}(\xi,\eta)\\ 
\frac{\partial}{\partial \eta}\Gamma^{-1}(\xi,\eta)=
v^u(\Gamma^{-1}(\xi,\eta))J^s\circ\Gamma^{-1}(\xi,\eta),
\end{align*}
where $J^s\circ \Gamma^{-1}(\xi,\eta)$ is the Jacobian of the holonomy
$H^u_\xi$ from the unstable fibre $\Gamma^{-1}\{(0,\cdot)\}$ to the fibre
$\Gamma^{-1}\{(\xi,\cdot)\}$ at the point $(0,\eta)$, and
$J^u\circ\Gamma^{-1}(\xi,\eta)$ is the Jacobian of the holonomy
$H^s_\eta$ from the  unstable fibre $\Gamma^{-1}\{(\cdot,0)\}$ to the fibre
$\Gamma^{-1}\{(\cdot,\eta)\}$ at the point $(\xi,0)$. While $v^s, v^u$
are the stable and unstable unit vector fields respectively.

The above formulae imply that
\begin{equation}
\label{eq:G-derivative}
D\Gamma^{-1}=A\begin{pmatrix}
J^s\circ \Gamma^{-1}&0\\
0&J^u\circ\Gamma^{-1}
\end{pmatrix}
\;;\quad A:=(v^u,\, v^s).
\end{equation}
As the same formulae apply to the change of coordinates $\tilde
\Gamma$ associated to the map $\tilde T$, we have
\begin{equation}
\label{eq:F-derivative}
DF=A\begin{pmatrix}
\frac{J^s\circ \Gamma^{-1}}{\tilde J^s\circ{\tilde\Gamma}^{-1}}&0\\
0&\frac{J^u\circ\Gamma^{-1}}{\tilde J^u\circ{\tilde\Gamma}^{-1}}
\end{pmatrix}\tilde A^{-1},
\end{equation}
where the tilde designate the quantities relative to $\tilde\Gamma$.
It is immediate to see that $A,\tilde A$ are $\Co^{1+\alpha}$ matrix
valued functions. Moreover, the fact that the distributions are
uniformly transversal, equation (\ref{eq:diststab}) and the fact  
that $\|F-\Id\|_\infty\leq C\ell^{\rho_0}$ show that there 
exists a constant $C>0$ such that 
\[
\begin{array}l
\|A\|_\infty+\|\tilde A\|_\infty+\|A^{-1}\|_\infty
+\|\tilde A^{-1}\|_\infty\leq C\\
\|A\tilde A^{-1}-\Eins\|_\infty\leq C\ell^{\rho_0}.
\end{array}
\]
Hence, to see that $DF$ is close to unity we need only to control the
ratio between the Jacobians of the corresponding holonomies. Since 
the stable and the unstable one are treated exactly in the same way 
we will consider only the unstable ones. We want to prove that there 
exist $C,\rho>0$ such that
\begin{equation}
\label{eq:hol-obj}
\left\|\frac{J^u\circ\Gamma^{-1}}{\tilde
J^u\circ{\tilde\Gamma}^{-1}}-1\right\|_\infty\leq C\ell^{\rho}.
\end{equation}
The above fact is a ready consequence of the well known formula for
the Jacobian of holonomies \cite{Ma}:
\[
J^u(x)=\prod_{n=0}^\infty\frac{D_{y_n}T|_{E^s(y_n)}}{D_{x_n}T|_{E^s(x_n)}},
\]
where $x_0:=x=\Gamma^{-1}(\xi,\eta)$, $y_0:=H^u_\xi(x)$ and
$x_n:=T^{-n}x$, $y_n:=T^{-n}y$. 
Since $x_0$ and $y_0$ are on the same unstable fibre,
$|x_n-y_n|\leq\const\cdot \lambda_u^{-n}|x_0-y_0|$, so that
there is some constant $C>0$ such that, for each $N\in\N$,
\[
\left|
J^u(x)\prod_{n=0}^N\frac{D_{x_n}T|_{E^s(x_n)}}{D_{y_n}T|_{E^s(y_n)}}
-1\right|\leq C\,\lambda_u^{-N}\ .
\]
Accordingly, letting $\tilde x:=F^{-1}(x)$,
\begin{displaymath}
\begin{split}
  \left|\frac{J^u(x)}{\tilde J^u(\tilde x)}-1\right| 
  &=
  \left|\prod_{n=0}^\infty\frac
  {D_{y_n}T|_{E^s(y_n)}\cdot D_{\tilde x_n}\tilde T|_{\tilde E^s(\tilde x_n)}}
  {D_{x_n}T|_{E^s(x_n)}\cdot D_{\tilde y_n}\tilde T|_{\tilde E^s(\tilde y_n)}}
  -1\right|\\
  &\leq 
  C\,\lambda_u^{-N}+\left|\prod_{n=0}^N\frac
  {D_{y_n}T|_{E^s(y_n)}\cdot D_{\tilde x_n}\tilde T|_{\tilde E^s(\tilde x_n)}}
  {D_{x_n}T|_{E^s(x_n)}\cdot D_{\tilde y_n}\tilde T|_{\tilde E^s(\tilde y_n)}}
  -1\right|
\end{split}
\end{displaymath}
Denote by $\Lambda$ the maximum of the derivatives of the maps in
the neighbourhood under consideration. Since we know already that 
$|x-\tilde x|=|F(\tilde x)-\tilde x|\leq C\ell^{\rho_0}$, it follows that
$|x_n-\tilde x_n|\leq \ell^{\rho_0}C\Lambda^n$,
and the same estimate holds for $|y_n-\tilde y_n|$. Remembering
(\ref{eq:diststab}) and the fact that the foliations are
$\Co^{1+\alpha}$ this yields immediately
\[
\left|\frac{J^u(x)}{\tilde J^u(\tilde x)}-1\right| \leq
C\,\lambda_u^{-N}+C\ell^{\rho_0}\Lambda^N\|T\|_{\Co^2}\ .
\]
The result follows then by choosing
$N=\frac{\rho_0|\ln\ell|}{\ln(\lambda_u\Lambda)}$,
which yields
$\rho=\rho_0\frac{\ln \lambda_u}{\ln(\lambda_u\Lambda)}$. 
\end{proof}

\subsection{Proofs: Spectral stability -- Ulam finite rank approximation}
\label{subsec:ulam-proof}

\begin{proof}[\bf Proof of Proposition~\ref{prop:ulam-resolvent}]
Our starting point is the formal identity
\begin{displaymath}
  (z-\Lp_{\Ac,\ve})^{-1}
  =
  {\underbrace{\left(1+(z-Q_\ve\Lp)^{-1}(1-\Pi_\Ac)Q_\ve\Lp\right)}_
  {=:{\Cal N}}}^{-1} (z-Q_\ve\Lp)^{-1}\ .
\end{displaymath}
We show that ${\Cal N}$ is indeed invertible:
For $f\in\Co^1(\To^2,\R)$ we have
\begin{displaymath}
\begin{split}
  &\hspace*{-1cm}\left\|(z-Q_\ve\Lp)^{-1}(1-\Pi_\Ac)Q_\ve\Lp f\right\|_1\\
  \leq&
  \left\|(z^{-1}+z^{-2}Q_\ve\Lp)(1-\Pi_\Ac)Q_\ve\Lp f\right\|_1
  +
  \left\|z^{-2}Q_\ve\Lp(z-Q_\ve\Lp)^{-1}Q_\ve\Lp
    (1-\Pi_\Ac)Q_\ve\Lp f\right\|_1\\
  \leq&
  (r^{-1}+r^{-2})\,\frac 2N\,\ve^{-1}\,\|f\|_1
  +
  \const\,\ve^{-1}
  \left\|(z-Q_\ve\Lp)^{-1}Q_\ve\Lp(1-\Pi_\Ac)Q_\ve\Lp f\right\|
\end{split}
\end{displaymath}
where we used (\ref{eq:ihes6}) and Lemma~\ref{lem:bvaverage}.
Using Proposition~\ref{prop:resolvent-estimates}
the second summand is further estimated by
\begin{displaymath}
\begin{split}
  \left\|(z-Q_\ve\Lp)^{-1}Q_\ve\Lp(1-\Pi_\Ac)Q_\ve\Lp f\right\|
  &\leq
  (a+b)\,\left\|Q_\ve\Lp(1-\Pi_\Ac)Q_\ve\Lp f\right\|\\
  &\leq
  \const\,\ve^{-1}\,N^{-1}\,\|Q_\ve\Lp f\|_\BV\\
  &\leq
  \const\,N^{-1}\,\ve^{-2}\,\|f\|_1
\end{split}
\end{displaymath}
where we also used (\ref{eq:ihes5}) and 
Lemmas~\ref{lem:buzzikeller} and \ref{lem:bvaverage}.
Combining both estimates we see that
\begin{displaymath}
  \left\|(z-Q_\ve\Lp)^{-1}(1-\Pi_\Ac)Q_\ve\Lp\right\|_1  
  \leq
  \const\,N^{-1}\,\ve^{-3}\ .
\end{displaymath}
This proves that ${\Cal N}$ is invertible with $\|{\Cal N}^{-1}\|_1\leq2$
if $N^{-1}\ve^{-3}$ is sufficiently small. Hence
\begin{displaymath}
\begin{split}
  \left\|(z-\Lp_{\Ac,\ve})^{-1}f\right\|_1
  &\leq
   2\,\left\|(z^{-1}+z^{-2}Q_\ve\Lp
   +z^{-2}Q_\ve\Lp(z-Q_\ve\Lp)^{-1}Q_\ve\Lp)f\right\|_1\\
  &\leq
  \const\,\left(\|f\|_1
   +\ve^{-1}\left\|(z-Q_\ve\Lp)^{-1}Q_\ve\Lp f\right\|\right)\\
  &\leq
  \const\,\left(\|f\|_1+\ve^{-1}\left\|Q_\ve\Lp f\right\|\right)
\end{split}
\end{displaymath}
where we used (\ref{eq:ihes6}). This proves (\ref{eq:ulam-resolvent-bound}).

We turn to the proof of (\ref{eq:ulam-resolvent-difference}).
\begin{displaymath}
\begin{split}
  &
  \left\|\left((z-Q_\ve\Lp)^{-1}-(z-\Lp_{\Ac,\ve})^{-1}\right)f\right\|_w\\
  =&
  \left\|(z-Q_\ve\Lp)^{-1}(\Pi_\Ac-1)Q_\ve\Lp(z-\Lp_{\Ac,\ve})^{-1}f
    \right\|_w\\
  \leq&
  \left\|z^{-1}(\Pi_\Ac-1)Q_\ve\Lp(z-\Lp_{\Ac,\ve})^{-1}f\right\|_1
  +
  \left\|z^{-1}(z-Q_\ve\Lp)^{-1}Q_\ve\Lp(\Pi_\Ac-1)Q_\ve\Lp
    (z-\Lp_{\Ac,\ve})^{-1}f \right\|\\
  \leq&
  \const\,N^{-1}\,\left\|Q_\ve\Lp(z-\Lp_{\Ac,\ve})^{-1}f\right\|_\BV
  +
  \const\,N^{-1}\,\ve^{-1}\,\left\|Q_\ve\Lp(z-\Lp_{\Ac,\ve})^{-1}f \right\|_\BV
  \\
  \leq&
  \const\,N^{-1}\,\ve^{-1}\,\ve^{-1}\,\left\|(z-\Lp_{\Ac,\ve})^{-1}f\right\|_1
  \\
  \leq&
  \const\,N^{-1}\,\ve^{-2}\,\left(\|f\|_1+\ve^{-1}
    \left\|Q_\ve\Lp f\right\|\right)\\
  \leq&
  \const\,N^{-1}\,\ve^{-2}\,\left(\|f\|_1+\ve^{-1}
    \left\|f\right\|\right)\\
  \leq&
  \const\,N^{-1}\,\ve^{-3}\,\|f\|_{1,1}\ .
\end{split}
\end{displaymath}
This is even a bit stronger than (\ref{eq:ulam-resolvent-difference}).
\end{proof}

\subsection{Proofs: Compact embedding}
\label{subsec:compactness-proof}

In this section we prove Proposition \ref{prop:compactnes}.  We start
with a simple remark on products of functions.
\begin{lem}
\label{lem:multiplication} For each $\phi\in \Co^{2}(\manif,\R)$
there exists $C_\phi$ such that, for each $f\in \Co^1(\manif,\R)$,
holds
\begin{displaymath}
\|f\phi\|_w\leq C_\phi \|f\|_w\ ,\quad
\|f\phi\|_s\leq C_\phi \|f\|_s\ ,\quad
\|f\phi\|_u\leq C_\phi \|f\|\ .
\end{displaymath}
\end{lem}
\begin{proof}
A direct computation yields, for each $\vf\in\D_{\tilde\beta}$, 
$\tilde\beta\in(0,1]$, $v\in\V_{\beta}$,
\[
  \frac{\phi\vf}{|\phi|_{\Co^{1}}}\in\D_{\tilde\beta};\quad
  \frac{\phi v}{|\phi|_{\Co^{1}}}\in\V_{\beta};\quad
  \frac{d\phi(v)}{|\phi|_{\Co^{2}}}\in\D_\beta.
\]
\relax From this the lemma follows immediately with
$C_\phi=2|\phi|_{\Co^{2}}$.
\end{proof}

The above Lemma allows us to use a smooth partition of unity
$\{\phi_i\}$ and write $f=\sum_i f\phi_i$. The advantage consists in
the possibility of working locally and using explicit coordinates
which makes the argument more transparent. 

\begin{proof}[\bf Proof of Proposition \ref{prop:compactnes}]
Let us set $\tau':=\min\{\tau,1\}$.
The idea of the proof is to represent the embedding 
$\B\to\B_w$ as a composition of compact and continuous linear maps:
\begin{displaymath}
\begin{array}c
  \B\\ \\
  f
\end{array}
\longrightarrow
\begin{array}c
  (\Co^\beta)^*\times(\Co^\beta)^*\\ \\
  (f,df)
\end{array}
\overset{\mbox{\scriptsize compact}}\hookrightarrow
\begin{array}c
  (\Co^{\gamma{\tau'}})^*\times(\Co^{\gamma{\tau'}})^*\\ \\
  (f,df)
\end{array}
\longrightarrow
\begin{array}c
  \B_w\\ \\
  f
\end{array}
\end{displaymath}
The first embedding is continuous, because
$\|f\|_{\Co^\beta}^*\leq \|f\|_s$, $\|df\|_{\Co^\beta}^*\leq\|f\|_u$.
The compactness of the second embedding is equivalent to the compact
embedding of $\Co^{\gamma{\tau'}}(\manif,\R)$ into 
$\Co^\beta(\manif,\R)$, and this is well known because
$\gamma{\tau'}>\beta$.
The continuity of the last embedding is an immediate consequence of 
the next lemma.
\end{proof}

\begin{lem}
\label{lem:objective} There exists $K>0$ such that, for each
$\vf\in\D_\gamma$, there are
$\phi^\vf\in \Co^{\gamma{\tau'}}(\manif,\R)$ and  
$V^{\vf}\in\Co^{\gamma{\tau'}}(\manif,\R)\cap \V_\beta$ with
\[
\|\phi^\vf\|_{\Co^{\gamma{\tau'}}}+\|V^\vf\|_{\Co^{\gamma{\tau'}}}\leq
K
\]
and such that for each $f\in\Co^{1}(\manif,\R)$
\begin{displaymath}
  \int_{\manif}f\vf\,dm
  =
  \int_\manif f\phi^\vf\,dm
  +
  \int_{\manif} df(V^{\vf})\,dm\ .
\end{displaymath}
\end{lem}

The rest of this section is devoted to the proof of this lemma. We need 
some more preparations. By Lemma \ref{lem:multiplication} it follows that 
we can restrict ourselves to functions supported in some small open set 
$U$. In particular we require that $U$ belongs to only one chart 
$\Phi : U\to\R^d$.  Using the Cartesian structure of the chart
we can identify all the tangent spaces in $U$ with the tangent space
of a chosen point $\bar z\in U$. We can do a linear change of coordinates 
in $\Phi(U)$ such that in the new coordinates 
$E^u(\bar z)=\{(x,0)\in\R^d\,:\;x\in\R^{d_u}\}$
and $E^s(\bar z)=\{(0,y)\in\R^d\,:\;y\in\R^{s_s}\}$. In addition, 
without loss of generality we can assume that $\Phi(\bar z)=(0,0)$ and 
$\Phi(U)$ is the product of two balls of radius $4r$ centred at zero.

\relax From now on we will use the Cartesian coordinates $(x,y)$
without further comments and we will identify $U$ with $\Phi(U)$. 
Note that, since $\Phi$ is smooth, the Riemannian distance is uniformly 
equivalent to the Cartesian distance in the chart.

Let us specify more precisely how small the set $U$ needs to be.
The first smallness assumption on $U$
results from the requirement that all unstable manifolds in $U$ can
be viewed as a graph of a function of $x$. We can then introduce the
maps $H:\R^{2d_u+d_s}\to\R^{d_s}$ defined by the following requirements
\begin{itemize}
\item $H(x,x,y)=y$;
\item for each $(x,y)\in B_{3r}(0)\times B_{3r}(0)$, 
      $\{(\xi, H(x,\xi,y))\,:\; \xi\in B_{3r}(0)\}$ is the local unstable 
      manifold containing the point $(x,y)$.
\end{itemize}
The point $(\xi,H(x,\xi,y))$ is nothing else than the intersection of the
unstable manifold of the point $(x,y)$ with the subspace
$\{(w,z)\in\R^{d_u+d_s}:\; w=\xi\}$. This implies the obvious
formula
\begin{equation}
\label{eq:inverseH}
H(\zeta,\xi,H(x,\zeta,y))=H(x,\xi,y).
\end{equation}

Depending on which of its three arguments we fix, 
$H$ can serve various purposes and have different smoothness properties, 
which can all be inferred from the information on stable and unstable 
foliation collected in the appendix.
\begin{equation}\label{eq:holprop0}
  \parbox{\myparwidth}
  {For each $(x,y)\in B_{3r}(0)\times B_{3r}(0)$ 
  the function $\xi\mapsto H(x,\xi, y)$ is
  of class $\CoT$, since it is just a straightforward parametrisation 
  of the unstable manifold through $(x,y)$.}
\end{equation}
This leads to a second smallness requirement, namely
$\sup_{(x,y)}\|H(x,\cdot,y)\|_{\Co^{1}(B_{3r})}\leq \frac16$.\footnote{This 
can be achieved since $|H(0,\xi,0)|\leq C|\xi|^2$,
by the $\Co^2$ smoothness of the unstable manifolds, and by the
${\tau'}$-H\"older continuity of the foliation, see the appendix.}
Accordingly, $\{H(x,\xi,y):\;\xi\in B_{3r}(0)\}\subset B_{4r}(0)$.
Moreover,
\begin{align}
&\parbox{\myparwidth}
{For fixed $(x,\xi)$ the map $y\mapsto H(x,\xi,y)$ is the holonomy
along unstable fibres from the subspace 
$\{(w,z)\in\R^{d_u+d_s}\,:\; w=x\}$ to the subspace
$\{(w,z)\in\R^{d_u+d_s}\,:\; w=\xi\}$.
Denote by $J_3H(x,\xi,y)$ the Jacobian of this holonomy.}
\label{eq:holprop1}\\
&\parbox{\myparwidth}
{$J_3H(x,\xi,y)=J_3H(x,\zeta,y)\,J_3H(\zeta,\xi,H(x,\zeta,y))$ 
by equation (\ref{eq:inverseH}).}
\label{eq:holprop2}\\
&\parbox{\myparwidth}
{The maps $y\mapsto H(x,\xi,y)$ and $y\mapsto J_3H(x,\xi,y)$ are
${\tau'}$-H\"older (see (\ref{eq:fol-bunching}) and \cite{Ma}).}
\label{eq:holprop3}\\
&\parbox{\myparwidth}
{There is $C>1$ such that 
$C\geq J_3H(x,\xi,y)\geq C^{-1}$ for each $x,\xi,y$ (see \cite{Ma}).}
\label{eq:holprop4}\\
&\parbox{\myparwidth}
{There is $C>0$ such that 
$\sup_y|1-J_3H(x,\xi,y)|\leq C\,|x-\xi|$ for all 
$x,\xi$ (Lemma \ref{lem:regularity}). }
\label{eq:holprop5}
\end{align}

The instrumental change of variables 
$\Psi:B_{3r}(0)\times B_{3r}(0)\subset \R^d\to U$ is defined by
\begin{equation}\label{eq:Psineu}
  \Psi(x,y):=(x,H(0,x,y))\ . 
\end{equation}
Note that equation (\ref{eq:inverseH})
implies that 
\begin{equation}\label{eq:Psineu-inv}
  \Psi^{-1}(x,y)=(x,H(x,0,y))\ . 
\end{equation}
Figure \ref{fig:boxes} shows how $\Psi$ is related to neighbourhoods 
of $(0,0)$ of various sizes.
\begin{figure}[ht]\ 
\put(-60,0){\line(1,0){30}} 
\put(-60,0){\line(0,1){30}}
\put(-30,30){\line(0,-1){30}}
\put(-30,30){\line(-1,0){30}} 
\multiput(20,-5)(2,0){20}{\line(1,0){1}} 
\multiput(20,-5)(0,2){20}{\line(0,1){1}}
\multiput(60,35)(-2,0){20}{\line(-1,0){1}}
\multiput(60,35)(0,-2){20}{\line(0,-1){1}} 
\multiput(25,0)(2,0){15}{\line(1,0){1}} 
\multiput(25,0)(0,2){15}{\line(0,1){1}}
\multiput(55,30)(-2,0){15}{\line(-1,0){1}}
\multiput(55,30)(0,-2){15}{\line(0,-1){1}} 
\multiput(30,5)(2,0){10}{\line(1,0){1}} 
\multiput(30,5)(0,2){10}{\line(0,1){1}}
\multiput(50,25)(-2,0){10}{\line(-1,0){1}}
\multiput(50,25)(0,-2){10}{\line(0,-1){1}} 
\multiput(35,10)(2,0){5}{\line(1,0){1}} 
\multiput(35,10)(0,2){5}{\line(0,1){1}}
\multiput(45,20)(-2,0){5}{\line(-1,0){1}}
\multiput(45,20)(0,-2){5}{\line(0,-1){1}} 
\qbezier(25,33)(40,35)(55,28) 
\qbezier(25,2)(40,1)(55,-4)
\put(25,2){\line(0,1){31}}
\put(55,-4){\line(0,1){32}}
\put(-20,15){\vector(1,0){30}}
\put(-8.5,18){$\Psi$}
\put(-53,-5){$B_{3r}\times B_{3r}$}
\put(33,-10){$B_{4r}\times B_{4r}$}
\put(51,39.5){\vector(-2,-1){12}}
\put(43,40){$\scriptstyle \Psi(B_{3r}\times B_{3r})$}
\put(61,15){$\scriptstyle 4r$}
\put(56,15){$\scriptstyle 3r$}
\put(51,15){$\scriptstyle 2r$}
\put(46,15){$\scriptstyle r$}
\put(0,-15){\ } 
\put(39,15){\line(1,0){2}}
\put(40,14){\line(0,1){2}}
\put(-46,15){\line(1,0){2}}
\put(-45,14){\line(0,1){2}}
\caption{Neighbourhoods related to the $\Psi$ coordinate change}
\label{fig:boxes}
\end{figure}
Other basic properties are described in the next Lemma.

\begin{lem}
\label{lem:changeova}
The change of coordinates $\Psi$ has the following properties:
\begin{enumerate}[a)]
\item In the new coordinates the unstable foliation consists of parallel
subspaces.
\item \label{item:changeova-b}
$\Psi$ is absolutely continuous and its Jacobian
$J\Psi(x,y)=J_3H(0,x,y)$.
\item $\sup_y|J\Psi(x,y)-J\Psi(\xi,y)|\leq C\,|x-\xi|$.
\item For each $(x,\xi,y)\in B_{3r}(0)\times B_{3r}(0)\times B_{3r}(0)$,
      $E^u(\xi,H(x,\xi,y))=\{(w,D_2H(x,\xi,y)w)\,:\;w\in\R^{d_u}\}$.\footnote{
      Here, and in the following, by $D_2$ we mean the derivative with 
      respect to the second variable.} 
\end{enumerate}
\end{lem}

\begin{proof}
a) is obvious, b) follows from Fubini's theorem, because $\Psi$ is a skew 
product over the identity, 
c) is a consequence of (\ref{eq:holprop2}) and (\ref{eq:holprop5}), and 
d) is again obvious, since, as already remarked,
$\xi\mapsto(\xi,H(x,\xi,y))$ parametrises an unstable manifold.
\end{proof}

Next we need an auxiliary computation.
\begin{lem}      
\label{lem:representation}
Let $f\in \Co^{0}(\R^{d_u+d_s})$ 
with $\supp(f)\subset B_r(0)$ be such that
$\xi\mapsto f(\xi,v)$ is of class $\Co^{1}$
for all $v$. Then, for all $(u,v)\in B_r(0)$,
\[
\frac 1{Z_{2r}}\int_{B_{r}(0)}f(\xi,v) d\xi
=
f(u,v)
+\sum_{i=1}^{d_u}\frac 1{Z_{2r}}\int_{B_{r}(0)}D_i
f(\xi,v)\Theta_i(\xi,u) d\xi,
\]
where $Z_{2r}$ denotes the $d_u$-dimensional volume of $B_{2r}(0)$ and
\[
\Theta_i(\xi,u):=\frac{(2r)^{d_u}-\|\xi-u\|^{d_u}}
{d_u\|\xi-u\|^{d_u}} (\xi_i-u_i).
\]
\end{lem}

\begin{proof}
The result follows by using polar coordinates around the point
$(u,v)$ and observing that $f=0$ outside the ball $B_r(0)$:
\begin{align*}
  \int_{B_{r}(0)}f(\xi,v)\,d\xi
  &=
  \int_{S^{d_u}}\sigma(d\theta)\, 
  \int_0^{2r}d\rho\, \rho^{d_u-1}f(\rho \theta+u,v)\\
  &=
  \int_{S^{d_u}}\sigma(d\theta)\, 
  \int_0^{2r}d\rho\, \rho^{d_u-1}
    \left(f(u,v)+\int_0^\rho Df_{|t\theta+u,v}\cdot\theta\,dt\right)\\
  &=
  Z_{2r}\,f(u,v)
  +\sum_{i=1}^{d_u}\int_{S^{d_u}}\sigma(d\theta)\,
  \int_0^{2r}dt\,t^{d_u-1}
  D_i f(t\theta+u,v)\theta_i
  \frac{(2r)^{d_u}-t^{d_u}}{d_u\, t^{d_u-1}}\\
  &=
  Z_{2r}\,f(u,v)
  +\sum_{i=1}^{d_u}\int_{B_r(0)}
  D_i f(\xi,v)(\xi_i-u_i)
  \frac{(2r)^{d_u}-\|\xi-u\|^{d_u}}{d_u\,\|\xi-u\|^{d_u}}
\end{align*}
where $\sigma(d\theta)$ is the measure on the $d_u-1$ dimensional
spherical surface $S^{d_u}$.
\end{proof}  

Lemma \ref{lem:representation} allows to write the integral of $f$
against a test function in a way that makes the proof of Lemma
\ref{lem:objective} particularly easy.

\begin{proof}[\bf Proof of Lemma \ref{lem:objective}]
Let $\{\psi_k\}$ be a smooth partition of unity such that each
$V_k:=\supp\psi_k$ is contained in an open set $U_k$ 
where $\{U_k\}$ satisfy the smallness
requirements previously discussed for the generic local
neighbourhood $U$. In addition $U_k$ and $V_k$ are chosen such that, 
if $\Phi_k$ denotes the chart associated to $U_k$ so that
$\Phi_kU_k= B_{4r}(0)\times B_{4r}(0)$, 
then $\Phi_k(V_k)\subset B_r(0)\times B_r(0)$.

Let us start by considering any $(U, V)\in\{(U_k, V_k)\}$ and consider
$f\in\Co^1$, $\supp f\subset V$.

Applying the coordinate change from Lemma~\ref{lem:changeova}
to the function $f\circ\Psi$ and using
the formula from Lemma~\ref{lem:representation} we have
\begin{align*}
  \int_V&f\vf\,dm\\
  =&
  \int_{B_{r}(0)}\int_{B_{r}(0)}du\,dv\; 
  f(\Psi(u,v))\,\vf(\Psi(u,v))\,J\Psi(u,v)\\
  =&
  \int_{B_{r}(0)}\int_{B_{r}(0)}dv\,d\xi\; f(\Psi(\xi,v))
  \frac 1{Z_{2r}}\int_{B_{r}(0)} du\;\vf(\Psi(u,v))J\Psi(u,v)\\
  &\hspace*{-10pt}
  -\sum_{i=1}^{d_u}\int_{B_{r}(0)}\int_{B_{r}(0)}
  \frac 1{Z_{2r}}\int_{B_{r}(0)} du\,dv\,d\xi\;
  d_{\Psi(\xi,v)}f(({\bf e}_i,D_2 H(0,\xi,v){\bf e}_i))
     \Theta_i(\xi,u)\vf(\Psi(u,v))J\Psi(u,v),
\end{align*}
where ${\bf e}_i$ $(i=1,\dots,d_u)$ are the canonical unit vectors in 
$\R^{d_u}$.
This suggests to define
\begin{align*}
\tilde\phi^\vf\circ\Psi(\xi,v)
&:=
J\Psi(\xi,v)^{-1}\frac 1{Z_{2r}}\int_{B_{r}(0)}du\; 
\vf(\Psi(u,v))J\Psi(u,v);\\
\tilde V_i^{\vf}\circ\Psi(\xi,v)
&:=
J\Psi(\xi,v)^{-1}\frac 1{Z_{2r}}\int_{B_{r}(0)}du\;\Theta_i(\xi,u)
\vf(\Psi(u,v))J\Psi(u,v)({\bf e}_i,D_2 H(0,\xi,v){\bf e}_i)\,.
\end{align*}
Notice that equation (\ref{eq:Psineu-inv}) and (\ref{eq:holprop3}) imply
\begin{displaymath}
  J\Psi(\Psi^{-1}(x,y))=J_3H(0,x,H(x,0,y))=(J_3H(x,0,y))^{-1}
\end{displaymath}
because $H(x,x,y)=y$, and
\[
J\Psi(u,H(x,0,y))=J_3H(0,u,H(x,0,y))=\frac{J_3H(x,u,y)}{J_3H(x,0,y)}\ .
\]
Hence, for each $(x,y)\in B_r(0)\times B_r(0)$,
\begin{equation}
\label{eq:newfunctions}
\begin{split}
  \tilde\phi^\vf(x,y)
  &=
  \frac 1{Z_{2r}}\int_{B_{r}(0)}du\;\vf(u,H(x,u,y))J_3H(x,u,y)\\ 
  \tilde V_i^{\vf}(x,y)
  &=
  \frac 1{Z_{2r}}\int_{B_{r}(0)}du\;\Theta_i(x,u)
  \vf(u,H(x,u,y))J_3H(x,u,y)({\bf e}_i,D_2 H(x,x,y){\bf e}_i)
\end{split}
\end{equation}
since $D_2H(0,x,H(x,0,y))=D_2H(x,x,y)$ \footnote{Just differentiate
the particular instance of (\ref{eq:inverseH}),
$H(0,\xi,H(x,0,y))=H(x,\xi,y)$, with respect to $\xi$.}
and 
\begin{equation}
\label{eq:ihes20}
  \int_V f\vf\,dm
  =
  \int_V f\tilde\phi^\vf\,dm
  -\sum_i\int_V df(\tilde V_i^\vf)\,dm\ .
\end{equation}
Let $\tilde V^\vf:=\sum_i\tilde V_i^\vf$.
\begin{lem}
\label{lem:localtest}
There exists $\tilde K>0$ such that, for each $\vf\in\D_\gamma$,
$\tilde\phi^\vf\in\Co^{\gamma{\tau'}}(V,\R)$, 
$\tilde V^\vf\in\Co^{\gamma{\tau'}}(V,\R^d)$
and
\[
\|\tilde\phi^\vf\|_{\Co^{\gamma{\tau'}}}+\|\tilde
V^\vf\|_{\Co^{\gamma{\tau'}}} \leq \tilde K.
\]
\end{lem}

Lemma \ref{lem:localtest} is clearly the local version of what we aim at; 
we postpone its proof to the end of the section and use it
beforehand to conclude the argument.

To go to the global level the first step is to define globally
$\tilde\phi^\vf$ and $\tilde V^\vf$. Let $\chi\in\Co^2(\manif,\R)$
be such that $\chi\equiv 1$ on $V$ and $\chi\equiv 0$ outside
$B_{2r}(0)\times B_{2r}\subset\Psi\left(B_{3r}(0)\times B_{3r}(0)\right)
\subset U$. Then $\hat \phi^\vf:=\chi\tilde\phi^\vf$ and $\hat
V^\vf:=\chi\tilde V^\vf$ are well defined on all of
$\manif$, $\hat K^{-1}\hat\phi^\vf\in\D_\beta$, $\hat K^{-1}\hat
V^\vf\in\V_\beta$ for suitable $\hat K>0$, and since $f$ is supported 
in $V$ it follows from (\ref{eq:ihes20}) that
\begin{equation}
\label{eq:almostthere}
  \int_\manif f\vf\,dm
  =
  \int_\manif f \hat\phi^\vf\,dm
  -\int_\manif df(\hat V^\vf)\,dm\ .
\end{equation}

We can now conclude the argument by going back to our partition of
unity. Given $f\in\Co^1(\manif,\R)$ and $\vf \in\D_\beta$ for each
$\psi_k$ we have $\hat\phi^\vf_k, \hat V_k^\vf$ such that
\begin{align*}
\int_\manif f\vf&=\sum_k\int_\manif (f\psi_k)\vf=\sum_k\int_\manif f
\psi_k\hat\phi^\vf-\int_\manif d(f\psi_k)(\hat V^\vf)\\
&=\sum_k\int_\manif
f(\psi_k\hat\phi^\vf+d\psi_k(\hat V_k^\vf))-\int_\manif df(\psi_k\hat
V^\vf_k)\\ 
&=\int_\manif f\phi^\vf-\int_\manif df(V^\vf),
\end{align*}
where
\begin{displaymath}
\phi^\vf:=\sum_k\psi_k\hat\phi^\vf+d\psi_k(\hat V_k^\vf)
\quad\text{and}\quad
V^\vf:=\sum_k\psi_k\hat V^\vf_k\ .
\end{displaymath}
\end{proof}

\begin{proof}[\bf Proof of Lemma \ref{lem:localtest}]
\relax From formula (\ref{eq:newfunctions}) it is clear that the supremum
norms of the above test functions are bounded, more care is needed to
control the H\"older norms.
\relax First of all, as stated in (\ref{eq:holprop3}), 
$H(x,\xi,y)$ and $J(x,\xi,y)$ are
${\tau'}$-H\"older in $y$ and the same for
$D_\xi H(x,\xi,y)$.\footnote{This last statement is nothing else than
the ${\tau'}$-H\"older continuity of the unstable foliation.}

We discuss $\tilde\phi^\vf$ first. The idea is to see how the function 
varies when moving along the unstable manifold and when varying $y$. Since
\begin{align*}
\tilde\phi^\vf(x',H(x,x',y))
&=
\frac 1{Z_{2r}}\int_{B_{r}(0)} du\;
\vf(u,H(x',u,H(x,x',y)))J_3H(x',u,H(x,x',y))\\
&=
\tilde\phi^\vf(x,y)J_3H(x',x,y),
\end{align*}
it follows
\[
|\tilde\phi^\vf(x,y)-\tilde\phi^\vf(x',H(x,x',y))|=
|\tilde\phi^\vf(x,y)|\,|1-J_3H(x',x,y)|\leq C|\vf|_\infty |x-x'|\ ,
\]
where we have used (\ref{eq:inverseH}) and the fact that the Jacobian
of the holonomy is close to the unity when the two transversal
manifolds are close, see (\ref{eq:holprop5}).
Next, given $y,y'\in\R^{d_s}$, we can define a map
$G:\R^{d_u}\to\R^{d_u}$ by the stable holonomy $H^s$ between the
unstable manifold of $(x,y)$ and of $(x,y')$. Namely, for each point
$(\xi, H(x,\xi,y))$ of the first unstable manifold we define $G$ such
that
\[
H^s(\xi, H(x,\xi,y))=(G(\xi),H(x,G(\xi),y')).
\]
The reader can look at Figure \ref{figure:holonomies} to have a quick
pictorial idea of all the quantities defined so far.
\begin{figure}[ht]\
\put(-60,0){\line(1,0){120}}
\put(-30,-20){\line(0,1){80}}
\multiput(0,-20)(0,2){40}{\line(0,1){1}}
\multiput(30,-20)(0,2){40}{\line(0,1){1}}
\qbezier(-40,16)(0,20)(20,15)
\qbezier(20,15)(40,10)(60,21)
\qbezier(-40,43)(0,40)(20,35)
\qbezier(20,35)(40,30)(60,34)
\qbezier(28,5)(33.5,30)(40,44)
\multiput(35.5,0)(0,2){17}{\line(0,1){1}}
\multiput(-30,17.6)(2,0){16}{\line(1,0){1}}
\multiput(-30,38.8)(2,0){15}{\line(1,0){1}}
\multiput(-30,13.5)(2,0){30}{\line(1,0){1}}
\multiput(-30,33)(2,0){30}{\line(1,0){1}}
\put(1,-3){$x$}
\put(31,-3){$\xi$}
\put(36.5,-3){$G(\xi)$}
\put(-33,38.5){$y'$}
\put(-33,18){$y$}
\put(-45,13){$H(x,\xi,y)$}
\put(-47,32.5){$H(x,\xi,y')$}
\put(50,13){$W^u(x,y)$}
\put(50,36){$W^u(x,y')$}
\put(35,48){$W^s(\xi,H(x,\xi,y))$}
\put(0,-22){\ }
\caption{ }\label{figure:holonomies}
\end{figure}
Accordingly,
\[
\tilde\phi^\vf(x,y')
=
\frac 1{Z_{2r}}\int_{G^{-1}(B_{r}(0))} d\xi\; \vf(G(\xi),H(x,G(\xi),y')
J_3H(x,G(\xi),y')\,JG(\xi).
\]
But $|1-JG|_\infty\leq C|y-y'|^{\tau'}$ since the holonomy is close to
unity and the distance between the two manifolds is
bounded by $|y-y'|^{\tau'}$.\footnote{This time we have an holonomy
between unstable manifolds, for this case the claimed estimate on the
Jacobian can be found also in \cite{Ma}.}
By the same reason, since the stable foliation is uniformly transversal
to the unstable one, $\|G(\xi)-\xi\|\leq C\,|y-y'|^{\tau'}$. 
Using (\ref{eq:holprop3}), (\ref{eq:holprop2}) and (\ref{eq:holprop5})
one shows that
$|J_3H(x,\xi,y)-J_3H(x,G(\xi),y')|\leq C|y-y'|^{\tau'}$.
Then it follows, for each $(x,y)\in B_r(0)\times B_r(0)$ with
$|y-y'|\leq\delta^{\frac 1{{\tau'}}}$,
\begin{align*}
|\tilde\phi^\vf(x,y)-\tilde\phi^\vf(x,y')|
&\leq
2C|\vf|_\infty|y-y'|^{\tau'}
+\frac C{Z_{2r}}\int_{B_{r}(0)}d\xi\;
|\vf(\xi,H(x,\xi,y))-\vf(H^s(\xi,H(x,\xi,y)))|\\
&\leq 2C|\vf|_\infty|y-y'|^{\tau'}+C\,H_\gamma^s(\vf)\,|y-y'|^{{\tau'}\gamma}
\leq 
3C\,|y-y'|^{\gamma{\tau'}}\ .
\end{align*}
This clearly implies that the test function  $\tilde\phi^\vf$ is
${\tau'}\gamma$-H\"older with H\"older norm bounded uniformly in 
$\vf\in\D_\gamma$.

The study of the vectors $\tilde V^\vf_i$ is completely similar apart for 
the presence of the singular kernel $\Theta_i$. To control it, we will 
use the following result.

\begin{sublem}       
\label{slem:syngker}
There exists a constant $C>0$ such that, for each $\phi\in L^\infty$,
\begin{displaymath}
\begin{split}
  &\frac 1{Z_{2r}}\left|\int_{B_{r}(0)}\Theta_i(x,u)\phi(u)\,du
      -\int_{B_{r}(0)}\Theta_i(x',u)\phi(u)\,du\right|
  \leq
  C|x-x'|\ln|x-x'|^{-1}\,|\phi|_\infty\\
  &\frac 1{Z_{2r}}\left|\int_{B_{r}(0)}\Theta_i(x,\xi)\phi(\xi)\,d\xi
    -\int_{B_{r}(0)}\Theta_i(x,G(\xi))\phi(\xi)\,d\xi\right|
  \leq
  C|y-y'|^{\tau'}\ln|y-y'|^{-1}\,|\phi|_\infty .
\end{split}
\end{displaymath}
\end{sublem}

\begin{proof}
An explicit computation yields
\begin{displaymath}
\begin{split}
  Z_{2r}^{-1}\,|\Theta_i(x,u)|
  &\leq 
  C{\|x-u\|^{-(d_u-1)}}\\
  Z_{2r}^{-1}\,|\Theta_i(x,u)-\Theta_i(x',u')|
  &\leq 
  C\,\frac{\|(x-u)-(x'-u')\|}
  {\min\{\|x-u\|^{d_u},\|x'-u'\|^{d_u}\}}
\end{split}
\end{displaymath}
for some fixed constant $C$. The estimate is then done by dividing the
integral into two parts. In the first inequality one considers
separately the integral in a ball of radius $2\|x-x'\|$ around $x$ and
the rest (where $\|x'-u\|\geq\frac12\|x-u\|$).

This yields
\[
\const\,|\phi|_\infty\,\int_0^{2\|x-x'\|} d\rho  
+\const\,|\phi|_\infty\,\int_{2\|x-x'\|}^r \rho^{-1}\,d\rho
\]
from which the wanted result follows. The other inequality is done in
the same way by separating a ball of radius $a\|y-y'\|^{\tau'}\geq
|1-G|_\infty$ around $u$ with $a$ large enough.
\end{proof}

The regularity of $\tilde V_i^\vf$ follows then trivially from the
same arguments  used to study $\tilde \phi^\vf$ and by Sub-Lemma
\ref{slem:syngker}. 
\end{proof}

\subsection{Proofs: Unstable seminorm}
\label{subsec:unorm-proof}

\begin{proof}[\bf Proof of Lemma \ref{lem:weak-unstable}]
In a sufficiently small neighbourhood $U$ we can use the   
coordinates $\Psi$ introduced in the proof of Lemma \ref{lem:objective}, 
\ie $\Psi(\xi,y)=(\xi,H(0,\xi,y))$.
Recall from Lemma~\ref{lem:changeova} that
$E^u(\Psi(\xi,y))=\{(w,D_2H(0,\xi,y)w)\,:\;w\in\R^{d_u}\}$.
We can then construct a vector field basis of the type requested in 
the lemma as  
$V_i\circ\Psi(\xi,y)=\frac\partial{\partial\xi}\Psi(\xi,y){\bf e}_i
=({\bf e}_i,D_2H(0,\xi,y){\bf e}_i)$.
Note that, according to
Lemma \ref{lem:changeova} and the regularity of the unstable
distribution stated in (\ref{eq:dist-bunching}),
the vector fields $C_0^{-1}V_i$ are in $\V_\beta$ for a suitable $C_0>0$.

Consider now some $\vf\in\Co^{\beta}(\manif,\R^{d_u})$ with
$\supp\vf\subset U$ and $|\vf|_\infty\leq1$, $H_\beta^s(\vf)\leq1$
which is $\Co^1$ when restricted to any 
unstable manifold, and let $V_\vf:=\sum_i\vf_iV_i$. 
Then $C^{-1}V_\vf\in\V_\beta$ for some $C>0$ that does not depend on $\vf$
and, denoting $\tilde U=\Psi^{-1} U$, we have
\begin{align*}
\int_\manif df(V_\vf)\,dm
&=
\sum_i\int_\manif \vf_i\, df(V_i)\,dm\\
&=\sum_i\int_{\tilde U} \vf_i(\Psi(\xi,y))\, d_{\Psi(\xi,y)}
  f(V_i(\Psi(\xi,y)))\, J\Psi(\xi,y)\, d\xi dy\\
&=\sum_i\int_{\tilde U} (\vf_i\circ\Psi)(\xi,y)\,J\Psi(\xi,y)\, 
  \frac{\partial}{\partial\xi_i}(f\circ\Psi)(\xi,y) \, d\xi dy\\
\end{align*}
In view of (\ref{eq:holprop0}) the functions $\vf_i\circ\Psi(\xi,y)$ are 
Lipschitz in the variable $\xi$, and in view of Lemma \ref{lem:changeova} 
also $J\Psi(\xi,y)$ is Lipschitz with respect to $\xi$. 
This allows to integrate by parts obtaining
\begin{equation}\label{eq:integration-by-parts}
\begin{split}
\int_\manif df(V_\vf)\,dm
&=
 -\sum_i\int_{\tilde U}\frac{\partial}{\partial\xi_i}
\big( (\vf_i\circ\Psi)(\xi,y)J\Psi(\xi,y)\big)\, f(\Psi(\xi,y))\,  d\xi dy\\
&=
-\sum_i\int_\manif f\,d\vf_i(V_i)\,dm
+\sum_i\int_\manif f\cdot\vf_i A_i\,dm
\end{split}
\end{equation}
where
\[
d_{(x,y)}\vf_i(V_i(x,y)):=
\frac{\partial}{\partial\xi_i}(\vf_i\circ\Psi)(\Psi^{-1}(x,y))
\quad\text{and}\quad
A_i:=-\left[ \frac{\frac{\partial}{\partial_{\xi_i}}J\Psi}{J\Psi}\right]\circ
\Psi^{-1}.
\]
In particular, 
\begin{equation}\label{eq:unstable-divergence}
  \sum_{i=1}^{d_u}d\vf_i(V_i)
  =
  \divg^u(\vf\circ\Psi)\circ\Psi^{-1}
  :=
  \sum_{i=1}^{d_u}\frac\partial{\partial\xi_i}
        \left(\vf_i\circ\Psi\right)\circ\Psi^{-1}\ .
\end{equation}    
In view of Lemma \ref{lem:regularity}, not only $|A_i|_\infty<\infty$, 
but also $H^{\beta'}_s(A_i)<\infty$.
\relax Finally, if $\tilde V_i$ is a different basis then
there must exist functions $\alpha_{ij}\in\Co^{{\tau'}}(\manif,\R)$, but
$\Co^1$ when restricted to any unstable manifold, such that
$V_i=\sum_j\alpha_{ij}\tilde V_j$. Then the test vector field
$v^u=\sum_i\vf_i V_i$ can be written as
$v^u=\sum_j\tilde\vf_j\tilde V_j$ with 
$\tilde\vf_j:=\sum_i\vf_i\alpha_{ij}$. Thus
\[
\sum_j d\tilde\vf_j(\tilde V_j)=\sum_{ij}\alpha_{ij}d\vf_i(\tilde
V_j)+\sum_{ij}d\alpha_{ij}(\tilde V_j)\vf_i
=\sum_{i}d\vf_i(V_j)+\sum_{ij}d\alpha_{ij}(\tilde V_j)\vf_i,
\]
which yields the same formula only with different functions $A_i$.
\end{proof}

\begin{proof}[\bf Proof of 
Proposition~\ref{prop:SRB-properties}(\ref{item:SRB-properties-d})]  
Let $h=\Pi_jf\in\Pi_j\B$. Without loss of generality we can assume that 
$f\in\Co^1(\manif,\R)$.
We must show that the measure $\mu_h$, locally\footnote{Globally, the 
$\sigma$-algebra generated by the unstable foliation is coarser than we 
intend it to be. For example, in the case of torus automorphisms, it 
contains only sets of Lebesgue measure $0$ or $1$.} conditioned to the
unstable foliation, is absolutely continuous on $\mu$-almost every fibre
with respect to the Riemannian measure on the fibre with a density of 
bounded variation. So let $U$ be a small neighbourhood on which, as in 
the last proof, we can use the change of coordinates
$\Psi$ from (\ref{eq:Psineu}), and recall that $\tilde U=\Psi^{-1}U$.
Since the maps $x\mapsto\Psi(x,y)$ for each fixed $y$ are absolutely 
continuous with Jacobians close to $1$ which are Lipschitz in the variable 
$x$ (see (\ref{eq:holprop5})), it suffices to prove the
same property for the measure $\tilde\mu_h:=\mu_h\circ\Psi$ and its 
conditional measures on the subspaces 
$N_y:=\{(x,v)\in\R^{d_u+d_s}\,:\;v=y\}$.
Hence we must show that there exists a constant $C>0$ such that
for each $\phi\in\Co^1(\R^{d_u+d_s},\R^{d_u})$
with $\supp(\phi)\subset\tilde U$, $(\xi,y)\mapsto\phi(\xi,y)$, 
\begin{equation}\label{eq:variation-to-show}
  \int_{\tilde U}\divg^u(\phi)(\xi,y)\,
  \tilde\mu_h(d\xi dy)
  \leq
  C\cdot\int_{\tilde U}|\phi|\,d\tilde\mu\ ,
\end{equation}
where $\tilde\mu:=\tilde\mu_{\Pi_11}$.
We will shortly see that this estimate implies for 
$\tilde\mu$-almost every $N_y$ 
\begin{equation}\label{eq:unstable-bounded}
  \int_{\tilde U\cap N_y}\divg^u(\phi)(\xi,y)\,
  (\tilde\mu_h)_y(d\xi)
  \leq
  C\cdot|\phi|_\infty\quad \forall \phi\in\Co^1
\end{equation}
where $(\tilde\mu_h)_y$ denotes the conditional measure
of $\tilde\mu_h$ on $N_y$. This implies the announced absolute continuity
of $(\tilde\mu_h)_y$ w.r.t. Lebesgue measure on $N_y$
with a density of bounded variation.\footnote{To see it, 
denote by $\nu$ the measure $(\tilde\mu_h)_y$ on $\R^{d_u}$.
By taking a convolution with a smooth mollifier $j_\ve$ we have
that $\nu*j_\ve=:\nu_\ve$ is absolutely continuous with respect
to Lebesgue measure $dm$ on $\R^{d_u}$. Moreover,
$\nu_\ve$ converges weakly to $\nu$ as $\ve\to 0$. In addition,
\begin{align*}
  \int\divg(\phi)\,d\nu_\ve
  &=
  \int\divg(\phi)*j_\ve\, d\nu
  =
  \int\divg(\phi*j_\ve)\,d\nu
  \leq
  C|\phi*j_\ve|_\infty
  \leq 
  C|\phi|_\infty.
\end{align*}
Hence the densities $p_\ve$ of $\nu_\ve$ with respect to Lebesgue
are uniformly bounded in bounded variation norm. Accordingly, they
form a relatively compact sequence in $L^1(\R^{d_u})$, \cite{EG}. Let
$p\in L^1(\R^{d_u})$ be an accumulation point and $\{\ve_k\}$ a
sequence such that $\lim_{k\to\infty}\|p_{\ve_k}-p\|_1=0$. 
Then $p\in\BV(\R^{d_u})$ and for each $\phi\in\Co^0(\R^{d_u})$
\[
  \int \phi p\, dm
  =
  \lim_{k\to\infty}\int\phi p_{\ve_k}\,dm
  =
  \lim_{k\to\infty}\int\phi\, d\nu_{\ve_k}
  = 
  \int \phi\,d\nu
\]
that is $p\,dm = d\nu$.}
To obtain (\ref{eq:unstable-bounded}) from (\ref{eq:variation-to-show}) 
it is convenient to introduce some 
natural notations: let $\F^u$ be the $\sigma$-algebra generated by the 
subspaces $N_y$; $\mu^u_h:={\mu_h}|_{\F^u}$ the measure restricted to the 
$\sigma$-algebra $\F^u$ (in our case this is 
simply a marginal of $\mu_h$); $\Co^1_u$ be the $\Co^1$ functions that are 
$\F^u$ measurable; and by $\E_\mu$ let us designate the 
expectation with respect to the measure $\mu$. Clearly, in this 
coordinate free language, $\int \phi\, d(\tilde\mu_h)_y$ corresponds 
to the conditional expectation $\E_{\tilde\mu_h}(\phi\;|\;\F^u)(N_y)$. 
\relax Finally, let us designate by 
$|\cdot|_p^{h,u}:=|\cdot|_{L^p(\tilde\mu_h^u)}$ the norms in the spaces 
$L^p(\tilde\mu_h^u)$.
Then
\begin{align*}
|\E_{\tilde\mu_h}(\divg^u \phi\;|\;\F^u)|_\infty^{h,u}
&=\sup_{ \psi\in L^1(\tilde\mu_h^u)\atop 
|\psi|_1^{h,u}\leq 1}\E_{\tilde\mu_h^u}(\psi\E_{\tilde\mu_h}
(\divg^u \phi\;|\;\F^u))\\
&=\sup_{\psi\in L^1({\tilde\mu_h^u})\atop 
|\psi|_1^{h,u}\leq 1}\E_{\tilde\mu_h}(\psi\divg^u \phi)\\
&=\sup_{ \psi\in \Co_u^1\atop 
|\psi|_1^{h,u}\leq 1}\E_{\tilde\mu_h}(\psi\divg^u \phi)\\
&=\sup_{ \psi\in \Co_u^1\atop 
|\psi|_1^{h,u}\leq 1}\E_{\tilde\mu_h}(\divg^u (\psi\phi))\\
&\leq C\cdot|\phi|_\infty\int_{\tilde U}|\psi|\,d\tilde\mu\ ,
\end{align*}
where the third equality is due to the fact that the $\F^u$ measurable 
functions are simply functions of only one coordinate (hence the obvious 
denseness of $\Co^1$ in the corresponding $L^1$), and the last inequality 
follows by (\ref{eq:variation-to-show}). 

This proves that for each function $\phi\in \Co^1$ there exists a set 
$A(\phi)$, $\tilde\mu(A(\phi))=0$, such that if $N_y\not\in A(\phi)$, 
(\ref{eq:unstable-bounded}) applies. Since $\Co^1$ is separable,
(\ref{eq:unstable-bounded}) follows by standard approximation arguments.

To conclude we are left with the proof of estimate 
(\ref{eq:variation-to-show}).
Let $\vf:=\phi\circ\Psi^{-1}$. Because of the continuity properties
of $\Psi$ the vector field 
$\vf$ is of the type considered in the previous proof,
and we define $V_\vf=\sum_i\vf_iV_i$ as there.
Then $\sum_i d\vf_i(V_i)\circ\Psi=\divg^u(\vf\circ\Psi)=\divg^u(\phi)$.

Since $\mu_h=\mu_{\Pi_jf}$ is defined in terms of $\Lp^kf$
(see (\ref{eq:Plambda})),
we start with the following observation: For
$v\in\V_\beta$ recall from Sub-lemma~\ref{vectorlem}
that $(R_nv)(x)=A^{-2}\lambda_u^{n}\,d_{T^nx}T^{-n}(v(T^nx))\in\V_\beta$
for $n=0,\dots,N$ where $N$ was chosen such that 
$(\sigma\cdot\min\{\lambda_u,\lambda_s^\beta\})^N>9A^2$.
Hence, applying identity (\ref{eq:scambio}) repeatedly (for $n=N$), we get
\begin{displaymath}
  \int_\manif d(\Lp^{kN}f)(V_\vf)\,dm
  =
  (A^2\lambda_u^{-N})^k\,\int_\manif df(R_N^kV_\vf)\,dm
    +\int_\manif f\cdot\sum_{l=0}^{k-1}
       (A^2\lambda_u^{-N})^l\Delta_N^{R_N^lV_\vf}\circ T^{(k-l)N}\,dm
\end{displaymath}
where $\Delta_N^v$ denotes the distortion defined in (\ref{eq:distorsione})
and where $\left|\int_\manif df(R_N^kV_\vf)\,dm\right|\leq C\cdot\|f\|_u$ 
for all $k\geq0$ because 
$C^{-1}V_\vf\in\V_\beta$ and $R_N(\V_\beta)\subset V_\beta$.
In the rest of the proof we will, for simplifying the estimate,
assume that $N=1$ and $A=1$
so that each integer $n$ is trivially a multiple of $N$
and write $R$ instead of $R_1$.\footnote{
If $N>1$ one has to use the decomposition $n=kN+j$, apply the subsequent 
estimate with $R_N$ instead of $R_1$, and treat the remaining $j$ iterates 
separately -- as we have done so often in previous sections.}
Observing (\ref{eq:Plambda}) and (\ref{eq:integration-by-parts}),
it follows that
\begin{displaymath}
\begin{split}
  &\hspace*{-1cm}\int_{\tilde U}\divg^u(\phi)(\xi,y)\,
  \tilde\mu_h(d\xi dy)
  =
  \int_\manif\sum_i d\vf_i(V_i)\,d\mu_h\\
  =&
  \limn\frac 1n\sum_{k=0}^{n-1}\lambda_j^{-k}\,
    \int_\manif\sum_id\vf_i(V_i)\,\Lp^kf\,dm\\
  =&
  \limn\frac 1n\sum_{k=0}^{n-1}\lambda_j^{-k}
  \left(-\int_\manif d(\Lp^kf)(V_\vf)\,dm
    +\int_\manif\Lp^kf\,\sum_i\vf_iA_i\,dm\right)\\
  =&
  \limn\frac 1n\sum_{k=0}^{n-1}\lambda_j^{-k}
  \left(-\int_\manif f\cdot\sum_{l=0}^{k-1}
       \lambda_u^{-l}\Delta_1^{R^lV_\vf}\circ T^{k-l}\,dm
    +\int_\manif\Lp^kf\,\sum_i\vf_iA_i\,dm\right)\\
  =&
  -\sum_{l=0}^{\infty}\left(\frac{\lambda_u^{-1}}{\lambda_j}\right)^l
  \limn\frac 1n
  \int_\manif\sum_{k=l+1}^{n-1  }\Lp^{k-l}f\cdot\Delta_1^{R^lV_\vf}\,dm
  +\limn\frac 1n\sum_{k=0}^{n-1}\lambda_j^{-k}
  \int_\manif\Lp^kf\,\sum_i\vf_iA_i\,dm\\
  =&-\int_\manif\sum_{l=0}^\infty
    \left(\frac{\lambda_u^{-1}}{\lambda_j}\right)^l\Delta_1^{R^lV_\vf}\,d\mu_h
    +\int\sum_i\vf_iA_i\,d\mu_h
\end{split}
\end{displaymath}
where, for interchanging the sum and the limit, we made use of the fact that
\begin{displaymath}
  \left|\int_\manif f\cdot
       \lambda_u^{-l}\Delta_1^{R^lV_\vf}\circ T^{k-l}\,dm\right|
  \leq
  \sigma^l\,\|f\|_\infty\,|\Delta_1^{R^lV_\vf}|_\infty
  \leq
  \sigma^l\,\|f\|_\infty\,|R^lV_\vf|_\infty
  \leq
  C\,\sigma^l\,\|f\|_\infty\ .
\end{displaymath}
Since 
$|\Delta_1^{R^lV_\vf(x)}|\leq\const\cdot|V_\vf(T^lx)|
\leq\const\cdot|\vf(T^lx)|$
with a constant that depends only on the Jacobian $g$ of $T$ and on the 
vector fields $V_i$, it follows that the first integral is bounded by
$\frac{\const}{1-\lambda_u^{-1}}\,\int|\vf|\,d\mu
\leq\const\cdot\int|\phi|\,d\tilde\mu$.
(Observe that $\mu_h$ has bounded density w.r.t. $\mu$ and that
$\mu$ is $T$-invariant.)
\relax For the second integral the corresponding bound follows from
the boundedness of the $A_i$.
This proves estimate (\ref{eq:variation-to-show}) and concludes the proof
of Proposition~\ref{prop:SRB-properties}(\ref{item:SRB-properties-d}).
\end{proof}   

\begin{proof}[\bf Proof of Lemma~\ref{lem:crucial}]
As usual it suffices to consider $f\in\Co^1(\manif,\R)$. By definition
there are $g_n\in\B$ such that 
\begin{displaymath}
  \|f-g_n\|_w\to0\quad\text{and}\quad \limsup_{n\to\infty}\|g_n\|
  \leq\newnorm{f}\ .
\end{displaymath}
Since $\Co^1(\manif,\R)$ is dense in $\B$, we may assume that 
$g_n\in\Co^1(\manif)$.

Our first goal is to estimate 
$\|f\|_s=\sup_{\vf\in\D_\beta}\int f\,\vf\,dm$ in terms of $\|g_n\|_s$.
To this end we approximate a given $\vf\in\D_\beta$ by functions 
$\tilde\vf_\rho$, $\frac12\tilde\vf_\rho\in\D_\beta$, $\rho>0$,
which are such that $H_\gamma^s(\tilde\vf_\rho)\leq K_\rho$ for some 
$K_\rho>0$ and 
$|\vf-\tilde\vf_\rho|_\infty\leq \rho^\beta$.\footnote{Such $\tilde\vf_\rho$
are constructed similarly as in Sub-lemma~\ref{sublem:phi-approx}, 
except that the construction was carried out for a fixed $\rho$.
Again we can carry out the construction leafwise, which now means that we
can deal with each one-dimensional stable fibre $W^s(x)$ separately.
We parametrise it by arc-length and define
$\tilde\vf_\rho(x):=(2\rho)^{-1}\int_{W^s_\rho(x)}\vf(t)\,dt$,
where $W^s_\rho(x):=\{y\in W^s(x):\,d^s(x,y)<\rho\}$.
Then $\tilde\vf_\rho$ is differentiable along each $W^s(x)$, so
$H_\gamma^s(\tilde\vf_\rho)\leq K_\rho$ for some $K_\rho>0$, 
$|\vf-\tilde\vf_\rho|_\infty\leq (2\rho)^\beta\,H_\beta^s(\vf)
\leq (2\rho)^\beta$, and $\frac 12\tilde\vf_\rho\in\D_\beta$.}
Hence
\begin{displaymath}
\begin{split}
  \int f\,\vf\,dm
  &=
  \int(f-g_n)\,\tilde\vf_\rho\,dm
   +\int g_n\,\tilde\vf_\rho\,dm
   +\int f\,(\vf-\tilde\vf_\rho)\,dm\\
  &\leq
  \|f-g_n\|_w\,K_\rho+2\|g_n\|_s+\|f\|_\infty\,\rho^\beta
\end{split}
\end{displaymath}
so that, letting first $n\to\infty$ and then $\rho\to0$, we see that
$\|f\|_s\leq\liminf_{n\to\infty}2\|g_n\|_s$.

Next we have to estimate $\|f\|_u=\sup_{v\in\V_\beta}\int df(v)\,dm$
in terms of $\|g_n\|_u$. 
We start with $v\in\V_\beta\cap\Co^{1+\alpha}(\manif)$ so that
$C_v^{-1}\,\divg(v)\in\D_\gamma$ for some $C_v>0$. Then
\begin{displaymath}
  \int df(v)\,dm
  =
  -\int(f-g_n)\,\divg(v)\,dm+\int dg_n(v)\,dm
  \leq
  \|f-g_n\|_w\,C_v+\|g_n\|_u\ ,
\end{displaymath}
from which $\int df(v)\,dm\leq\liminf_{n\to\infty}\|g_n\|_u$.
If we can extend this estimate to all $v\in\V_\beta$, it follows that
$\|f\|_u\leq\liminf_{n\to\infty}\|g_n\|_u$ and hence
$\|f\|\leq\liminf_{n\to\infty}\|g_n\|\leq2\newnorm{f}$.
As we will have to work in local coordinates we end up with a weaker
estimate $\|\cdot\|\leq C\,\newnorm{\cdot}$ where $C$ is a 
constant determined by
a suitable smooth partition of unity, \cf Lemma~\ref{lem:multiplication}
and the remarks thereafter. Since we are in dimension $d=2$, there is a 
local $\Co^{1+\alpha}$ change of coordinates that straightens both 
foliations simultaneously. Therefore it suffices to show that each vector 
field $v(\xi,\eta)=\vf(\xi,\eta)\,({1\atop0})$ with $\vf\in\D_\beta$ can 
be approximated in the $L^1_m(\To^2)$-sense by $\Co^{1+\alpha}$-vector 
fields $\tilde\vf(\xi,\eta)\,({1\atop0})$ with $\tilde\vf\in\D_\beta$.
But this can be done by smoothening $\vf$ with convolution kernels
in both coordinate directions independently.
\end{proof}
We remark that in the case of linear automorphisms there is no need
to pass to local coordinates in this proof. 
Moreover, instead of $\frac12\tilde\vf\in\D_\beta$ the same
proof yields $\tilde\vf\in\D_\beta$ (in the estimate of $\|f\|_s$).
Hence $\|f\|=\newnorm{f}$
for all $f\in\B$ in this case. Furthermore, also
automorphisms of higher dimensional tori can be treated in the same
way, because the averaging procedures that produce 
$\tilde\vf_\rho$ and the $C^{1+\alpha}$-approximations to $v$ can 
can be easily adapted.

\appendix
\section{On splitting and foliation regularity}
\setcounter{equation}{0}
\renewcommand\theequation{\thesection.\arabic{equation}}

In this appendix we collect, for the reader's convenience, some
information on the regularity properties of the invariant foliations
in Anosov systems that are used in the paper.

First of all, as already mentioned, for $\Co^2$ Anosov diffeomorphisms the
invariant distributions (sometimes called {\em splittings}) are known
to be uniformly H\"older continuous. Let us be more precise. 

For each invertible linear map $L$ let $\theta(L):=\|L^{-1}\|^{-1}$.
We define $\|d^s_xT\|=\|d_xT|_{E^s_x}\|$,
$\|d^u_xT\|=\|d_xT|_{E^u_x}\|$, $\theta(d^s_xT)=\theta(d_xT|_{E^s_x})$ and
$\theta(d^u_xT)=\theta(d_uT|_{E^u_x})$.  
Then the following holds (\cite{KH}, \cite{Hasselblatt})\footnote{In
terms of the expansion rates used in the paper we have: $\|d^sT\|\sim
\lambda_s$; $\|d^uT\|\sim \mu_u$; $\theta(d^sT)\sim \mu_s$ and
$\theta(d^uT)\sim \lambda^u$.}
\begin{equation}\label{eq:dist-bunching}
\begin{array}l
\bullet\quad\hbox{ If there exists $\tau>0$ such that, for each $x\in\manif$,
$\|d^s_xT\|\,\|d^u_xT\|^\tau<\theta(d^u_xT)$,}\\
\quad\quad\hbox{then $E^s\in\Co^\tau$.}\\
\bullet\quad\hbox{ If there exists $\tau>0$ such that, for each $x\in\manif$,
$\theta(d^s_xT)^\tau\, \theta(d^u_xT)>\|d^s_xT\|$,}\\
\quad\quad\hbox{then $E^u\in\Co^\tau$.}
\end{array}\hfill
\end{equation}

Moreover the H\"older continuity is uniform (that is the $\tau$-H\"older 
norm of the distributions is bounded). The above conditions are often 
called {\em $\tau$-pinching} or {\em bunching} conditions.

\begin{arem}\label{rem:bunching}
Note that since the map is Anosov and $\Co^2$ we have 
$\sup\|d^s_xT\|<1$, $\inf \theta(d^u_xT)>1$ and $ \sup\|d^u_xT\|<\infty$.
Hence there always exists $\tau>0$ for which the bunching conditions are 
satisfied, from which the announced H\"older continuity follows. Moreover, 
if the stable foliation is one dimensional then $\|d^s_xT\|=\theta(d^s_xT)$,
hence the corresponding bunching condition holds for some $\tau>1$,
that is $E^s\in\Co^{1+\alpha}$; the same consideration applies to the
unstable foliation.
\end{arem}

The next relevant fact is that the above splittings are
integrable. The integral manifolds are the stable and unstable
manifolds, respectively. They form invariant continuous foliations.
Each leaf of such foliations is as smooth as the map and it is
tangent, at each point, to the corresponding distribution, \cite{KH}. In
addition, for $\Co^r$ maps, the $\Co^r$ derivatives of such manifolds
(viewed as graphs over the corresponding distributions) are uniformly
bounded, \cite{HPS}. In our case this implies that the curvature is uniformly
bounded. Finally, the foliations are uniformly transversal and
regular. For our purposes we need to detail these last two properties
more precisely.

\begin{alem}[\cite{KH}, Proposition 6.4.13]\label{lem:product}
Denote by $W^s_\epsilon(x)$ and $W^u_\epsilon(x)$ the $\epsilon$-ball
in the stable and unstable manifolds, respectively, centred at
$x$. There exists and $\epsilon>0$ such that for any $x,y\in\manif$
the intersection $W^u_\epsilon(x)\cap W^s_\epsilon(y)$ consists of at
most one point that will be called $[x,y]$. In addition, there
exists a $\delta>0$ such that whenever $d(x,y)<\delta$, then
$W^u_\epsilon(x)\cap W^s_\epsilon(y)\neq \emptyset$.
\end{alem}

In general it is also true that $[\cdot, \cdot]$ is continuous where it
is defined. The above property is
sometimes called {\em product structure}. Given $x\in\manif$, if we
consider any smooth chart of its stable and unstable manifolds and,
with an harmless abuse of notations, identify them with $\R^{d^s}$
and $\R^{d^u}$, we can define, in a $\delta$--neighbourhood of $x$, the
map $\Gamma(y)=([y,x],[x,y])$. This is the map we use in section
\ref{subsec:smooth-random-proof} and it has the obvious property of
transforming, locally, both foliations into a collection of parallel
hyperplanes. By the above comments the map $\Gamma$ is continuous but
we need to know much more about its regularity. 

In the case in which both distribution are $\Co^{1+\alpha}$, it
follows by Frobenius' theorem that the map
$\Gamma$ is $\Co^{1+\alpha}$ (see Lemma
\ref{lem:regularity-lemma} or section six of \cite{PSW}). If the
splitting is only H\"older the situation is more subtle. 

Probably for historical reasons the regularity of the foliations is
usually stated in terms of the corresponding {\em holonomy maps}. Given
a foliation $\mathcal F$ and two transversal manifolds $W_1,W_2$, the
associated holonomy $H_{W_1,W_2}:W_1\to W_2$ is defined by
$\{H_{W_1,W_2}(x)\}:=F(x)\cap W_2$, where $F(x)$ is the fibre of the
foliation that contains $x$. We will call {\em stable holonomy} any
holonomy constructed via the stable foliations and {\em unstable
holonomy} holonomies constructed by the unstable foliation. The basic
result on holonomies is given by the following \cite{PSW}.
\begin{equation}\label{eq:fol-bunching}
\begin{array}l
\bullet\quad \hbox{ If there exists $\tau>0$ such that, for each $x\in\manif$,
$\|d^s_xT\|\,\|d^u_xT\|^\tau<\theta(d^u_xT)$,}\\
\quad\quad\hbox{then the stable holonomies are
uniformly $\Co^\tau$.}\\
\bullet\quad \hbox{ If there exists $\tau>0$ such that, for each $x\in\manif$,
$\theta(d^s_xT)^\tau\, \theta(d^u_xT)>\|d^s_xT\|$,}\\
\quad\quad\hbox{then the unstable holonomies are
uniformly $\Co^\tau$.}
\end{array}
\end{equation}

Thus the same considerations of Remark \ref{rem:bunching} apply to
the holonomies.

The relation between regularity of holonomies and regularity of the
foliation (in the sense that the local foliation charts are regular) is
discussed in detail in \cite[section 6]{PSW}. Here we restrict
ourselves to what is needed in the present paper.

Consider any $\bar x\in \manif$ and a sufficiently small neighbourhood $U$ 
such that Lemma \ref{lem:product} applies. Next choose a smooth chart and
use it to identify all the tangent spaces with the tangent space of
$\bar x$. We will not distinguish between points 
on the manifold and corresponding
points in the chart since this causes no ambiguity, hence from now on
we will identify $\bar x$ with $0$.  We can also
assume, without loss of generality that the hyperplanes
$\Sigma_x:=\{(x,\xi)\}$ are uniformly transversal to the unstable
foliation. Then, for each two points $(x,0), (z,0)\in U$, we can
introduce the unstable holonomies $H(x,z,\cdot):\Sigma_x\to\Sigma_z$
such that $(z,H(x,z,y))$ is the intersection of the unstable manifold
of $(x,y)$ with the hyperplane $\Sigma_z$. Now (\ref{eq:fol-bunching})
asserts that all the maps $H(x,z,\cdot)$ are uniformly
$\tau$-H\"older.

This it is not yet enough for our purposes: we need to talk about the
Jacobian of the functions $\Gamma$ and $H$. Since we use $\Gamma$ only
in the case where the foliations are $\Co^{1+\alpha}$, no
difficulty arises in this case, 
in fact the regularity of the holonomies implies
that $\Gamma$ is $\Co^{1+\alpha}$ in each variable and then it is
$\Co^{1+\alpha}$ by \cite{Journe}.\footnote{In fact, the hypothesis of
higher regularity is not really necessary since it is
true that the Jacobian of $\Gamma$ always exists and is H\"older,
\cite{BL}.} The situation for $H$ is in principle less clear since a
H\"older function may very well fail to have a Jacobian. Nevertheless,
in our case $H$ has a Jacobian and such a Jacobian is itself a
H\"older function, \cite{Ma}.

The last information that we need is that the Jacobian of holonomies
between close manifolds are close to one, they are differentiable with
respect to $z$ and the derivative is a regular test function.
This last properties 
should be obvious to experts, yet we could not locate a clear cut 
reference applying directly to our case (typically unstable 
holonomies between stable manifolds are considered). Here is the 
sketch of a simple proof. Remember the change of coordinates
$\Psi(\xi,\eta)=(\xi, H(0,\xi,\eta)$ introduced in section
\ref{subsec:compactness-proof}. 

\begin{alem}\label{lem:regularity} 
There exists a constant $C>0$ such that the Jacobian $J(x,z,\cdot)$ of
$H(x,z,\cdot)$ satisfies the property
\[
|1-J(x,z,\cdot)|_\infty\leq C|x-z|.
\]
In addition there exists $\beta'\in(0,\beta)$ such that, 
calling $A_i\circ\Psi(z,y):=\partial_{z_i}\ln J(0,z,y)$,
\begin{displaymath}
  H^s_{\beta'}(A_i)<\infty\ .
\end{displaymath}
\end{alem}
\begin{proof}
Considering instead of $T$ a sufficiently high iterate, we may
assume that the constant $A=1$ in (\ref{eq:hyperbolicity}), \ie that
$T$ expands and contracts strictly along the unstable resp.\ stable
leaves.\footnote{Note just that all powers of $T$ have the same stable and 
unstable foliations.}
Let $\Sigma_x=\{(x,y)\}_{y\in B_r(0)}$. As already mentioned
$J(x,z,\cdot)$ is just the Jacobian of the holonomy between $\Sigma_x$
and $\Sigma_z$. Calling $E_x$  the tangent space to $\Sigma_x$, and
setting $y':=H(x,z,y)$, the 
following formula holds, \cite{Ma}, 
\begin{equation}\label{eq:jacformula}
\begin{aligned}
  J(x,z,y)
  =&
  \prod_{n=1}^\infty\frac{\det(D_{\bar x_n}T|_{(D_{\bar x_0}T^{-n})E_x})}
       {\det(D_{\bar z_n}T|_{(D_{\bar z_0}T^{-n})E_z})}\\
  =&
  \exp\left(\sum_{n=1}^\infty
    \ln(\det(D_{\bar x_n}T|_{(D_{\bar x_0}T^{-n})E_x}))
    -\ln(\det(D_{\bar z_n}T|_{(D_{\bar z_0}T^{-n})E_z}))\right),
\end{aligned}
\end{equation}
where, for convenience, we have set $\bar x_n:=T^{-n}(x,y)$ and $\bar
z_n:=T^{-n}(z,y')$. 

Since $(x,y)$ and $(z,y')$ belong to the same unstable manifold it
follows that $d(\bar x_n,\bar z_n)\leq\lambda_u^{-n}$. 
Thus the problem of
estimating the size of the Jacobian is reduced to the 
one of understanding the ``distance'' between the tangent spaces
$(D_{\bar x_0}T^{-n})E_x$ and $(D_{\bar z_0}T^{-n})E_z$. This requires a 
little preparation. First of all, as usual in the construction of stable and
unstable manifolds, it is convenient to choose appropriate coordinates
in the tangent spaces at the points $\bar x_n$. We will choose
coordinates such that $\{(\xi,0):\,\xi\in\R^{d_u}\}$ 
is the unstable subspace and
$\{(0,\eta):\,\eta\in\R^{d_s}\}$ the stable one. 
Clearly, by using such adapted coordinates, the maps $D_{\bar x_n}T$ are 
all represented by matrices of the form
\[
\begin{pmatrix}
A_n&0\\
0&D_n
\end{pmatrix}
\]
where $\|A_n^{-1}\|\leq \lambda_u^{-1}$ and $\|D_n\|\leq \lambda_s$. Since
we want to compare nearby matrices it is convenient to use coordinates 
related to $(x,y)$ also for the quantities associated to $(z,y')$.
To do so just use the exponential map to extend the above coordinates
in the tangent space of $\bar x_n$ to coordinates in a
neighbourhood of $\bar x_n$. Clearly, such neighbourhoods can be
chosen all of uniform size, hence if $d(x,z)$ is small enough,
$\bar z_n$ always belongs to the corresponding
neighbourhood. Since the above coordinates are uniformly smooth and
the map $T$ is $\Co^2$ it follows that the derivatives $D_{\bar z_n}T$
are all represented by matrices of the form
\begin{equation}\label{eq:super-final}
  \begin{pmatrix}
    A_n'&B'_n\\
    C'_n&D_n'
  \end{pmatrix}
  :=
  \begin{pmatrix}
    A_n&0\\
    0&D_n
  \end{pmatrix}
  +\Lambda_n
\end{equation}
where $\|\Lambda_n\|\leq c d(\bar x_n,\bar z_n)$.

Using the above coordinates the spaces $E_x^n(y):=(D_{\bar x_0}T^{-n})E_x$ 
and $E_z^n(y):=(D_{\bar z_0}T^{-n})E_z$ can be represented by $d^s\times d^s$
matrices, namely $E_x^n(y)=\{(U_n(y)\eta,\eta):\,\eta\in\R^{d_s}\}$ and
$E_z^n=\{(U_n'(y)\eta,\eta):\,\eta\in\R^{d_s}\}$.
The basic estimate on the above subspaces is given in the next
lemma the proof of which is postponed to the end of the appendix.
\begin{asublem}\label{sublem:last}
There exists a constant $c>0$ such that
\[
\|U_n(y)-U'_n(y)\|\leq c\lambda_u^{-n}|x-z|
\]
\end{asublem}

By this sub-lemma it follows that there exists a 
constant $c_1>0$ such that 
\[ 
\left|\ln[\det(D_{\bar x_n}T|_{E^n_x(y)})]-
\ln[\det(D_{\bar z_n}T|_{E^n_z(y)})]\right|\leq
c_1 \lambda_u^{-n}|x-z|,  
\] 
from which the first statement of Lemma~\ref{lem:regularity} follows.

In addition, since each term of the sum in (\ref{eq:jacformula}) is
clearly $\Co^1$, the above estimate on the Lip\-schitz constant is
indeed an estimate on the corresponding derivative. In order to show 
$\beta'$-H\"older continuity of this derivative we note first the formula
\begin{equation}\label{eq:jacderiv}
\begin{aligned}
  \partial_{z_i}\ln J(x,z,y)
  =&
  -\sum_{n=1}^\infty
  \partial_\Lambda\ln(\det(\Lambda))|_{\Lambda=D_{\bar z_n}T|_{E^n_z(y)}}
  \cdot\partial_{z_i}\left(D_{\bar z_n}T|_{E^n_z(y)}\right)\\
  =&
  -\sum_{n=1}^\infty
  \partial_\Lambda\ln(\det(\Lambda))|_{\Lambda=D_{\bar z_n}T|_{E^n_z(y)}}
  \cdot\sum_j\left[(\partial_{\xi_j}D_\xi T|_{E^n_z(y)})|_{\xi=\bar z_n}
  \frac{\partial z^n_j}{\partial z_i}\right.\\
  &\left.
  \ +\partial_V(D_{\bar z_n}T|_{E_V})|_{V=U_n(y)}\cdot 
   \frac{\partial U_n'(y)}{\partial z_i}\right] 
\end{aligned}
\end{equation}
where $E_V:=\{(V\eta,\eta):\,\eta\in\R^{d_s}\}$.
Since the maps $D_{\bar z_n}T|_{E^n_z(y)}$ are all uniformly invertible 
(because of hyperbolicity), all the derivatives appearing in the above 
formula are well defined. Varying $z$ the points $\bar z_0$ move along 
the unstable manifold, by definition. It follows that 
\[
\left|\frac{\partial z^n_j}{\partial z_i}\right|\leq c\lambda_u^{-n}
\]
and, together with Sub-Lemma \ref{sublem:last}, this implies that the
above series converges exponentially fast.

Next, let $x=0$, $\bar z:=\Psi(z,y)$, $\bar \zeta:=\Psi(\xi, \eta)$
and, as before, $y':=H(0,z,y)$, $\eta':=H(0,\xi,\eta)$. Accordingly
\[
  (|z-\xi|+|y'-\eta'|)\leq c\,d(\bar z,\bar\zeta)\ .
\]
Now each term in (\ref{eq:jacderiv}) is clearly Lipschitz with respect
to the variables $z, y'$ with Lipschitz constant bounded by $\mu_s^{-n}$.
It follows\footnote{The following estimate can be improved, but since
we are not able
to strengthen it sufficiently to get $\beta'=\tau'$ in the lemma,
we content ourselves with what follows.}
\[
\begin{aligned}
  |\partial_{z_i}\ln J(0,z,y)-\partial_{z_i}\ln J(0,\xi,\eta)|
  &\leq
  \sum_{i=1}^{\bar n}C
  \mu_s^{-i}(|y'-\eta'|+|z-\xi|)+C\lambda_u^{-\bar n}\\
  &\leq 
  c\,d(\bar z,\bar\zeta)^{\frac{\ln\lambda_u}{\ln\mu_s^{-1}\lambda_u}}
  =:c\,d(\bar z,\bar\zeta)^{\beta'} , 
\end{aligned}
\]
where we have chosen $\bar n$ appropriately.
\end{proof}
\begin{proof}[\bf Proof of Sub-Lemma \ref{sublem:last}]
To start we need to obtain a recursive equation for the $U_n$. 
Since
\[
\begin{pmatrix} A&B\\C&D\end{pmatrix}
\begin{pmatrix}U\eta\\ \eta\end{pmatrix}=
\begin{pmatrix}(AU+B)\eta\\ (CU+D)\eta\end{pmatrix}
\]
it follows that the space $\{(U\eta,\eta):\,\eta\in\R^{d_s}\}$ 
is sent into the space
$\{(\tilde U\eta,\,\eta):\,\eta\in\R^{d_s}\}$ with 
$\tilde U:=(AU+B)(CU+D)^{-1}$, \ie
$U=(A-\tilde UC)^{-1}(\tilde UD-B)$. Accordingly,
\[
\begin{aligned}
U_{n}&=A_n^{-1}U_{n-1}D_n\\
U_{n}'&=(A_n'-U_{n-1}'C_n')^{-1}(U_{n-1}'D_n'-B_n')
\end{aligned}
\]
where $U_0=U'_0=0$.
\relax From this follows immediately (recall that we assumed $A=1$)
\[
\|U_n\|\leq \lambda_u^{-n}\lambda_s^n,
\]
and, observing (\ref{eq:super-final}),
\[
\begin{aligned}
  \|U_n-U_n'\|
  &\leq 
  \|A_n^{-1}(U_{n-1}-U_{n-1}')D_n\|+c\, d(\bar x_n,\bar z_n)\\
  &\leq 
  \lambda^{-1}_u\lambda_s\|U_{n-1}-U'_{n-1}\|+c\, d(\bar x_n,\bar z_n)\\
  &\leq c\sum_{i=0}^{n-1}(\lambda_u^{-1}\lambda_s)^i\,
  d(\bar x_{n-i},\,\bar z_{n-i})\\
  &\leq 
  \const\cdot\lambda_u^{-n}|x-z|.
\end{aligned}
\]
\end{proof}



\begin{thebibliography}{99}

\bibitem{Bachtin} V.I. Bakhtin, {\em A direct method of constructing
an invariant measure on hyperbolic attractor}, Izv. AN SSSR, ser.
mat., {\bf 56}:5 (1992), 934-957.

\bibitem{Babook} V. Baladi, {\em Positive Transfer Operators and Decay of
Correlations}, Advanced Series in Nonlinear Dynamics, {\bf 16}, World
Scientific, Singapore (2000).

\bibitem{BaYo} V. Baladi, L.-S. Young, {\em On the spectra of randomly
perturbed expanding maps}, Comm. Math. Phys., {\bf 156}:2 (1993),
355-385; {\bf 166}:1 (1994), 219--220.

\bibitem{BlKe} M. Blank, G. Keller, {\em Random perturbations of
chaotic dynamical systems: stability of the spectrum},
Nonlinearity, {\bf 11} (1998), 1351-1364.

\bibitem{Blank} M. Blank, {\em Discreteness and continuity in problems of
chaotic dynamics}, Amer. Math. Soc., Providence, Rhode Island (1997).

\bibitem{BL} X. Bressaud, C. Liverani, {\em Anosov diffeomorphisms and
coupling},  Ergod. Th. \& Dynam. Sys., to appear.

\bibitem{Boo} W.M. Boothby, {\em An Introduction to Differential manifolds
and Riemannian Geometry}, Academic Press, New York, San Francisco, 
London (1975).

\bibitem{BSTV} F. Brini, S. Siboni, G. Turchetti, S. Vaienti,
{\em Decay of correlations for the automorphisms of the torus $\To^2$},
Nonlinearity, {\bf 10} (1997), 1257-1268.

\bibitem{BuK} J. Buzzi, G. Keller, {\em Zeta functions and Transfer 
operators for multidimensional piecewise affine and expanding maps}, 
Ergod. Th. \& Dynam. Sys., to appear.

\bibitem{FKS} I.P. Cornfeld, S.V. Fomin, Ya.G. Sinai, {\em Ergodic Theory},
Grundlehren der Mathem. Wiss. 245, Springer-Verlag New York, Heidelberg, 
Berlin (1982).

\bibitem{DJ1} M. Dellnitz, O. Junge, {\em On the approximation of 
complicated dynamical behavior}, SIAM J.~Numerical Analysis, 
{\bf 36} (1999), 491-515.

\bibitem{DJ2} M. Dellnitz, O. Junge, {\em Set Oriented Numerical Methods 
for Dynamical Systems}, B. Fiedler, G. Iooss and N. Kopell (eds.): 
Handbook of Dynamical Systems III: Towards Applications, World Scientific 
(2001), to appear.

\bibitem{DoCa} M.P. Do Carmo, {\em Riemannian Geometry,} Birkh\"auser, 
Boston (1992).

\bibitem{DS} N. Dunford, J.T. Schwartz, {\em Linear Operators, Part I},
Wiley, New York (1957).

\bibitem{EG} L.E. Evans, R.F. Gariepy, {\em Measure Theory and Fine
Properties of Functions}, CRC Press, Boca Raton, New York, London,
Tokyo (1992).

\bibitem{Fe} H.Federer, {\em Geometric Measure Theory}, Springer-Verlag,
Berlin, Heidelberg (1969).

\bibitem{Froyland1} G. Froyland, {\em Computer-assisted bounds for the rate 
of decay of correlations}, Commun. Math. Phys., {\bf 189} (1997), 237-257.

\bibitem{Froyland2} G. Froyland, {\em Extracting dynamical behaviour via 
Markov models}, Alistair Mees (eds.): ``Nonlinear Dynamics and Statistics: 
Proceedings, Newton Institute, Cambridge, 1998'', 283-324, Birkh\"auser (2000).

\bibitem{Hasselblatt} B. Hasselblatt, {\em Regularity of the Anosov
splitting and of horospheric foliations}, Ergod. Th.\& Dynam. Sys., 
{\bf 14} (1994), 645-666.

\bibitem{He} H. Hennion, {\em Sur un th\'eor\`eme spectral et son
application aux noyaux Lipchitziens}, Proceedings of the American
Mathematical Society, {\bf 118} (1993), 627--634.

\bibitem{HPS} M. Hirsch, C. Pugh, M. Shub, {\em Invariant Manifolds},
Lecture Notes in Math., {\bf 583} (1977).

\bibitem{ITM} C.T. Ionescu-Tulcea, G. Marinescu, {\em Th\'eorie
ergodique pour des classes d'op\'erations non compl\`etement continues}, 
Ann. Math., {\bf 52} (1950), 140-147.

\bibitem{Journe} J.-L. Journ\'e, {\em A regularity lemma for functions
of several variables}, Rev. Mat. Iberoamericana, {\bf 4} (1988), 187-193.

\bibitem{Junge} O. Junge, {\em Mengenorientierte Methoden zur numerischen 
Analyse dynamischer Systeme}, PhD thesis, University of Paderborn, 
Paderborn (1999).

\bibitem{KH} A. Katok, B. Hasselblatt, {\em Introduction to the Modern
Theory of Dynamical Systems}, Encyclopedia of Mathematics and its
Applications, {\bf 54}, G.-C.Rota editor, Cambridge University Press,
Cambridge (1995).

\bibitem{Keller1} G. Keller, {\em Stochastic stability in some chaotic 
dynamical systems}, Monatsh. Math., {\bf 94} (1982), 313-333.

\bibitem{KL} G. Keller, C. Liverani, {\em Stability of the spectrum for
transfer operators}, Annali della Scuola Normale Superiore di Pisa,
Scienze Fisiche e Matematiche, (4) {\bf XXVIII} (1999), 141-152.

\bibitem{Kifer1} Yu. Kifer, {\em Random Perturbations of Dynamical Systems},
Progress in Probability and Statistics 16, Birkh\"auser, Boston (1988).

\bibitem{Kifer2} Yu. Kifer, {\em Computations in dynamical systems via 
random perturbations}, Discr. Cont. Dynam. Sys., {\bf 3} (1997), 457-476.

\bibitem{Kitaev} A.Yu. Kitaev, {\em Fredholm determinants for hyperbolic
diffeomorphisms of finite smoothness}, Nonlinearity, {\bf 12} (1999), 141-179.

\bibitem{Li} C. Liverani, {\em Decay of Correlations}, Annals of Mathematics, 
{\bf 142} (1995), 239-301.

\bibitem{Liverani-paderborn} C. Liverani, {\em Rigorous numerical 
investigation of the statistical properties of piecewise expanding 
maps -- A feasibility study}, Nonlinearity, to appear.

\bibitem{Ma} R. Ma\~ne, {\em Ergodic Theory and Differentiable Dynamics},
Springer-Verlag, Berlin Heidelberg (1987).

\bibitem{Nussbaum} R.D. Nussbaum, {\em The radius of the essential spectrum},
Duke Math. J., {\bf 37} (1970), 473-478.

\bibitem{PSW} C. Pugh, M. Shub, A. Wilkinson, {\em H\"older Foliations}, 
Duke Math. J., {\bf 86} (1997), 517-546.

\bibitem{Ru}  W. Rudin, {\em Real and Complex Analysis}, Mc Graw-Hill
series in higher Mathematics, Second Edition (1974).

\bibitem{Rugh} H.H. Rugh, {\em Generalized Fredholm determinants and
Selberg zeta functions for Axiom A dynamical systems}, Ergod.
Th.\& Dynam. Sys., {\bf 16} (1996), 805-819.

\end{thebibliography}
\end{document}